\documentclass[12pt]{article}
\usepackage[utf8]{inputenc}
\usepackage[export]{adjustbox}
\usepackage{authblk}
\usepackage{bbm}
\usepackage{pgfpages}
\usepackage{graphicx}
\usepackage{multicol}
\usepackage{csquotes}
\usepackage{multirow}
\usepackage{graphicx} 
\usepackage{booktabs} 
\usepackage{authblk}
\usepackage{caption}
\usepackage{subcaption}
\usepackage{stackengine}
\usepackage{amsmath}

\usepackage[maxfloats=256]{morefloats}

\usepackage{comment}
\usepackage{lscape}
\usepackage{hyperref}
\hypersetup{
    colorlinks=true,
    linkcolor=blue,
    filecolor=blue,      
    urlcolor=blue,
    citecolor=blue,
}
\usepackage[capposition=top]{floatrow}

\usepackage[spanish,english]{babel}

\usepackage{amssymb}
\usepackage{stackrel}
\usepackage{natbib} % bibliography

\usepackage[a4paper, margin=0.85in]{geometry}

\title{\Large \textbf{Global Banks' Spillovers to Emerging Markets} \\ \Large Macro to Micro Transmission}

\author[1]{Luis Rodrigo Arnabal}

\author[2]{Santiago Camara}

\author[3]{Cecilia Dassatti}

\affil[1]{Banco Central del Uruguay}

\affil[2]{McGill University \& Red-NIE}

\affil[3]{Banco Central del Uruguay}

\begin{document}
    
\maketitle

\date{}
\maketitle
\begin{center}
%First version: November 1st. This version: \today
\normalsize
This version: \today. For the latest version \href{https://www.dropbox.com/scl/fi/n9ty5xda9zoii0isc936c/Latest_Manuscript.pdf?rlkey=6n3ujq7c9zs27e52qloyig8g4&st=m2tq9kxs&dl=0}{Click here!}
\end{center}

\begin{abstract}
This paper studies how shocks to global banks’ net worth transmit to Emerging Market Economies. Using the identification strategy of \cite{ottonello2022financial}, which isolates high-frequency surprises to banks’ credit-supply capacity, we show that positive shocks appreciate local currencies, lower external borrowing costs, increase capital flows to domestic banking sectors, and raise investment, credit, and real activity across EMEs. These effects are highly robust across specifications and samples. Using administrative credit-registry data from Uruguay, we find that better-capitalized banks transmit global credit easing more strongly. At the firm level, responses are weaker for more leveraged firms, especially those with foreign-currency debt, short maturities, or collateral not priced to market.

\bigskip

\noindent   
    \textbf{Keywords:} Global Banks; Financial intermediaries;  Emerging markets; International Spillovers; Exchange Rate; Firm Transmission; 

    \medskip
    
    \medskip

    \noindent
    \textit{JEL Codes:} F40, F41, E44, E51.
\end{abstract}

%%%%%%%%%%%%%%%%%%%%%%%%%%%%%%%%%%%%%%%%%%%%%%%%%%%%%%%%%%%%%%%%%%%%%%%%%%%%%%%%%%%%%%%%%%%%%%%%%%%%%%%%%%%%%%%%%%%%%%%%%%%%%%%%%%%%%%%%%%%%
%% Introduction
\newpage
\section{Introduction}

Understanding how financial shocks originating in advanced economies affect emerging market economies is a central question in international macroeconomics. A large literature argues that global financial intermediaries play an important role in transmitting fluctuations in global financial conditions across borders, including through leverage cycles and changes in risk bearing capacity as emphasized by \cite{rey2015dilemma}, \cite{bruno2015cross}, and \cite{miranda2020us}. Yet, direct evidence on how changes in global banks' balance sheets influence emerging-market credit and real activity remains limited.

\begin{figure}[t!]
    \centering

    \begin{subfigure}[b]{0.55\textwidth}
        \centering
        \includegraphics[width=\textwidth]{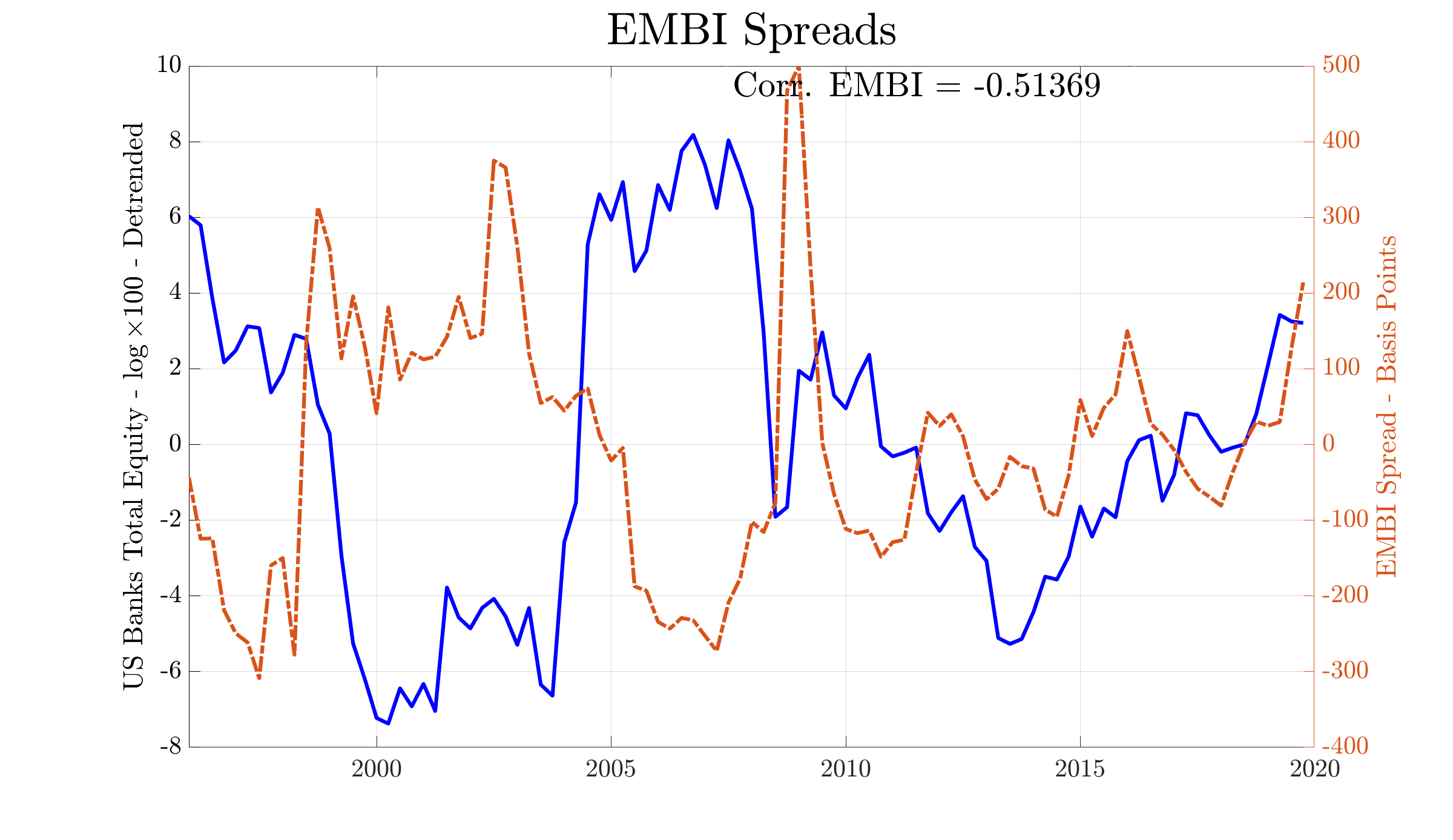}
        \caption{US Banks Net Worth and EMBI Spreads}
        \label{fig:embi}
    \end{subfigure}
    \vspace{0.5cm}

    \begin{subfigure}[b]{0.55\textwidth}
        \centering
        \includegraphics[width=\textwidth]{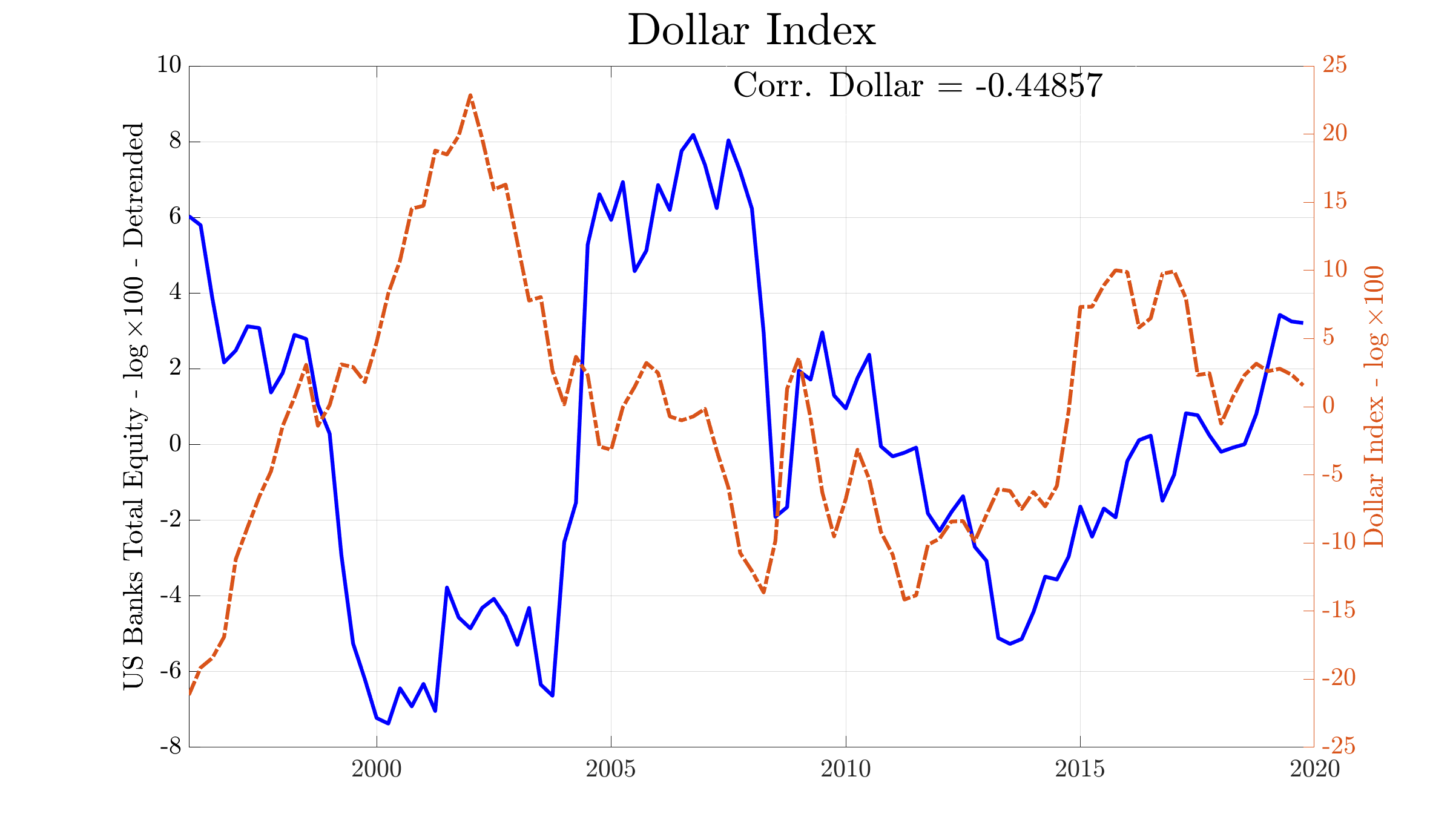}
        \caption{US Banks Net Worth and the Dollar Index}
        \label{fig:dollar}
    \end{subfigure}
    \vspace{0.5cm}

    \begin{subfigure}[b]{0.55\textwidth}
        \centering
        \includegraphics[width=\textwidth]{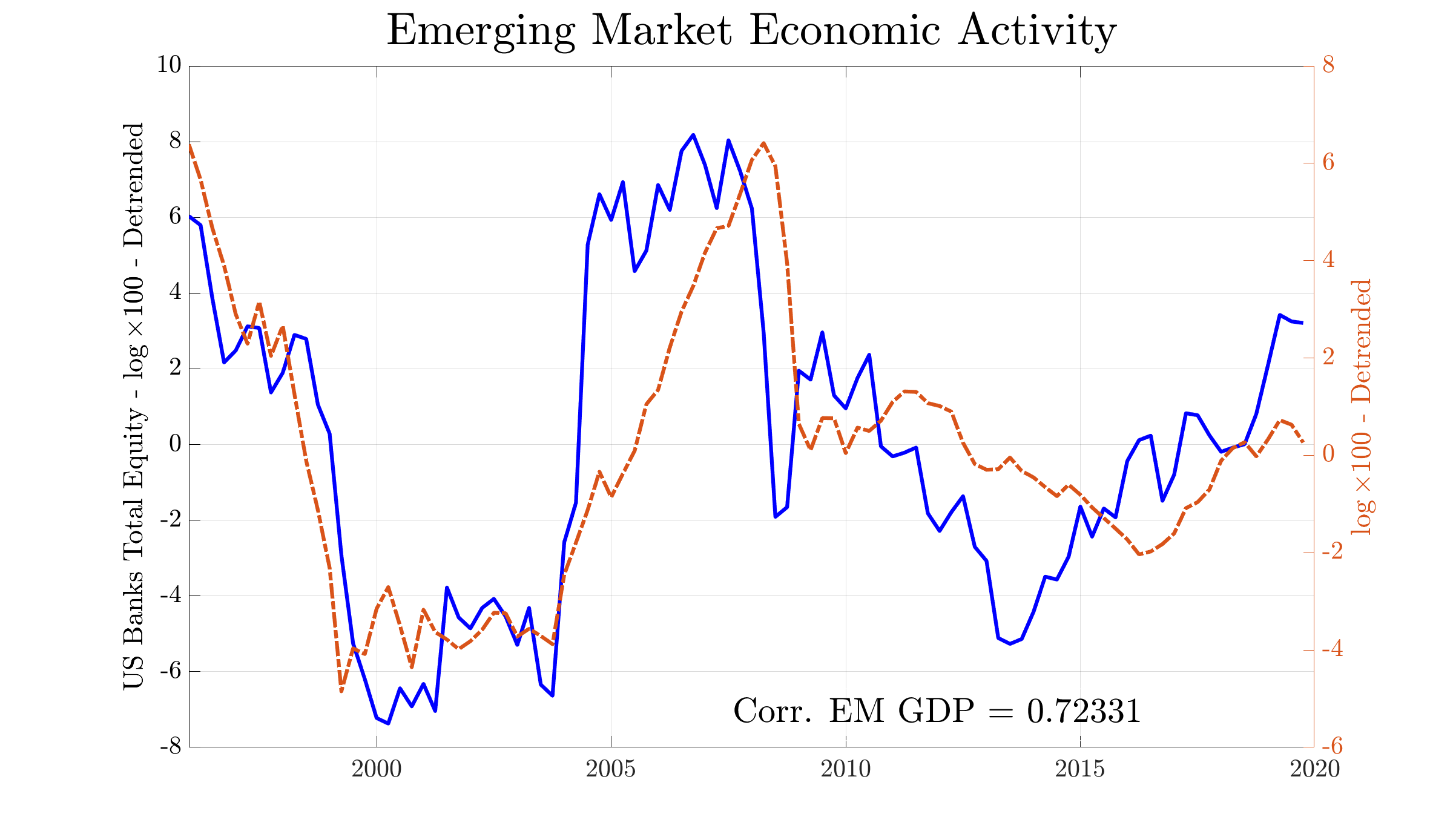}
        \caption{US Banks Net Worth and EME Economic Activity}
        \label{fig:econ}
    \end{subfigure}

    \caption{Motivating Correlations}
    \floatfoot{Each panel plots a detrended measure of United States global banks total equity against a key emerging market indicator. Panel (a) plots EMBI sovereign spreads, panel (b) the broad United States dollar index, and panel (c) aggregate emerging market economic activity. All three series comove strongly with the global bank net worth measure. Correlation coefficients are reported within each panel.}
    \label{fig:motivating_vertical}
\end{figure}

Figure \ref{fig:motivating_vertical} provides a preliminary look at the data. The figure plots a detrended measure of United States global banks equity, our proxy for global banks net worth, against sovereign spreads, the broad United States dollar index, and a measure of emerging market economic activity. All three series display notable comovement with the global bank net worth measure. Improvements in banks balance sheets are associated with lower EMBI spreads, a weaker dollar, and expansions in emerging market activity. These patterns suggest that fluctuations in global banks financial conditions are correlated with the cost of external finance and with business cycle developments in emerging markets.

These correlations motivate the main question of the paper, namely, to what extent shocks to global banks' net worth spill over into emerging markets, and through which domestic channels these shocks are transmitted. Despite substantial interest in the global financial cycle, two empirical challenges complicate this analysis. At the aggregate level, commonly used proxies for global financial conditions, such as the VIX, broker-dealer leverage, or capital flow episodes, combine movements in credit supply with shifts in risk appetite and demand. At the micro level, existing credit registry studies focus on domestic financial frictions, such as \cite{baskaya2017capital} for Turkey or \cite{paravisini2015dissecting} for Peru, but they do not directly link firm-level borrowing to shocks originating from global financial intermediaries. As a result, we lack a unified view of how global bank balance sheet shocks propagate across countries, through domestic intermediaries, and ultimately to firms.

The paper provides such an analysis. We study the transmission of global bank credit supply shocks into emerging markets using a macro-to-micro empirical strategy. Our global shock measure follows \cite{ottonello2022financial} and is based on high-frequency stock price reactions around earnings announcements of large United States global banks. The narrow event windows isolate revisions to banks' net worth that are orthogonal to monetary policy surprises and other macro news, following recent work on high-frequency identification, such as \cite{gertler2015monetary} and \cite{miranda2021transmission}. We focus on the component of these surprises associated with changes in risk-bearing capacity, which provides a plausibly exogenous source of fluctuations in global credit supply.

We combine these shocks with two complementary datasets. First, we estimate the aggregate responses of a panel of emerging markets using local projection and vector autoregression methods, focusing on exchange rates, sovereign spreads, capital flows, and real activity. Second, we trace the domestic transmission mechanism using the universe of firm-bank currency monthly credit registry data from Uruguay, which includes detailed information on loan amounts, loan currencies, collateral, and maturity structures. This approach allows us to link the same global shock to developments at the macro, bank, and firm levels, in the spirit of macro-to-micro frameworks such as \cite{di2022international}.

Our findings are as follows. At the aggregate level, positive shocks to global banks' net worth are followed by appreciations, declines in sovereign spreads, increases in economic activity, and sizable capital inflows. These inflows are concentrated in the domestic banking system rather than in the public sector, which highlights the central role of financial intermediaries in absorbing global liquidity, consistent with the view in \cite{bruno2015cross}. At the bank level, institutions with stronger balance sheets, including those with higher capital ratios, greater liquidity, and lower leverage, expand credit more in response to global easing episodes. At the firm level, borrowing responses depend on ex ante financial characteristics. Firms with lower leverage, more pledgeable collateral, and smaller foreign currency exposures increase borrowing by more than financially constrained firms.

\bigskip
\noindent\textbf{Related literature.}
Our paper contributes to the literature on the international transmission of financial shocks and the role of global intermediaries in shaping emerging market business cycles. We are closely related to \cite{cesa2018international}, who show that shocks to U.S.\ broker-dealer leverage affect credit, exchange rates, and consumption across EMEs. Our paper shares their emphasis on global intermediaries but differs in three central respects: we identify shocks directly from high-frequency movements in global banks’ net worth rather than relying on leverage-based proxies; we link these shocks to both cross-country macro responses and credit-registry micro data within a single emerging economy; and we exploit rich within-country heterogeneity in balance-sheet and collateral characteristics for both banks and firms.

A second strand of work studies how the Global Financial Cycle shapes domestic credit conditions, including \cite{di2022international} for Turkey. While they proxy global conditions using the VIX, we treat global banks’ net-worth shocks—purged of credit-demand and information components following \cite{ottonello2022financial}—as the primitive disturbance. Our micro evidence further complements this literature by showing that firm-level heterogeneity in leverage, maturity structure, currency denomination, and collateral pledgeability critically shapes the transmission of global credit supply shocks. Relatedly, \cite{camararamirez2022} show that more indebted Chilean firms contract investment more when U.S.\ monetary policy tightens. We differ by focusing on shocks to global banks’ balance sheets rather than U.S.\ policy shocks, and by tracing how bank–firm matching and collateral quality jointly govern the pass-through of global credit easing.

More broadly, our paper relates to macro–micro studies of bank balance-sheet channels, including \cite{khwaja2008tracing}, \cite{schnabl2012international}, and \cite{morais2019international}, as well as to recent work emphasizing cross-border funding and international lending channels. We bridge these strands by identifying a global bank net-worth shock, quantifying its macroeconomic effects across EMEs, and documenting how domestic bank and firm balance-sheet heterogeneity shapes the microeconomic transmission of global credit supply shocks within an emerging market.

\bigskip
\noindent
\textbf{Organization.}
The remainder of the paper is organized as follows. Section 2 describes the construction of global bank credit supply shocks and outlines the empirical framework. Section 3 presents the aggregate results for emerging markets. Section 4 examines the domestic transmission using credit registry data. Section 5 reports robustness checks. Section 6 concludes.

%%%%%%%%%%%%%%%%%%%%%%%%%%%%%%%%%%%%%%%%%%%%%%%%%%%%%%%%%%%%%%%%%%%%%%%%%%%%%%%%%%%%%%%%%%%%%%%%%%%%%%%%%%%%%%%%%%%%%%%%%%%%%%%%%%%%%%%%%%%%
%% Identification Strategy
\section{Identification Strategy} \label{sec:identification_strategy}

%%%%%%%%%%%%%%%%%%%%%%%%%%%%%%%%%%%%%%%%%%%%%%%%%%%%%%%%%%%%%%%%%
\subsection{A Stylized Model of Global Banks' Net Worth \& EME Credit} \label{subsec:stylized_model}

The goal of this subsection is not to develop a full structural model of global intermediation, but to isolate the minimal ingredients needed to discipline our identification strategy and to interpret the heterogeneous firm level responses documented below. The model is intentionally stylized and draws on the mechanisms in \cite{morelli2022global} and \cite{ottonello2022financial}. The full model details and derivations are presented in Appendix \ref{sec:appendix_stylized_model}.

A global bank allocates its balance sheet between a representative United States borrower and a continuum of emerging market firms. The United States borrower has high pledgeability, so collateral recoverability is high. Each emerging market firm \(i\) chooses capital \(k_i\) and new borrowing \(d_i\), carries legacy debt \(D_{0i}\), and produces
\[
y_i = \varepsilon_i A_i k_i^\alpha,
\qquad \varepsilon_i \sim U[0,1], \quad 0<\alpha<1.
\]
The firm repays its debts if
\[
\varepsilon_i A_i k_i^\alpha \ge \frac{d_i}{q_i} + D_{0i},
\]
which defines the default threshold
\[
\varepsilon_i^* = \frac{(d_i/q_i) + D_{0i}}{A_i k_i^\alpha}.
\]

If the firm defaults, the global bank recovers a fraction \(\theta_i \in (0,1)\) of output. The parameter \(\theta_i\) therefore summarizes collateral quality and pledgeability. Firms with more liquid collateral, longer maturity liabilities, or stronger balance sheets have higher \(\theta_i\), while firms with high leverage, short maturity structures, or collateral not priced to market have lower pledgeability. These dimensions correspond directly to the heterogeneity observed in our credit registry, where we measure leverage in total and in foreign currency, the maturity structure of outstanding loans, and the share of collateral that is valued at market prices.

The term \(D_{0i}\) represents the firm legacy debt burden. Although the model is written in a single good and a single currency, \(D_{0i}\) should be interpreted broadly as the effective repayment obligation faced by the firm. In practice, this burden reflects several balance sheet characteristics that we observe directly in the credit registry. Firms with a larger share of liabilities denominated in foreign currency face higher effective repayment risk because exchange rate movements increase the volatility of their debt burden relative to domestic revenues. Firms whose debts have short maturities or large rollover needs face higher effective repayment requirements because short term obligations must be refinanced frequently. Firms with high overall leverage also have larger effective legacy burdens. Treating \(D_{0i}\) as a sufficient statistic for the effective debt burden allows the model to accommodate these empirical differences without introducing additional goods, currencies, or long horizon dynamics. These empirical dimensions increase the default threshold and reduce the marginal equity value of additional borrowing, thereby reducing the sensitivity of such firms to improvements in global bank net worth.

The global bank finances itself with deposits and is subject to a leverage constraint,
\[
R^d d_0 \le \theta_{US} q^{US} b^{US} + \int_0^1 \theta_i q_i b(i) \, di + \lambda n_0,
\]
where \(n_0\) denotes bank net worth. When the constraint binds with multiplier \(\mu>0\), optimality implies the spread condition
\begin{equation}
\label{eq:spread}
\frac{1}{q_i} - R^d = \mu (1 - \theta_i).
\end{equation}
Equation \eqref{eq:spread} is the key equilibrium relationship linking the balance sheet strength of global intermediaries to cross-firm variation in borrowing costs. An increase in bank net worth, \(n_0\), relaxes the leverage constraint, reduces \(\mu(n_0)\), and compresses loan spreads. Differentiating,
\[
\frac{\partial}{\partial n_0}\Big(\frac{1}{q_i} - R^d\Big)
= (1 - \theta_i)\,\mu'(n_0) < 0.
\]
Thus, a positive net worth shock reduces spreads for all emerging market firms. This mirrors the sign restriction used in \cite{ottonello2022financial}; credit supply shocks move intermediaries' net worth and borrowing costs in opposite directions.\footnote{This logic is analogous to the decomposition in \cite{camara2025spillovers}, who shows that separating the fundamental and information components of high-frequency monetary surprises is essential for estimating international spillovers.}

Given the rate \(1/q_i\), the firm chooses \(k_i = d_i\) to maximize expected equity. Let \(k_i^*(n_0)\) denote the optimum. The model predicts
\[
\frac{\partial k_i^*}{\partial n_0} > 0,
\]
so aggregate emerging market credit expands following a positive net worth shock.

The relative response across firms, however, is ambiguous and depends on leverage, maturity, and collateral quality. Differentiating the firm's first-order condition yields
\[
\frac{\partial k_i^*}{\partial n_0}
= -\,\frac{(1 - \theta_i)\,\mu'(n_0)\,H_i(k_i^*;D_{0i},\theta_i)}{\Pi_{ii}},
\]
where \(\Pi_{ii}<0\) and \(H_i(\cdot)\) depends on the default threshold, leverage, and debt overhang. Two forces interact. Firms with lower pledgeability, due to high leverage, short maturities, foreign currency liabilities, or collateral not priced to market, experience larger declines in spreads when \(\mu\) falls. At the same time, firms with very poor collateral or large effective legacy burdens have high default thresholds, so marginal borrowing yields little equity value. Depending on which force dominates, either more leveraged or less leveraged firms may respond more to improvements in global banks' net worth.

In summary, the model provides two implications that guide our empirical design. First, a positive global bank credit supply shock reduces spreads and increases aggregate emerging market credit. Second, cross-firm responses depend on leverage, currency mismatch, maturity, and the pledgeability of collateral, which are precisely the dimensions we exploit in our micro-level analysis.

%%%%%%%%%%%%%%%%%%%%%%%%%%%%%%%%%%%%%%%%%%%%%%%%%%%%%%%%%%%%%
\subsection{Measuring Global Bank Credit Supply Shocks} \label{subsec:Identification_OS}

To identify exogenous shifts in global banks' net worth and credit supply, we follow the high-frequency identification strategy in \cite{ottonello2022financial}. Their approach isolates shocks as unanticipated changes in the large United States intermediaries market value of net worth around earnings announcements. Within a narrow intraday window, these movements reflect balance sheet news specific to intermediaries, not macroeconomic or policy information.

For intermediary \(i\) reporting earnings at time \(t\), they compute
\[
v^F_t = \theta_{i,q(t)} \cdot \Delta p_{F,i,t},
\]
where \(\Delta p_{F,i,t}\) is the sixty-minute change in log stock price and \(\theta_{i,q(t)}\) is the market capitalization weight among large financial firms. The shock \(v^F_t\) measures innovations to intermediaries' net worth.

Because earnings contain both credit supply and credit demand information, \cite{ottonello2022financial} decomposes
\[
v^F_t = v^{CS}_t + v^{CD}_t,
\]
where
\[
\text{corr}(v^{CS}_t, \Delta \text{EBP}_t) < 0,
\qquad
\text{corr}(v^{CD}_t, \Delta \text{EBP}_t) > 0.
\]
Shocks that reduce intermediaries' net worth and raise the excess bond premium are classified as adverse credit supply shocks. The decomposition uses a Givens rotation procedure following \cite{jarocinski2020deconstructing}, ensuring orthogonality between \(v^{CS}_t\) and \(v^{CD}_t\) and preserving the variance of \(v^F_t\).

The resulting series \(v^{CS}_t\) captures exogenous movements in global banks' effective net worth and risk-bearing capacity. We take these shocks from the authors' publicly available replication files. In the empirical analysis below, we use \(v^{CS}_t\) as our baseline measure, normalize it to one standard deviation, and aggregate it to quarterly frequency to match our datasets.
gs contain both credit supply and credit demand information, \cite{ottonello2022financial} decomposes
\[
v^F_t = v^{CS}_t + v^{CD}_t,
\]
where
\[
\text{corr}(v^{CS}_t, \Delta \text{EBP}_t) < 0,
\qquad
\text{corr}(v^{CD}_t, \Delta \text{EBP}_t) > 0.
\]
Shocks that reduce intermediaries' net worth and raise the excess bond premium are classified as adverse credit supply shocks. The decomposition uses a Givens rotation procedure following \cite{jarocinski2020deconstructing}, ensuring orthogonality between \(v^{CS}_t\) and \(v^{CD}_t\) and preserving the variance of \(v^F_t\).

The resulting series \(v^{CS}_t\) captures exogenous movements in global banks' effective net worth and risk-bearing capacity. We take these shocks from the authors' publicly available replication files. In the empirical analysis below, we use \(v^{CS}_t\) as our baseline measure, normalize it to one standard deviation, and aggregate it to quarterly frequency to match our datasets.

%%%%%%%%%%%%%%%%%%%%%%%%%%%%%%%%%%%%%%%%%%%%%%%%%%%%%%%%%%%%%%%%%%%%%%%%%%%%%%%%%%%%%%%%%%%%%%%%%%%%%%%%%%%%%%%%%%%%%%%%%%%%%%%%%%%%%%%%%%%%
%%
\section{Macroeconomic Impact} \label{sec:macro_impact}

This section documents how improvements in global banks’ net worth affect macroeconomic activity in Emerging Market Economies (EMEs). We quantify the responses of exchange rates, output, investment, and trade to exogenous shifts in global bank balance-sheet strength, using the identified credit-supply shocks described in Section~\ref{sec:identification_strategy}. Our benchmark specification is a pooled panel VAR that captures the joint dynamics of domestic and global variables across EMEs. As a complementary and less parametric approach, we also report panel Local Projection (LP) estimates. Across both methods, the impulse responses can be interpreted as the average effect of a positive global credit-supply shock—that is, an improvement in global banks’ balance sheets that relaxes external financing conditions and eases credit for EMEs.

\subsection{Data Description} \label{subsec:macro_data}

We construct a quarterly panel of macro-financial variables for a broad set of EMEs over 2002Q3--2019Q4, consistent with the availability of the identified global credit-supply shocks and excluding the COVID-19 period. The data are drawn from the IMF, OECD, BIS, and U.S.\ regulatory sources.

Our benchmark analysis focuses on five EME-specific variables---(i) the nominal exchange rate against the U.S.\ dollar, (ii) real GDP, (iii) gross fixed capital formation, (iv) total exports, and (v) total imports---which together capture external adjustment and domestic activity. To account for global conditions, we also include a small set of U.S.\ variables: the federal funds rate, industrial production, the PCE price index, and the excess bond premium (EBP).

The benchmark panel consists of 18 EMEs with at least 40 quarterly observations. A subset of countries with full coverage for all 70 quarters forms the balanced sample used in baseline estimates, while the full unbalanced panel is used for robustness (country lists are reported in Appendix~\ref{sec:appendix_data_details}). This ensures that the sample spans both tranquil and stress periods, including the 2008--2009 Global Financial Crisis.

To study the transmission of global credit-supply shocks through financial flows, we incorporate several key financial quantities in an extended specification:
\begin{itemize}
    \item[(i)] U.S.\ banks' consolidated claims on foreign obligors,\footnote{Sourced from the FFIEC's \textit{E.16 Country Exposure Lending Survey}. See \url{https://www.ffiec.gov/data/e16}. We focus on ``Claims on Foreign Obligors Held by U.S.\ Banks on an Ultimate Risk Basis (Adjusted to Reflect Guarantees and Other Risk Transfers) \& Off-Balance-Sheet Items,'' available from 2006Q1.}
    \item[(ii)] private-sector foreign liabilities,\footnote{Data from \cite{avdjiev2017gross}.}
    \item[(iii)] public-sector foreign liabilities,\footnote{Same source as in (ii).}
    \item[(iv)] real domestic credit to the private non-financial sector.\footnote{From BIS ``Total Credit'' statistics. See \url{https://data.bis.org/topics/TOTAL_CREDIT}.}
\end{itemize}
Coverage for these series is more limited---particularly early in the sample---so these variables are examined using a reduced sample of countries.

Finally, where available, we also collect additional macro--financial indicators such as EMBI spreads, a multilateral real exchange rate, nominal exchange rates vis-à-vis the euro, equity indexes, and domestic lending rates. All series are drawn from official datasets to ensure consistency in definitions and measurement.

%%%%%%%%%%%%%%%%%%%%%%%%%%%%%%%%%%%%%%%%%%%%%%%%%%%%
\subsection{Econometric Specification} \label{subsec:macro_econometric_specifications}

We quantify the dynamic effects of global banks’ credit-supply shocks on Emerging Market Economies (EMEs) using a pooled panel Vector Autoregression (VAR) as our benchmark specification (see \cite{canova2013panel}).\footnote{See \cite{camara2025international} for another application of this model in international macroeconomics.} The pooled VAR provides a parsimonious structure to capture the joint dynamics of domestic and global variables in response to exogenous financial shocks.

\medskip
\noindent
\textbf{Benchmark pooled panel VAR.}  
For each country $i$, the vector of endogenous variables includes five EME-specific and four U.S.-specific variables:
\[
Y_{i,t} =
\begin{bmatrix}
\ln \text{NER}_{i,t} \\
\ln \text{GDP}_{i,t} \\
\ln \text{GFC}_{i,t} \\
\ln \text{EXPO}_{i,t} \\
\ln \text{IMPO}_{i,t} \\
\text{FedFunds}_{t} \\
\ln \text{INDPRO}_{t} \\
\ln \text{PCEPI}_{t} \\
\text{EBP}_{t}
\end{bmatrix}.
\]
The identified global bank net-worth shock, $\widehat{\text{Shock}}_t$, enters the system as an exogenous variable. Stacking observations yields the compact representation
\[
Y = X\beta + \varepsilon, \qquad \varepsilon \sim \mathcal{N}(0,\,\Sigma_c \otimes I_T),
\]
where $X$ collects two lags of all endogenous variables and the shock. The shock is standardized to unit variance so that impulse responses correspond to a one–standard–deviation improvement in global banks’ balance-sheet strength. The VAR requires a balanced panel, so baseline results are estimated on the sample of EMEs with complete data coverage.

\medskip
\noindent
\textbf{Complementary Local Projection estimates.}  
To exploit all available country–time observations, we also estimate dynamic responses using panel Local Projections (LPs), which naturally accommodate an unbalanced panel. For each horizon $h$, we estimate
\[
y_{i,t+h} = \alpha_q
+ \sum_{p=1}^{2} \Gamma^{\text{EME}}_{p} X^{\text{EME}}_{i,t-p}
+ \sum_{p=1}^{2} \Gamma^{\text{US}}_{p} X^{\text{US}}_{t-p}
+ \sum_{p=0}^{2} \theta_{p}\,\widehat{\text{Shock}}_{t-p}
+ u_{i,t+h}, \label{eq:LP_Reg}
\]
where $y_{i,t}$ is the outcome of interest and $\alpha_q$ denotes quarter-of-year fixed effects. Standard errors are clustered by date. LPs provide a flexible, horizon-specific alternative to the VAR and, importantly, allow us to estimate responses for the full unbalanced panel of EMEs. Consistency between VAR and LP estimates strengthens the robustness of our findings.

\subsection{Macroeconomic Results} \label{subsec:macro_main_results}

This subsection presents the responses of Emerging Market Economies (EMEs) to a one--standard--deviation improvement in global banks’ net worth. We begin with five core macroeconomic variables: nominal exchange rate, real GDP, gross fixed capital formation, exports, and imports, which summarize the main margins of external adjustment and domestic activity. Impulse responses are obtained from the pooled panel VAR described above; LP estimates for both the balanced and unbalanced panels are reported in Appendix~\ref{sec:appendix_macro_results} and confirm the robustness of the results.

\begin{figure*}[p]
    \centering
    \includegraphics[width=0.95\textwidth]{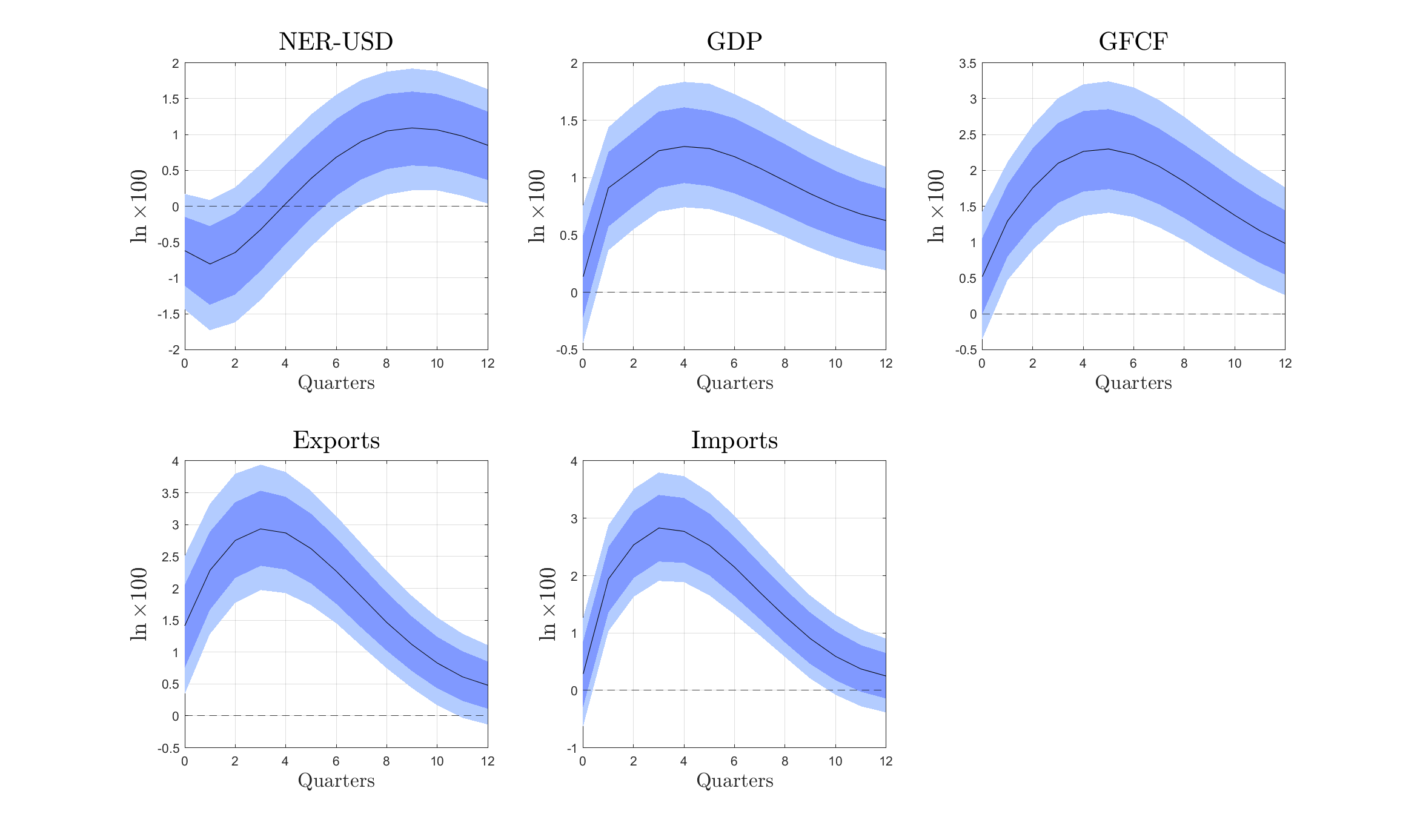}
    \caption{Macroeconomic Responses of EMEs to a Global Bank Net-Worth Shock}
    \label{fig:Pooled_EME}
    \floatfoot{\textbf{Note:} The figure displays impulse responses of five key macroeconomic variables to a one--standard--deviation increase in global banks’ net worth, estimated using the pooled panel Bayesian VAR. Each row corresponds to a different variable, and the left and right columns report results for the balanced and unbalanced panels, respectively. The black line shows the median IRF, while the dark and light blue shaded areas denote the 68\% and 90\% credible intervals.}
\end{figure*}

Figure~\ref{fig:Pooled_EME} presents the benchmark VAR responses of five key macroeconomic variables in EMEs to a positive global bank net-worth shock. The nominal exchange rate (Panel~(1,1)) appreciates sharply on impact and continues to strengthen modestly in the following quarter. After this initial appreciation, the exchange rate gradually depreciates toward its pre-shock level over the medium term. This pattern reflects the front-loaded FX response typical of a decline in global risk premia: capital inflows and improved external financial conditions generate an immediate appreciation that subsequently unwinds as the initial impulse fades.

Real GDP and investment (Panels~(1,2) and (1,3)) rise quickly and substantially. GDP peaks around four to six quarters after the shock, while gross fixed capital formation displays an even stronger response, peaking at roughly twice the size of the output response. This stronger reaction is consistent with the sensitivity of investment to external financing conditions and the relaxation of borrowing constraints when global bank balance sheets strengthen.

Exports and imports (Panels~(2,1) and (2,2)) both increase meaningfully, with broadly similar magnitudes at their peaks. Exports rise rapidly and peak around two to three quarters after the shock, whereas imports peak slightly later, around three to four quarters, and exhibit a somewhat more front-loaded initial increase. The relative timing reflects the immediate boost to domestic demand and currency appreciation on the import side, while export adjustments follow more gradually as production and external demand respond. The initially stronger import response generates a temporary widening of the trade balance before both series converge back toward trend.

Overall, the benchmark impulse responses reveal a coherent pattern: an improvement in global banks’ net worth reduces risk premia, eases external financing conditions, and stimulates domestic activity in EMEs, with especially pronounced effects on investment and import demand.

\begin{figure*}[t]
    \centering
    \includegraphics[width=0.85\textwidth]{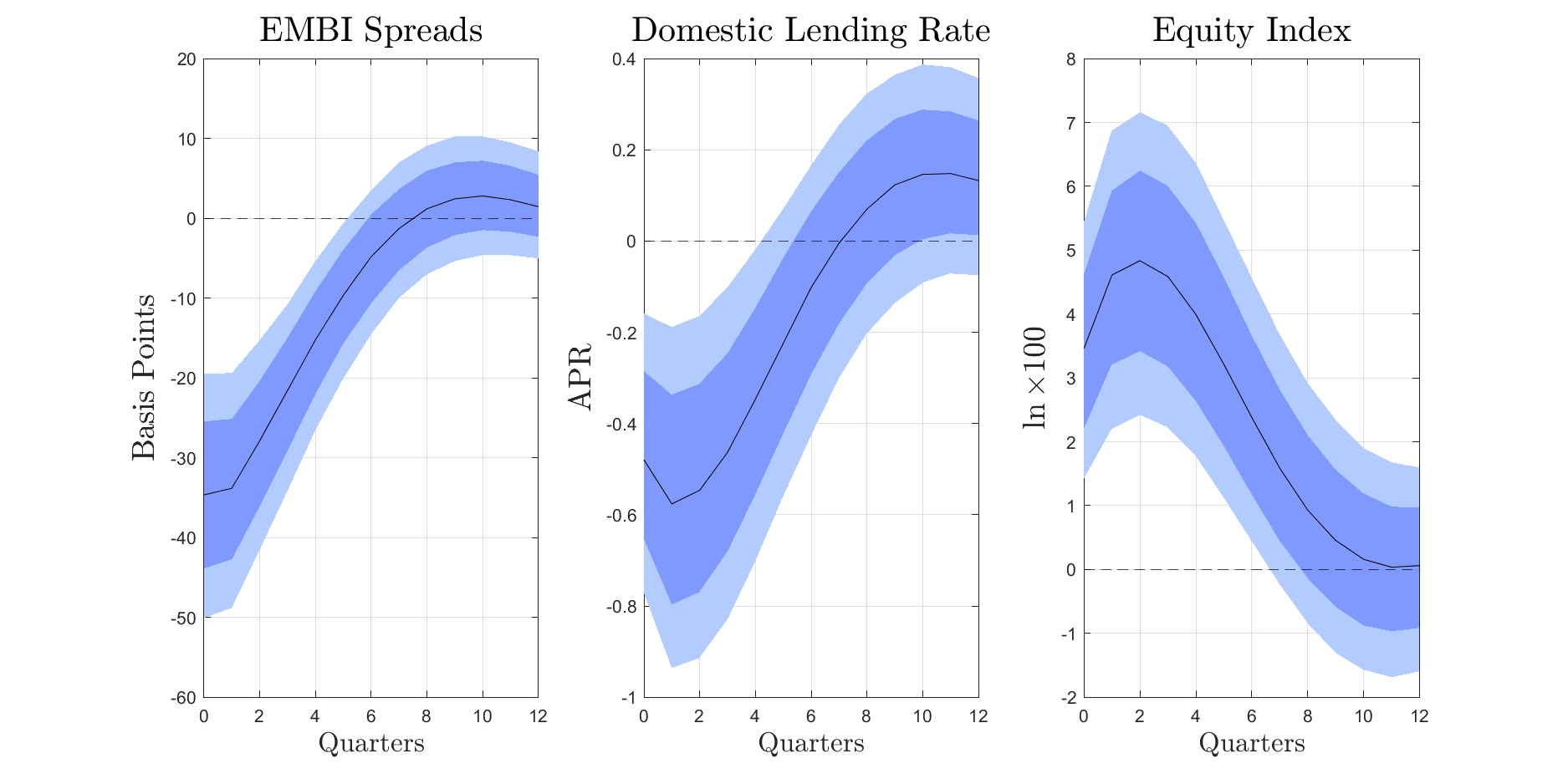}
    \caption{Responses of Financial Prices in EMEs to a Global Bank Net-Worth Shock}
    \label{fig:Pooled_EME_Financial_Prices}
    \floatfoot{\textbf{Note:} The figure displays impulse responses of EMBI spreads, domestic lending rates, and equity prices to a one--standard--deviation improvement in global banks’ net worth. Results are shown for both the balanced and unbalanced panels.}
\end{figure*}

\medskip
\noindent\textbf{Financial price responses.}
Figure~\ref{fig:Pooled_EME_Financial_Prices} shows the responses of three financial price variables that capture the cost of external and domestic financing. EMBI spreads fall sharply and persistently following the shock, consistent with a broad reduction in sovereign and corporate risk premia. The decline is gradual but long-lasting, suggesting that improvements in global bank balance sheets ease perceptions of EME credit risk for an extended period.

Domestic lending rates also fall meaningfully, reflecting the pass-through of global financial easing into domestic monetary and credit conditions. The decline is sizable on impact and remains significant for several quarters, consistent with domestic banks facing lower external funding costs and improved balance-sheet strength.

Equity prices rise strongly, indicating higher firm valuations and improved investor sentiment. This response is consistent with reduced uncertainty, lower discount rates, and the expectation of stronger real activity. Taken together, the behavior of financial prices points to a clear and sustained easing of both external and domestic financial conditions.

\begin{figure*}[t]
    \centering
    \includegraphics[width=0.75\textwidth]{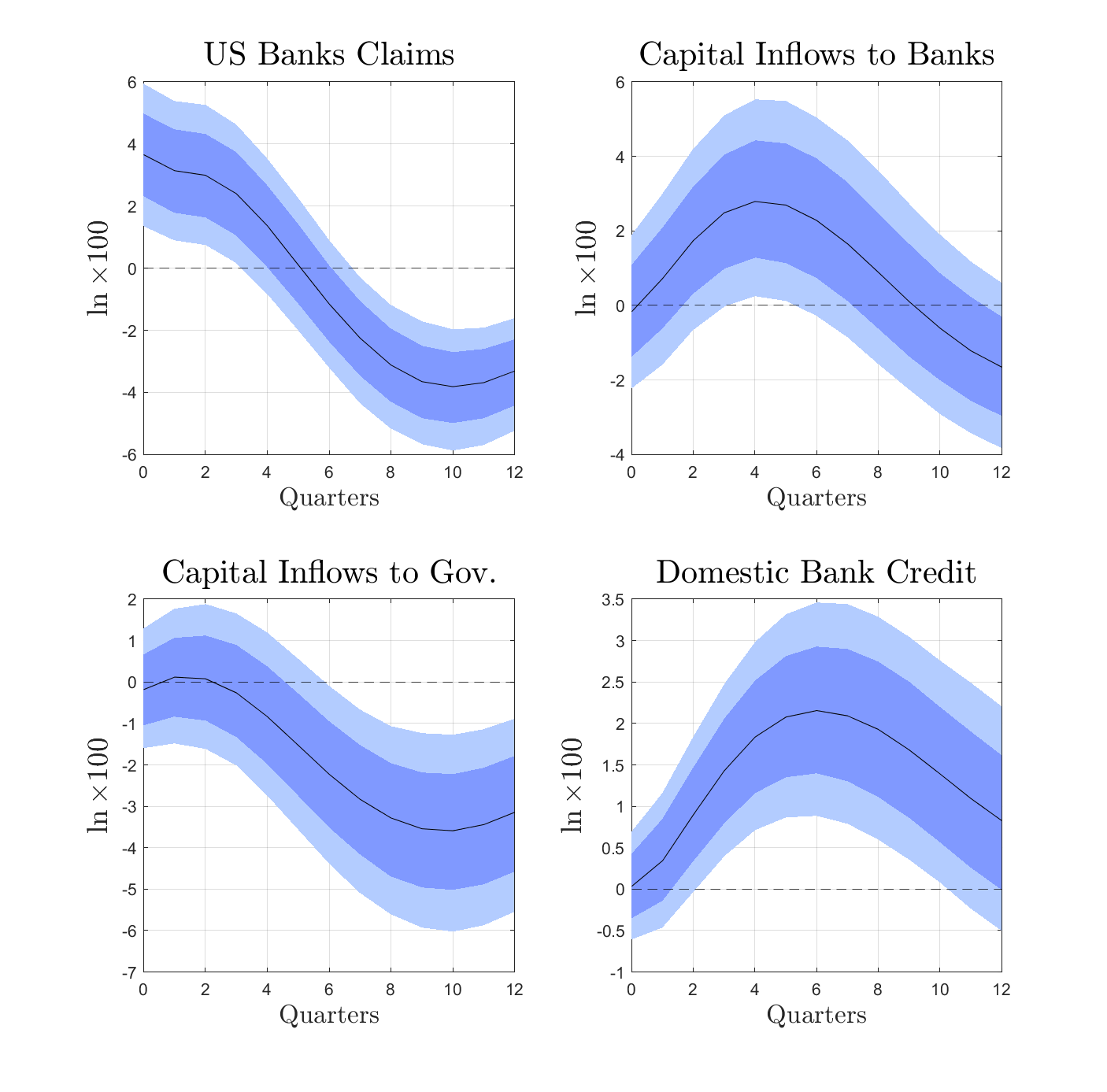}
    \caption{Responses of Financial Quantities in EMEs to a Global Bank Net-Worth Shock}
    \label{fig:Pooled_EME_Financial_Quantities}
    \floatfoot{\textbf{Note:} The figure displays impulse responses of public-sector foreign liabilities, banking-sector foreign liabilities, domestic private credit, and U.S.\ banks’ consolidated claims to a global bank net-worth shock.}
\end{figure*}

\medskip
\noindent\textbf{Financial quantity responses.}
Figure~\ref{fig:Pooled_EME_Financial_Quantities} examines the response of key financial flows and credit aggregates. U.S.\ banks’ consolidated claims on EMEs rise sharply on impact, indicating that the improvement in global bank balance sheets translates directly into higher cross-border exposures. The response is immediate and economically sizable, consistent with global banks expanding their international lending when their net worth strengthens.

Domestic EME banks also experience a strong increase in external liabilities, with a hump-shaped response peaking after several quarters. This suggests a gradual accumulation of foreign borrowing as global conditions ease. In contrast, public-sector foreign liabilities show little short-term reaction and even a mild decline at medium horizons, indicating that additional external financing is directed mainly toward private intermediaries rather than sovereigns.

Finally, domestic credit to the non-financial private sector increases persistently. This confirms that the easing of external constraints spills over into the domestic financial system, relaxing credit conditions for firms and households. The joint behavior of flows and credit aggregates supports the interpretation that the transmission operates primarily through global and domestic banking channels.

\medskip
\noindent\textbf{Robustness: Local Projection estimates.}
Local Projection responses for both the balanced and unbalanced panels closely match the VAR results in magnitude, timing, and persistence. LP estimates show similar appreciations of local currencies, comparable increases in GDP and investment, and nearly identical movements in financial prices and quantities. Confidence bands are wider, as expected, but the qualitative patterns remain unchanged. Because LPs naturally accommodate the unbalanced panel, these estimates confirm that the results are not driven by sample selection or VAR structure.

To further assess robustness, we estimate specifications with country fixed effects, time trends, and alternative lag structures. All yield responses that are qualitatively and quantitatively similar to those in the baseline, underscoring the stability of the results.

\medskip
\noindent\textbf{Macro evidence for Uruguay.}
Uruguay, the economy for which detailed credit-registry data are available, exhibits aggregate responses that are highly consistent with the broader EME sample. Using monthly macro--financial series compiled from the Central Bank of Uruguay (with quarterly flow variables interpolated via cubic spline), we estimate the same VAR specification at monthly frequency. A positive global bank net-worth shock leads to a peso appreciation, lower lending rates and EMBI spreads, an expansion of domestic credit, and a rise in industrial production (Figure~\ref{fig:Uruguay_Macro_Responses}).

When U.S.\ banks’ consolidated claims on Uruguay became available in 2006, we also observed a marked increase in cross-border bank flows following the shock. Capital inflows to domestic banks rise strongly, while inflows to the government sector remain modest and short-lived (Figure~\ref{fig:Uruguay_Macro_Responses_Flows}). These results confirm that Uruguay mirrors the broader EME transmission mechanism.

\medskip
\noindent\textbf{Summary and transition.}
Taken together, the macro evidence shows that improvements in global banks’ balance-sheet strength ease financing conditions, appreciate local currencies, and generate sustained expansions in credit, investment, and real activity. These effects operate primarily through cross-border and domestic banking channels. The next section turns to micro-level evidence using credit registry data to examine which firms benefit most from these shifts in global credit supply.

%%%%%%%%%%%%%%%%%%%%%%%%%%%%%%%%%%%%%%%%%%%%%%%%%%%%%%%%%%%%%%%%%%%%%%%%%%%%%%%%%%%%%%%%%%%%%%%%%%%%%%%%%%%%%%%%%%%%%%%%%%%%%%%%%%%%%%%%
%%%%%%%%%%%%%%%%%%%%%%%%%%%%%%%%%%%%%%%%%%%%%%%%%%%%%%%%%%%%%%%%%%%%%%%%%%%%%%%%%%%%%%%%%%%%%%%%%%%%%%%%%%%%%%%%%%%%%%%%%%%%%%%%%%%%%%%%
\section{Microeconomic Transmission} \label{sec:micro_transmission}

To examine how global bank credit-supply shocks are transmitted within an emerging market, we draw on two high-frequency administrative datasets from Uruguay that provide comprehensive information on loans, banks, and firms. Both datasets are maintained by the Central Bank of Uruguay, which supervises all domestic financial institutions. The data span April~2003 to December~2019 and cover the universe of formal lending activity.

%%%%%%%%%%%%%%%%%%%%%%%%%%%%%%%%%%%%%%%%%%%
\subsection{Bank-Level Analysis} \label{subsec:micro_banks}

\noindent
\textbf{Data description.}
The bank-level dataset contains monthly balance sheet and income statement information for all financial institutions operating in Uruguay.\footnote{\url{https://www.bcu.gub.uy/Servicios-Financieros-SSF/Paginas/Boletin-SSF.aspx}} We use these data for two purposes. First, we characterize how banks with different balance-sheet positions respond to global credit-supply shocks, providing direct evidence on the intermediary balance-sheet channel. Second, we incorporate bank characteristics into the firm-level regressions to account for heterogeneity in lender behavior.

Figure~\ref{fig:Aggregate_Evolution} in Appendix \ref{sec:appendix_data_details} shows the evolution of aggregate firm borrowing, separating loans in foreign and local currency. Both series trend upward, but exhibit pronounced slowdowns during global financial stress episodes—most notably the 2008--2010 Global Financial Crisis and the 2016--2018 Fed tightening cycle. These patterns already indicate that Uruguay’s credit cycle is sensitive to external financial conditions.

Uruguay’s financial system includes 14 banks. Foreign-owned banks account for roughly two-thirds of total firm credit, while domestically owned banks supply the remaining third (see Figure~\ref{fig:Bank_Shares} in Appendix \ref{sec:appendix_data_details}). This mix of foreign and domestic intermediaries generates sizable variation in size, capitalization, liquidity, and funding structures, making Uruguay a highly informative setting for studying balance-sheet-driven transmission.

\medskip
Table~\ref{tab:bank_summary_statistics} reports descriptive statistics for key bank-level variables. Foreign banks operate with substantially higher leverage: their median liabilities-to-net-worth ratio is about 11, compared with 7 for domestic banks. They also extend a much larger share of credit in foreign currency (78\% versus 28\%), reflecting tighter integration with global capital markets. Despite their higher leverage, foreign banks exhibit markedly lower non-performing loan ratios (4\% versus 18\%), consistent with stronger asset quality and more conservative risk management. Reliance on demand deposits is similar across groups.

\begin{table}[ht]
    \centering
    \caption{Bank-Level Summary Statistics}
    \small
    \label{tab:bank_summary_statistics}
    \begin{tabular}{l | c c c c | c c }
        & \multicolumn{4}{c}{\textbf{All Banks}} & \textbf{Foreign} & \textbf{Domestic} \\ \hline
        & p10 & Median & p90 & Mean & Mean & Mean \\ \hline \hline
        Liabilities to Assets & 0.764 & 0.918 & 0.940 & 0.894 & 0.920 & 0.779 \\
        Liabilities to Net Worth & 3.435 & 10.594 & 12.674 & 9.820 & 11.010 & 6.970 \\
        Share of Dollar Credit & 0.024 & 0.774 & 0.886 & 0.675 & 0.780 & 0.280 \\
        Share of Non-Performing Loans & 0.030 & 0.053 & 0.277 & 0.084 & 0.040 & 0.180 \\
        Share of Demand Deposits & 0.617 & 0.736 & 0.892 & 0.745 & 0.740 & 0.730 \\ \hline \hline
    \end{tabular}
    \floatfoot{\textbf{Note:} Statistics are computed as averages over the sample period. Columns $p10$ and $p90$ report the 10th and 90th percentiles; group means are unweighted.
    }
\end{table}

\medskip
These cross-sectional differences, combined with substantial time-series variation in leverage, liquidity, and funding structure within banks, motivate the empirical strategy below. We test whether banks with stronger balance sheets—those with lower leverage, higher liquidity, or more stable funding—expand credit more aggressively when global financial conditions improve. By exploiting both cross-bank and within-bank variation, we provide a comprehensive picture of how intermediary financial health shapes the domestic transmission of global credit-supply shocks.

\medskip
\noindent
\textbf{Econometric specification.}
Our benchmark specification is a local-projection regression of the form:
\begin{align}
    \Delta \ln(\text{Loan}_{i,c,t+h}) 
    &= \beta_h \, \text{Shock}^{\text{Supply}}_t 
    + \delta_h \, \text{Shock}^{\text{Supply}}_t \times \text{Leverage}_{i,t-1} \nonumber \\
    &\quad + \mu_{i,h} + \mu_{c,h} 
    + \Gamma^{\text{Bank}}_{i,t-1} 
    + \Gamma^{\text{Macro}}_{i,t-1} 
    + \epsilon_{i,c,t+h},
    \label{eq:Regression_Bank_Level}
\end{align}
where $\Delta \ln(\text{Loan}_{i,c,t+h})$ is the log change in total loans extended by bank $i$ in currency $c$ at horizon $h \in \{0,\dots,36\}$. The terms $\mu_{i,h}$ and $\mu_{c,h}$ are bank and currency fixed effects, respectively.

Equation~\ref{eq:Regression_Bank_Level} includes three sets of controls: (i) \textit{Bank-level controls:} lagged log assets, lagged log liabilities, share of dollar loans, leverage ratio, and liquidity ratio (assets maturing within 30 days over total assets); (ii) \textit{Uruguay macro controls:} EMBI spreads, log industrial production, log nominal UYU/USD exchange rate, and log CPI; (iii) \textit{Global controls:} the Excess Bond Premium (EBP), Federal Funds Rate, log U.S.\ PCE price index, and log U.S.\ industrial production.

Standard errors are two-way clustered at the bank and date levels. All regressions are weighted by each bank's average size, following \citet{di2022international}. The coefficient $\beta_h$ captures the average lending response to the shock, while $\delta_h$ captures how this response varies with bank leverage. A negative $\delta_h$ implies that less-leveraged (better-capitalized) banks expand lending more.

\begin{figure}[ht]
\centering
\includegraphics[width=0.95\linewidth]{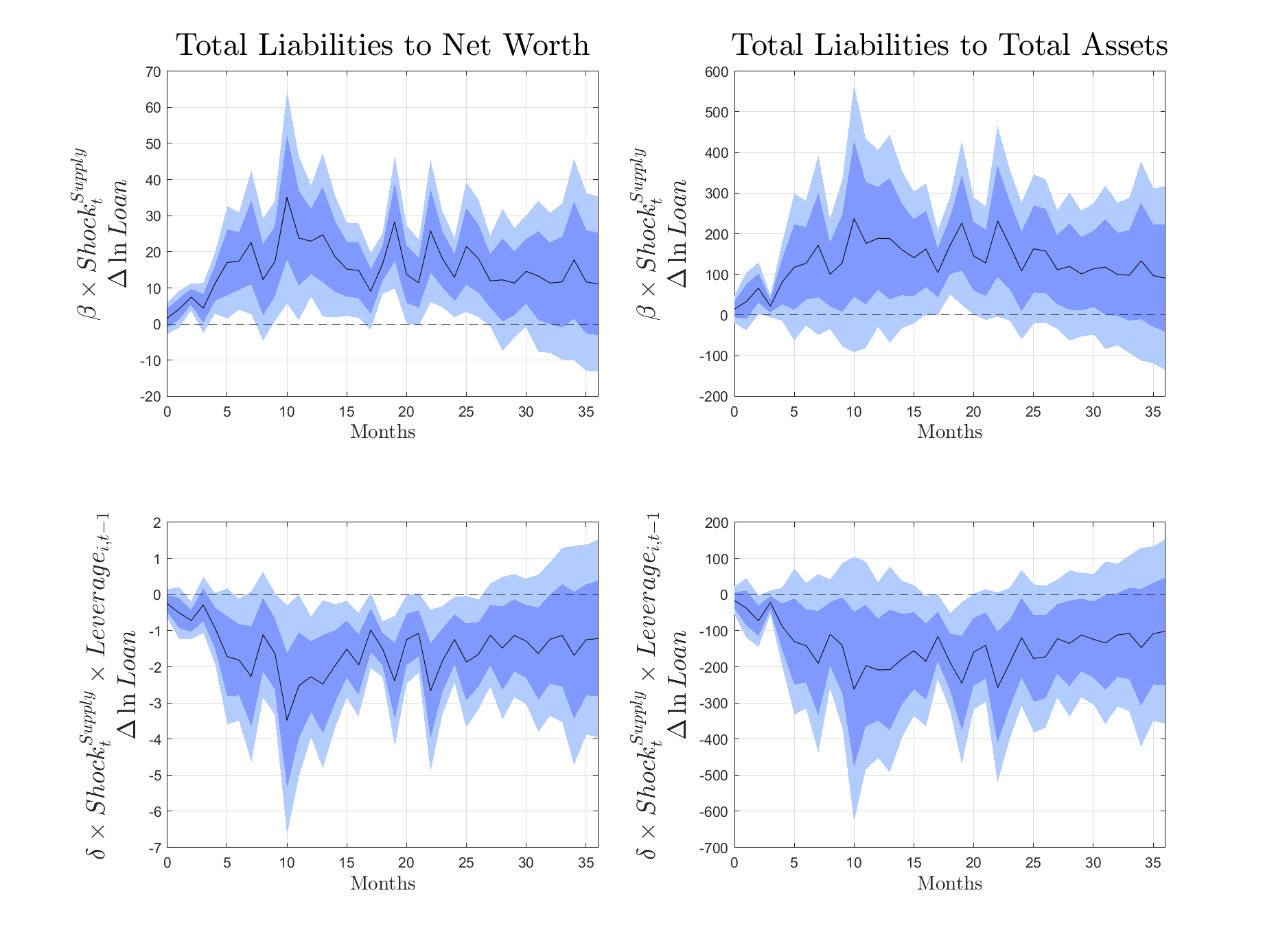}
\caption{Bank-Level Responses to Global Credit Supply Shocks by Leverage}
\label{fig:Ratios_Leverage_2by2}
\floatfoot{\textbf{Note:} The left column uses liabilities-to-net-worth; the right column uses liabilities-to-assets. Top panels plot the average response $\beta_h$; bottom panels plot the interaction coefficient $\delta_h$. Solid lines denote point estimates; dark and light blue bands show 68\% and 90\% confidence intervals, respectively.}
\end{figure}

%%%%%%%%%%%%%%%%%%%%%%%%%%%%%%%%%%%%%%%%%%%%%%%%%%%%%%%%%%%%%%%%%%
\bigskip
\noindent
\textbf{Bank-level results.}
Figure~\ref{fig:Ratios_Leverage_2by2} reports the estimated coefficients $\beta_h$ and $\delta_h$ from Equation~\ref{eq:Regression_Bank_Level}. The top panels show that a positive global credit-supply shock generates a strong and persistent expansion in lending. The response is hump-shaped—peaking between 10 and 24 months—and remains statistically significant over multiple horizons. This mirrors the macro-level results in Section~\ref{sec:macro_impact}, where global financial easing translated into sustained increases in domestic credit.

The bottom panels display the interaction coefficient $\delta_h$, which captures how the lending response varies with bank leverage. Across both leverage definitions, $\delta_h$ is positive and statistically significant over a broad range of horizons. Less-leveraged (better-capitalized) banks systematically expand lending more following a favorable global shock. The dynamic profile of $\delta_h$ closely tracks that of $\beta_h$, indicating that financial slack affects both the magnitude and timing of credit expansion.

The economic magnitude is sizeable. Sixteen months after the shock, the average lending response is approximately 14.8\%. Based on an interaction coefficient of 1.95 and the observed leverage distribution, a bank at the 10th percentile of leverage (3.4) expands lending by roughly 21.5\%, whereas a bank at the 90th percentile (12.7) expands lending by only 0.1\%. Balance-sheet strength therefore plays a first-order role in mediating the pass-through of global credit-supply shocks.

\bigskip
\noindent
\textbf{Robustness.}
A first robustness check adds time fixed effects to Equation~\ref{eq:Regression_Bank_Level} to absorb all domestic and global shocks common to all banks. Identification then relies solely on cross-bank differences in leverage within a given month. The resulting interaction coefficients, reported in Appendix Figure~\ref{fig:Ratios_Leverage_TimeFE}, remain positive, statistically significant, and display a nearly identical hump-shaped pattern. This confirms that our baseline results are not driven by omitted aggregate factors.

A second robustness exercise exploits within-bank variation in leverage over time. We standardize each bank’s leverage by subtracting its bank-specific mean and dividing by its bank-specific standard deviation, thereby isolating the effect of temporary deviations from a bank’s typical balance-sheet position. Appendix Figure~\ref{fig:Ratios_Leverage_2by2_Norm} shows that the interaction coefficient remains positive and significant across horizons. Sixteen months after the shock, banks operating one standard deviation below their historical leverage expand lending by 4–6 percentage points more than banks one standard deviation above. This demonstrates that leverage acts as both a cross-sectional and dynamic constraint on credit supply.

\bigskip
In Section \ref{subsec:additional_results_robustness_micro} we present several robustness exercises that reinforce the validity of our main bank-level findings. First, we show that alternative bank-level characteristics related to liquidity, capitalization, funding structure, and credit risk are fully consistent with our main results: banks that are less leveraged, less indebted, or more liquid exhibit stronger lending responses to global credit supply shocks. Second, we demonstrate that the role of leverage remains robust even when these alternative dimensions are considered jointly. Finally, we examine the sign dependence of the transmission mechanism, documenting that expansionary credit supply shocks play a particularly prominent role in driving bank-level lending dynamics.

%%%%%%%%%%%%%%%%%%%%%%%%%%%%%%%%%%%%%%%%%%%%%%%%%%%%%%%%%%%%%%%%%%%%%%%%%%%%%%%%%%%%%%%%%%%
\subsection{Firm-Level Analysis} \label{subsec:micro_firms}

\noindent
\textbf{Data description.}
Data description.  
We use administrative loan-level data from Uruguay’s credit registry covering April 2003 to December 2019. The dataset provides monthly information for all formal financial institutions operating in the country. For each firm–bank–currency relationship, we observe outstanding credit, currency of denomination, maturity at origination, collateral usage and characteristics (including type, currency, and whether it is marked to market), the bank’s internal risk rating of the borrower, loan status, and borrower and lender identifiers. This rich structure allows us to construct detailed measures of firms’ leverage, currency exposure, collateral coverage and liquidity, maturity structure, banking relationships, and credit dynamics.

Table~\ref{tab:summary_statistics_firm} reports summary statistics for key firm-level characteristics, computed as time averages for each firm and then summarized across firms.
\begin{table}[ht!]
\centering
\caption{Firm-Level Summary Statistics}
    \small 
    \label{tab:summary_statistics_firm}
\begin{tabular}{lcccc}
\hline\hline
 & p10 & Mean & Median & p90 \\ 
\hline
\textbf{Currency Composition} \\[1mm]
Share of Foreign-Currency Debt & 0.09 & 0.47 & 0.47 & 0.90 \\[2mm]

\textbf{Leverage} \\[1mm]
Debt-to-Collateral Ratio & 0.23 & 2.16 & 1.41 & 4.75 \\[2mm]

\textbf{Maturity Structure} \\[1mm]
Share of Short-Term Debt  & 0.00 & 0.18 & 0.14 & 0.45 \\
Share of Medium-Term Debt & 0.00 & 0.14 & 0.09 & 0.35 \\
Share of Long-Term Debt   & 0.00 & 0.14 & 0.05 & 0.42 \\[2mm]

\textbf{Collateral Characteristics} \\[1mm]
Share of Collateral in Foreign Currency & 0.76 & 0.96 & 1.00 & 1.00 \\
Share of Illiquid Collateral            & 0.11 & 0.72 & 0.89 & 1.00 \\[2mm]

\textbf{Banking Relationships} \\[1mm]
Average Number of Banks per Month & 1.00 & 1.21 & 1.00 & 2.00 \\[2mm]

\textbf{Loan Growth Rates (\%)} \\[1mm]
Month-Ahead LC Loan Growth & -17.75 & -2.96 & -1.59 & 9.90 \\
Month-Ahead FC Loan Growth & -21.86 & -5.92 & -3.11 & 3.81 \\
\hline\hline
\end{tabular}
\floatfoot{\textbf{Note:} Each row reports the cross-sectional distribution of firm-level time-averaged characteristics over 2003--2019. Growth rates are log differences multiplied by 100. ``Short-Term'' denotes maturities under 90 days; ``Medium-Term'' 90 days to one year; and ``Long-Term'' above one year. Illiquid collateral refers to collateral that is not marked to market.}
\end{table}
Several features stand out. Foreign-currency borrowing is pervasive: the median firm holds 47 percent of its debt in foreign currency, with the upper decile exceeding 90 percent. Leverage is high and heterogeneous. The median debt-to-collateral ratio is 1.41, while the top decile exceeds 4.7. The maturity structure is short-term oriented, with short-term credit representing roughly 14 percent of total borrowing for the median firm and 45 percent for firms in the 90th percentile.

Collateral characteristics also reveal a distinctive pattern. Although nearly all pledged collateral is denominated in foreign currency, most of it is not marked to market. The median firm has 89 percent of its collateral in illiquid form, and even firms at the 10th percentile rely on illiquid collateral for more than 10 percent of pledged assets. This rigidity in collateral valuation is an important feature of the financial environment and interacts closely with firms’ ability to borrow.

The intensity of banking relationships is limited. The median firm borrows from only one bank in a typical month, and even the 90th percentile borrows from just two, indicating a highly concentrated financial ecosystem. Finally, loan growth rates display substantial dispersion. At the monthly horizon, local-currency borrowing contracts by 1.6 percent for the median firm, but firms in the upper decile expand credit by nearly 10 percent; foreign-currency borrowing shows similarly wide variation. This cross-sectional and time-series heterogeneity provides a rich environment for studying the transmission of global credit supply shocks.

Together, these patterns highlight the main dimensions along which firms differ—currency exposure, leverage, maturity structure, collateral liquidity, and access to banks—and motivate the heterogeneous responses examined in the firm-level empirical analysis that follows.

\bigskip
\noindent
\textbf{Econometric specification.  }
To study heterogeneous firm-level responses to global credit-supply shocks, we estimate local-projection regressions at the firm–bank–currency level. The benchmark specification is
\begin{align} \label{eq:Benchmark_LP_Firm}
    \Delta \ln(\text{Loan}_{i,b,c,t+h}) 
    &=  \delta_h \, \big( \text{Shock}^{\text{Supply}}_t \times X_{i,b,c,t-1} \big)
    + \mu_c + \mu_{i,t} + \mu_{b,t} + \Gamma_{i,b,c,t-1} + \varepsilon_{i,b,c,t+h},
\end{align}
where $\Delta \ln(\text{Loan}_{i,b,c,t+h})$ denotes the log change in outstanding credit between firm $i$ and bank $b$ in currency $c$ at horizon $h$. The variable $\text{Shock}^{\text{Supply}}_t$ is the global bank credit-supply shock, normalized so that positive values correspond to easing episodes. The interaction term $\text{Shock}^{\text{Supply}}_t \times X_{i,b,c,t-1}$ captures differential credit responses based on lagged firm- or loan-level characteristics $X_{i,b,c,t-1}$, such as leverage, collateral liquidity, maturity, or currency denomination.

The specification includes currency fixed effects ($\mu_c$), firm–time fixed effects ($\mu_{i,t}$) that absorb contemporaneous borrower-specific demand shocks and macro exposures, and bank–time fixed effects ($\mu_{b,t}$) that control for time-varying differences in banks’ lending policies and funding conditions. Because the global shock is common across all firms in a given month, its main effect is absorbed by the fixed effects. Identification of $\delta_h$ therefore comes entirely from cross-sectional heterogeneity in $X_{i,b,c,t-1}$. Lagged controls ($\Gamma_{i,b,c,t-1}$) include firm- and bank-level variables that may jointly influence credit outcomes. Standard errors are clustered at the firm and date levels.

\begin{figure}[ht]
\centering
\includegraphics[width=0.75\linewidth]{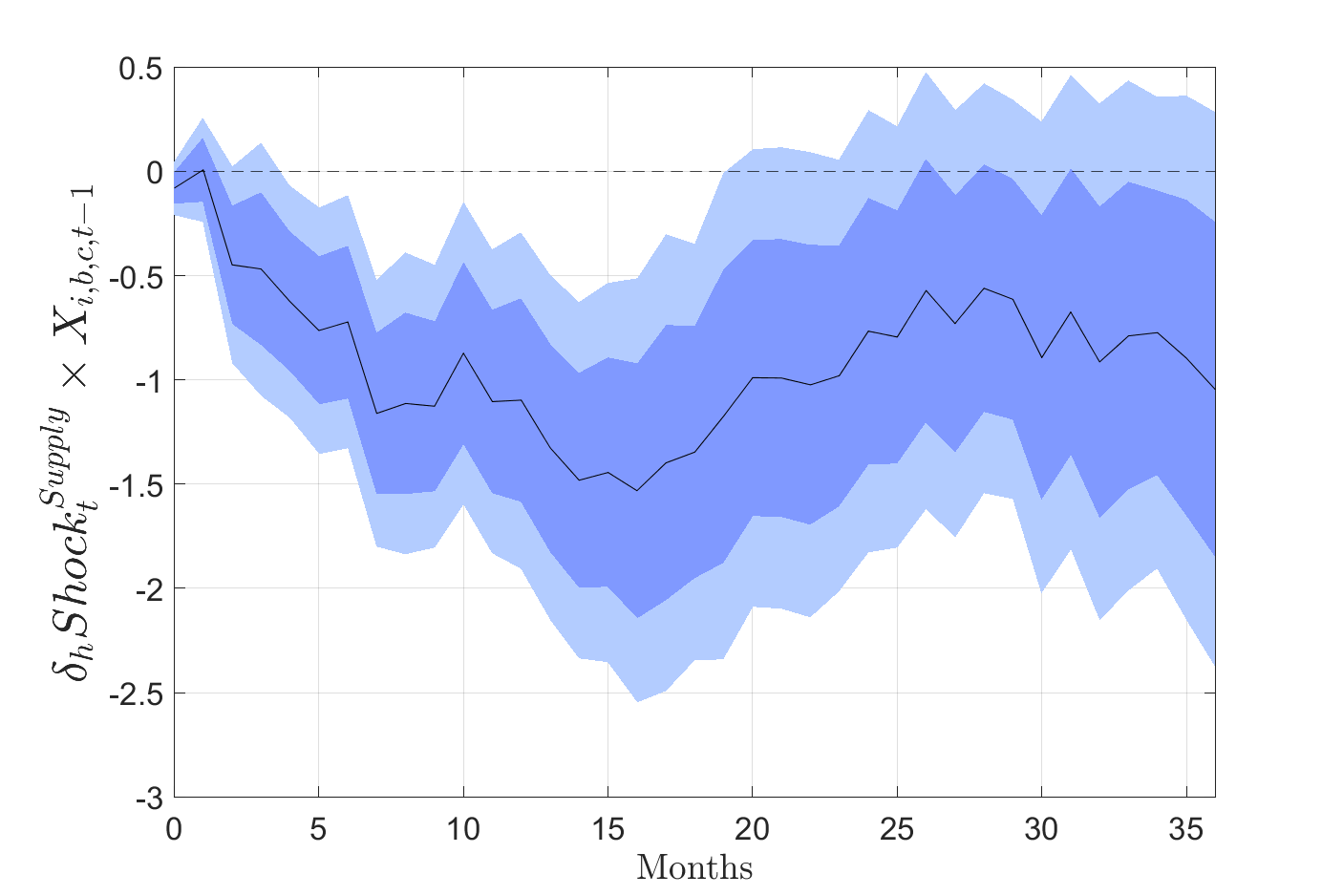}
\caption{Firm-Level Results -- Role of Total Leverage}
\label{fig:Total_Leverage}
\floatfoot{\textbf{Note:} The figure reports the estimated differential coefficient $\delta_h$ from Equation~\ref{eq:Benchmark_LP_Firm}, using total leverage (total loans to total collateral) as the interaction term. The solid black line denotes the point estimate; shaded areas show 68\% and 90\% confidence intervals.}
\end{figure}

%%%%%%%%%%%%%%%%%%%%%%%%%%%%%%%%%%%%%%%%%%%%%%%%%%%%%%%%%%%%%%%%%%%%%%%%%%%%%%%%%%%%%%%%%%%
\bigskip
\noindent 
\textbf{Result \#1: The role of total leverage.} \\
\textit{More leveraged firms benefit less from global credit supply expansions.}  
We begin by examining how the transmission of global credit supply shocks varies with firms’ initial leverage. Figure~\ref{fig:Total_Leverage} plots the heterogeneous responses when total leverage—defined as total loans over total collateral, is used as the interaction term in Equation~\ref{eq:Benchmark_LP_Firm}.

The coefficients are uniformly negative across horizons and statistically significant for most of the projection window, indicating that firms with higher leverage expand borrowing less following a positive global credit supply shock. The effect grows in magnitude over time, peaking between 12 and 20 months after the shock.

The implied economic magnitudes are sizable. Moving from the 10th percentile of leverage (0.26) to the 90th percentile (2.66) materially reduces loan growth following a positive shock. Over a one-year horizon, the difference in predicted credit growth spans several percentage points. Firms operating with larger collateral buffers therefore benefit much more from favorable global credit conditions, while those near their leverage limits experience a markedly muted pass-through.

This pattern is consistent with binding firm-level borrowing constraints: improvements in global intermediaries’ balance sheets ease external financing conditions, but firms closer to their collateral limits are unable to scale up borrowing to the same extent. The micro evidence thus mirrors the bank-level results, where better-capitalized intermediaries transmit global credit easing more forcefully. This result is consistent with our stylized model, where a higher effective repayment burden \(D_{0i}\) raises the default threshold and lowers the marginal value of new borrowing, reducing the sensitivity \(\partial k_i^*/\partial n_0\).

\begin{figure}[ht]
\centering
\includegraphics[width=1.\linewidth]{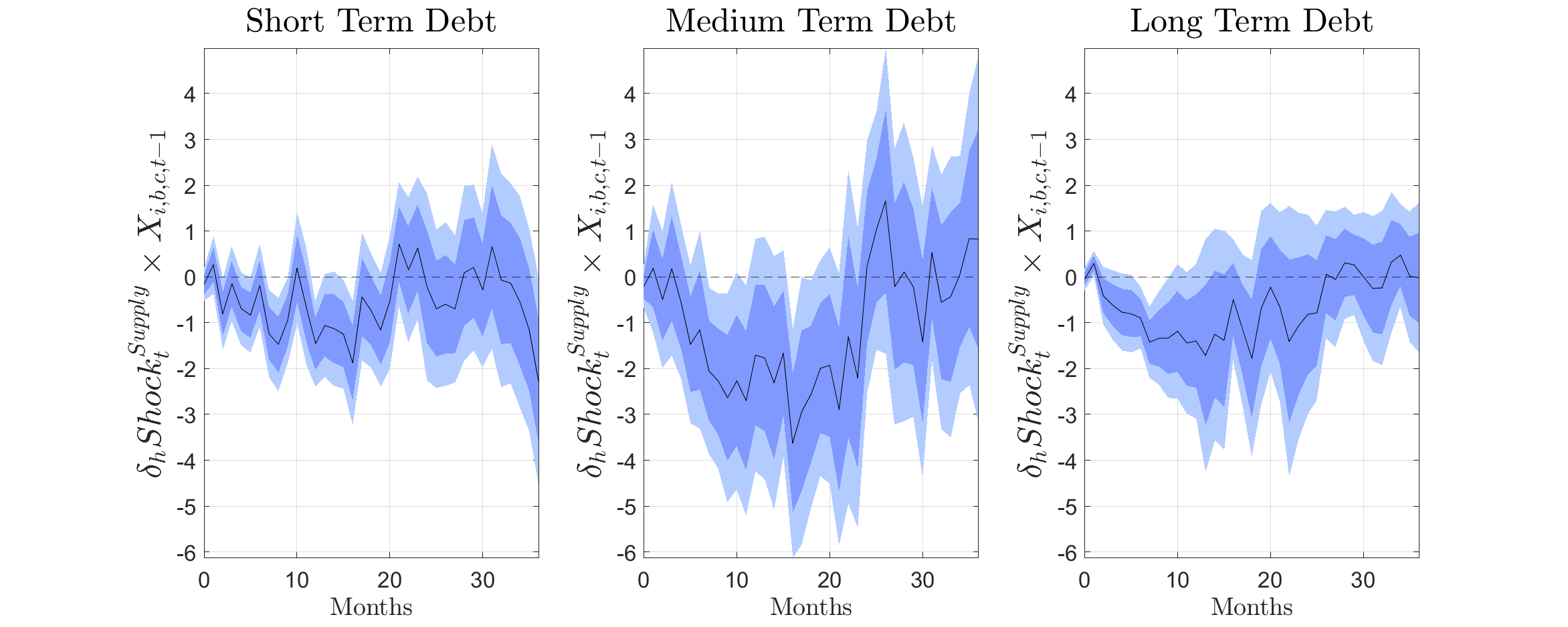}
\caption{Firm-Level Results -- Maturity Structure of Leverage}
\label{fig:Comparison_Leverage_STMTLT}
\floatfoot{\textbf{Note:} Each panel plots the estimated interaction coefficient $\delta_h$ using (left to right) short-, medium-, and long-term leverage ratios. The solid line denotes the point estimate; shaded areas show 68\% and 90\% confidence intervals.}
\end{figure}

\bigskip
\noindent
\textbf{Result \#2: Maturity structure of leverage.} \\
\textit{Medium-term leverage appears most sensitive to credit supply shocks.}  
We next decompose total leverage into short-, medium-, and long-term components, each defined as the ratio of loans in a given maturity bucket to total collateral. Figure~\ref{fig:Comparison_Leverage_STMTLT} plots the resulting differential responses.

While all three leverage measures yield negative coefficients across much of the horizon, medium-term leverage produces the largest and most persistent attenuation. Its coefficients fall further into negative territory and remain sizable over the 10–20-month window. Short-term leverage displays weaker and more volatile effects, while long-term leverage shows a similar differential impact but is more precise.

These patterns suggest that refinancing and rollover risks at intermediate maturities play a central role. Firms with large medium-term exposures appear especially constrained, leading to a stronger attenuation of the shock’s transmission relative to firms whose leverage is concentrated at very short or very long maturities. In relation to the stylized model, the results suggest that greater legacy debt, \(D_{0i}\), increases the effective burden of firms, weakening the response to improvements in global bank net worth through a smaller decline in the multiplier \(\mu\).

\begin{figure}[ht]
\centering
\includegraphics[width=0.85\linewidth]{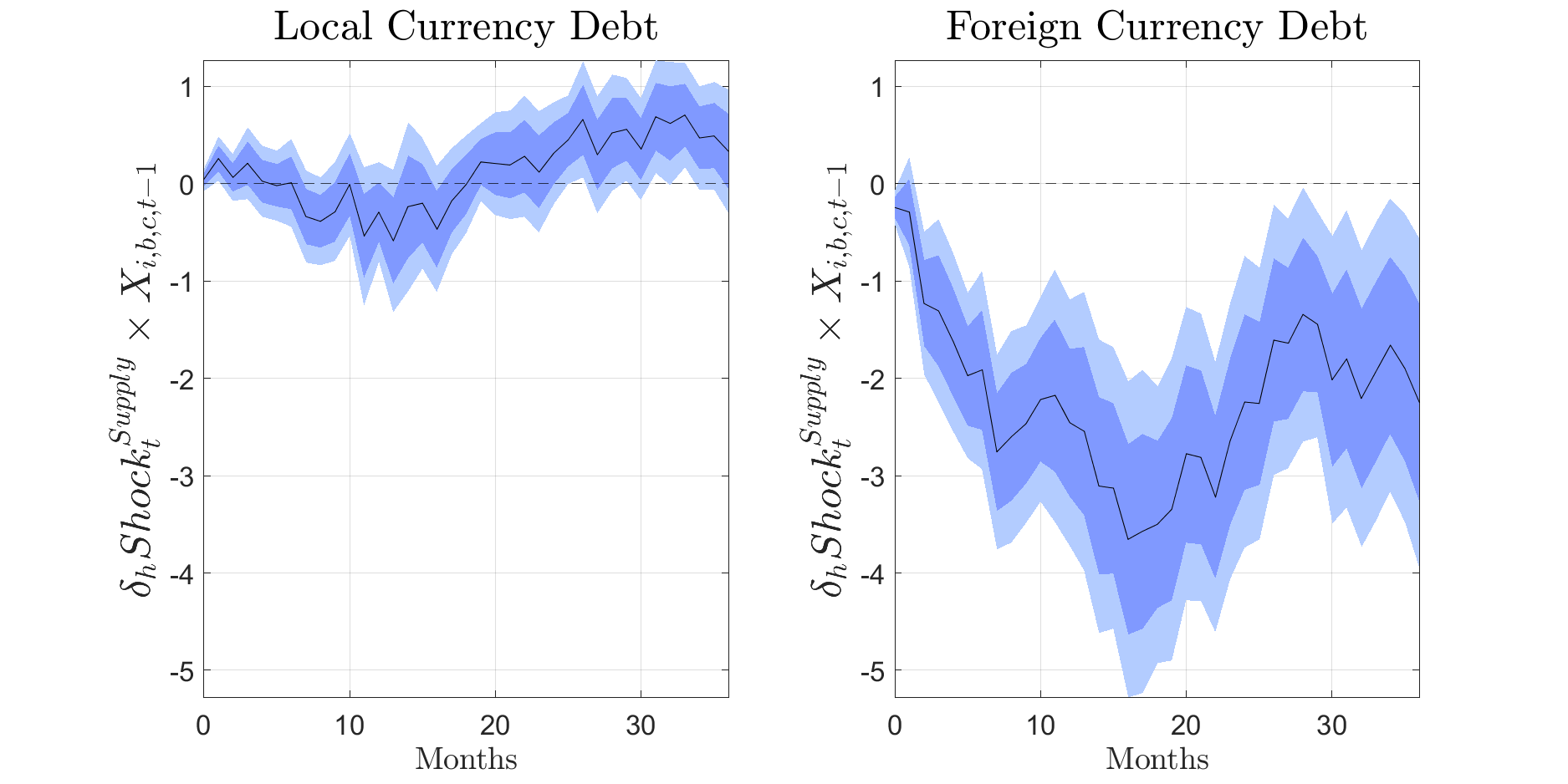}
\caption{Firm-Level Results -- Currency Structure of Leverage}
\label{fig:Comparison_Leverage_LCFC}
\floatfoot{\textbf{Note:} Each panel plots the estimated differential coefficient $\delta_h$ using (left) local-currency leverage and (right) foreign-currency leverage. Shaded areas show 68\% and 90\% confidence intervals.}
\end{figure}

\bigskip
\noindent
\textbf{Result \#3: Currency composition of leverage.} \\
\textit{Heterogeneous transmission is driven primarily by foreign-currency leverage.}  
We next examine whether the currency denomination of debt affects the strength of the leverage channel. Figure~\ref{fig:Comparison_Leverage_LCFC} reports results using local- and foreign-currency leverage ratios.

A sharp asymmetry emerges. Local-currency leverage yields small and mostly insignificant coefficients beyond the very short run, suggesting that local-currency liabilities play a limited role in shaping exposure to global credit conditions. By contrast, foreign-currency leverage produces large, negative, and persistent coefficients across nearly all horizons. Firms with elevated foreign-currency leverage expand credit materially less following a positive global credit supply shock, with the differential effect peaking between 10 and 20 months. These magnitudes are comparable to those observed using total leverage.

The results indicate that the heterogeneous transmission of global financial conditions is driven primarily by foreign-currency borrowing—precisely the segment where global banks are the principal lenders and where valuation and default risks are more sensitive to global conditions.

\begin{figure}[ht]
\centering
\includegraphics[width=0.7\linewidth]{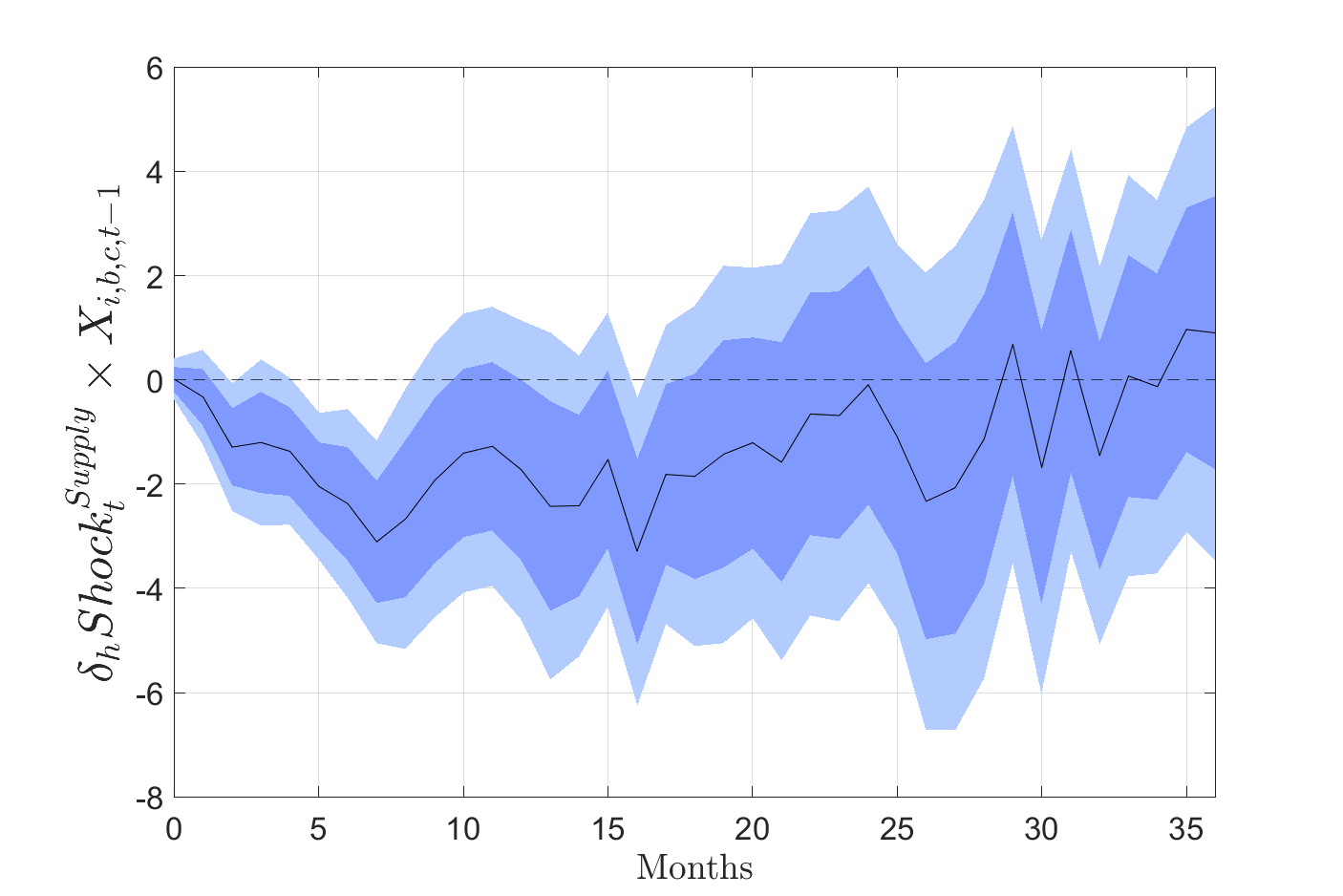}
\caption{Firm-Level Results -- Within-Firm Leverage Fluctuations}
\label{fig:Norm_Leverage}
\floatfoot{\textbf{Note:} The interaction term uses the firm-specific standardized leverage measure. Shaded areas show 68\% and 90\% confidence intervals.}
\end{figure}

%%%%%%%%%%%%%%%%%%%%%%%%%%%%%%%%%%%%%%%%%%%%%%%%%%%%%%%%%%%%%%%%
\bigskip
\noindent
\textbf{Result \#4: Within-firm leverage fluctuations.} \\
\textit{Leverage remains a binding constraint even within firms over time.}  
To assess whether the leverage channel reflects persistent cross-sectional differences or also operates within firms, we standardize each firm’s leverage relative to its own historical mean and standard deviation. Figure~\ref{fig:Norm_Leverage} presents the results.

The coefficients are consistently negative, with magnitudes between $-2$ and $-4$ at horizons of 5–15 months. When a firm is temporarily more leveraged than usual, it expands borrowing significantly less following a positive global credit supply shock. This result indicates that leverage is a dynamic, state-dependent constraint: even within the same firm, temporary increases in leverage meaningfully weaken the pass-through of global credit easing.

%%%%%%%%%%%%%%%%%%%%%%%%%%%%%%%%%%%%%%%%%%%%%%%%%%%%%%%%%%%%%%%%
\bigskip
\noindent
\textbf{Result \#5: Collateral computability.} \\
\textit{Collateral that is not priced to market sharply attenuates the transmission of global credit easing.}  
We next examine whether lenders’ ability to value pledged collateral affects firms’ responses. We use the share of collateral not priced to market as the interaction term and then include both leverage and computability jointly. 

\begin{figure}[ht]
\centering
\includegraphics[width=1.\linewidth]{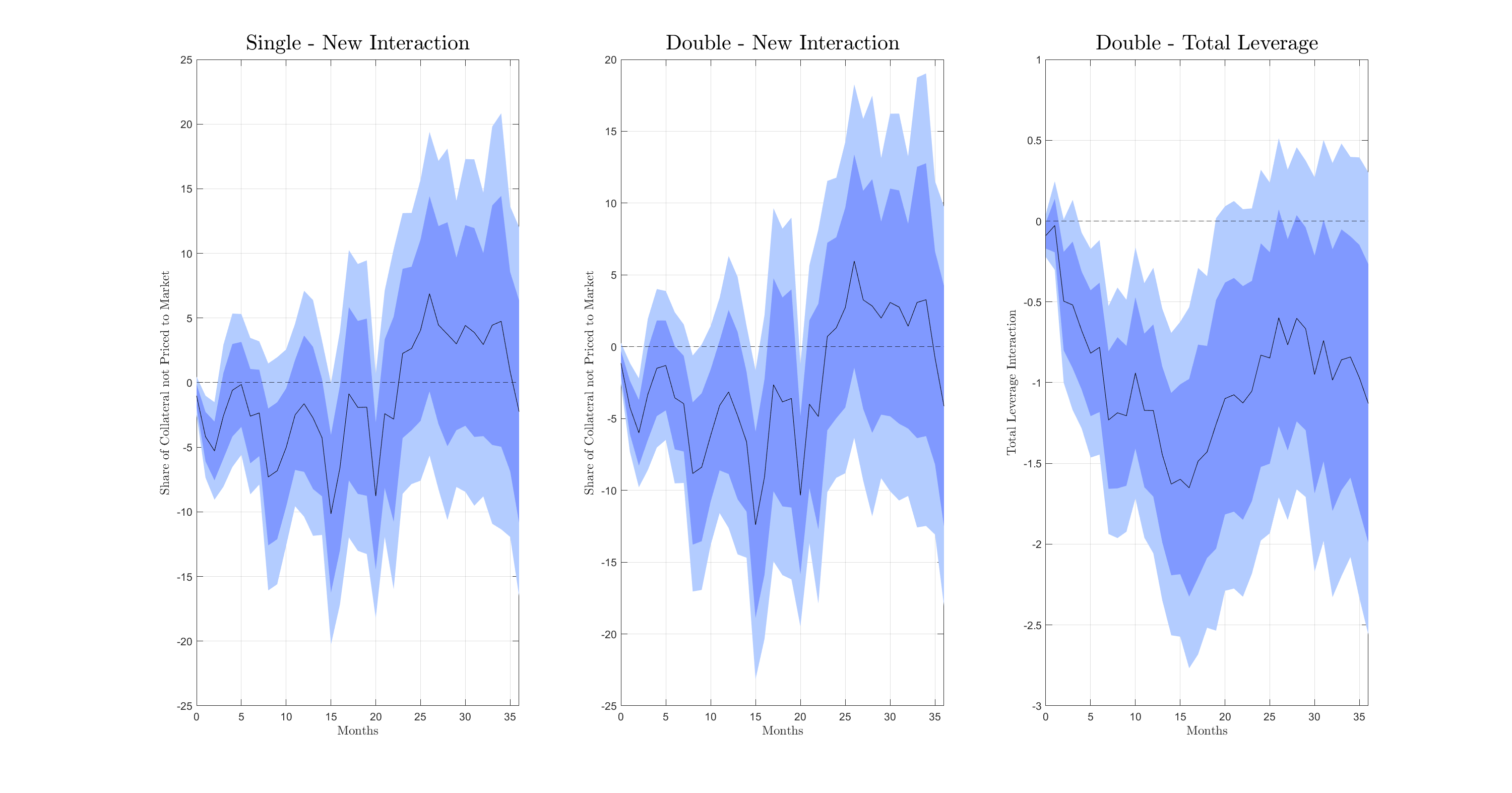}
\caption{Firm-Level Results -- Leverage \& Collateral Quality}
\label{fig:Leverage_Quality_NoComp}
\floatfoot{\textbf{Note:} The first column shows specifications replacing the leverage interaction with the collateral-quality interaction. The second and third columns show double-interaction specifications including both leverage and collateral quality. Shaded areas denote 68\% and 90\% confidence intervals.}
\end{figure}

Figure~\ref{fig:Leverage_Quality_NoComp} reports the results.
In the single-interaction specification, firms pledging a higher share of non-computable collateral exhibit markedly weaker borrowing responses, with large negative coefficients across horizons. When both leverage and computability are included, the computability coefficient remains large and negative, while the leverage coefficient retains the expected sign but is smaller. Thus, valuation risk constitutes an independent constraint on the transmission of global credit easing. In the stylized model, this corresponds to lower pledgeability, represented by a smaller \(\theta_i\), which raises the pricing term \(\mu(1-\theta_i)\) and limits the decline in borrowing costs when global bank net worth rises.

%%%%%%%%%%%%%%%%%%%%%%%%%%%%%%%%%%%%%%%%%%%%%%%%%%%%%%%%%%%%%%%%
\bigskip
\noindent
\textbf{Summary.}
Across all specifications, a clear pattern emerges: improvements in global banks’ balance sheets stimulate firm borrowing, but the strength of the response depends critically on firms’ initial financial positions. Firms with lower leverage, more liquid and easily valued collateral, and smaller foreign-currency exposures benefit substantially more from global credit easing. These findings hold both across firms and within firms over time, underscoring the state-dependent nature of borrowing constraints.

\medskip
Additionally, the firm-level heterogeneity mirrors the bank-level evidence documented above. Just as better-capitalized banks transmit global credit easing more powerfully, financially stronger firms are better positioned to take advantage of improved external conditions. The domestic transmission of global bank net-worth shocks thus operates through a balance-sheet amplification mechanism on both sides of the credit relationship, with the largest effects arising when both lenders and borrowers enter the shock in strong financial positions.

%%%%%%%%%%%%%%%%%%%%%%%%%%%%%%%%%%%%%%%%%%%%%%%%%%%%%
\section{Additional Results \& Robustness Checks} \label{sec:Additional_Robustness}

To complement the benchmark findings, this section presents additional results and robustness checks at both the macro and micro levels. Our goal is to demonstrate that the paper's main conclusions—namely, that increases in global banks’ net worth ease external financing conditions in EMEs and stimulate real activity—are not driven by specific modeling choices, sample selection, or estimation strategies. At the macro level, we 
show how crucial it is to exploit the co-movement between the net worth surprises and the EBP, and the robustness of our results to alternative econometric methods. At the micro level, we assess the robustness of the firm-bank-currency regressions to alternative measures of firm heterogeneity, specification changes, fixed-effects structures, and subsample restrictions. Across exercises, the results consistently reinforce the paper's core message: global bank net-worth shocks have economically meaningful and statistically robust effects on both aggregate and firm-level borrowing conditions in Emerging Market Economies.

%%%%%%%%%%%%%%%%%%%%%%%%%%%%%%%%%%%%%%%%%%%%%%%%%%%%
\subsection{Macro Level Results \& Robustness } \label{subsec:additional_results_robustness_macro}

This subsection presents a broad set of macro-level robustness checks designed to assess the stability of our benchmark results to alternative identification strategies, shock definitions, VAR configurations, and Local Projection specifications. Across all exercises, the key findings remain unchanged: increases in global banks’ net worth that comove negatively with the Excess Bond Premium (EBP) lead to an appreciation of the exchange rate, a decline in external borrowing costs, and an expansion in real activity across Emerging Market Economies (EMEs). We show below that these results are not driven by modeling assumptions, functional-form restrictions, or the specific construction of the shock.

\medskip
\noindent
\textbf{Spillovers of an increase in net worth associated with a positive co-movement with the EBP.}
In Section~\ref{sec:identification_strategy} we argued that a global bank credit supply shock is characterized by a negative co-movement between banks’ net-worth surprises and the EBP. We now show that exploiting this co-movement is crucial for correctly identifying a global bank credit supply shock.

Figure~\ref{fig:Pooled_EME_CD} presents the impulse responses to a one–standard–deviation increase in global banks' net worth associated with a positive co-movement with the EBP.\footnote{We construct this series as the difference between the raw net-worth surprise and its component associated with a negative co-movement with the EBP.} The figure makes clear that the co-movement is highly informative: the impulse responses are qualitatively opposite to those in Figure~\ref{fig:Pooled_EME}. A positive co-movement shock leads to a depreciation of the exchange rate and a contraction in investment, industrial production, and trade flows. These results highlight that only the component of net-worth surprises displaying a negative co-movement with the EBP captures an expansion in global credit supply, underscoring the necessity of our identification strategy.

\medskip
\noindent
\textbf{Impulse responses to the total net-worth surprise.}
We also estimate impulse responses to the raw net-worth surprise, which combines both components. Figure~\ref{fig:Pooled_EME_Outside} shows that the resulting responses are small and often statistically insignificant, consistent with the raw surprise mixing expansionary and contractionary components that partly offset one another. This reinforces our interpretation that only the negative co-movement component isolates a global credit-supply expansion.

\medskip
\noindent
\textbf{Average or group mean VAR specification.}
We next assess robustness to an alternative VAR specification. In an average, or group mean, VAR, variables are de-meaned by country and then averaged across EMEs. A Bayesian VAR with Minnesota priors is estimated on these aggregated data. Figure~\ref{fig:Average_EME} shows that a global credit-supply shock again leads to an appreciation of the exchange rate and an expansion of economic activity, driven by investment. Figures~\ref{fig:Average_EME_Financial_Prices} and \ref{fig:Average_EME_Financial_Quantities} show analogous results for financial prices and quantities. The similarity of these results to the benchmark pooled VAR indicates that our findings are not sensitive to the panel structure or cross-country weighting.

\medskip
\noindent
\textbf{Robustness to lag structure in the VAR specification.}
Figures~\ref{fig:Pooled_EME_Lags1} and \ref{fig:Pooled_EME_Lags4} show that our baseline results are robust to using either a single lag or four lags in the pooled VAR. In all cases, a positive global credit-supply shock leads to an appreciation of the exchange rate and an expansion in economic activity, particularly through investment and trade. The stability of these responses indicates that our conclusions are not driven by dynamic misspecification or by the choice of lag length.

\medskip
\noindent
\textbf{Alternative LP specifications.}
We also examine several alternatives to the benchmark Local Projection specification in Equation~\ref{eq:LP_Reg}. Specifically, we estimate LPs with: (i) time fixed effects that absorb global shocks and seasonal patterns, (ii) country-specific linear and quadratic trends that flexibly capture slow-moving domestic dynamics, and (iii) specifications that omit lagged values of the credit-supply shock. Figures~\ref{fig:CountryFE_Regressions_Comparison}, \ref{fig:Trends_Regressions_Comparison}, and \ref{fig:NoShockLags_Regressions_Comparison} show that the estimated responses remain highly stable across all alternatives. In each case, a global credit-supply expansion leads to an appreciation of the exchange rate and a broad-based expansion in real activity.

\medskip
\noindent
\textbf{IV Local Projection specification.}
As an additional robustness exercise, we estimate an instrumental-variables Local Projection (IV-LP) specification in which the Excess Bond Premium (EBP) is instrumented with the global credit-supply shock, following \cite{ottonello2022financial}.\footnote{The first-stage F-statistic exceeds 5 on average across horizons, indicating sufficient instrument relevance.} Figure~\ref{fig:IV_Regressions_Comparison} shows that the resulting impulse responses closely mirror those from the benchmark VAR and OLS LPs. The main difference is a hump-shaped appreciation of the nominal exchange rate under the IV-LP specification. Investment and trade flows display strong and persistent increases, and Figures~\ref{fig:IV_Additional_Variables_Prices_Comparison} and~\ref{fig:IV_Additional_Variables_Quantities_Comparison} confirm that responses of financial prices and quantities are likewise robust. These results demonstrate that our findings do not hinge on functional-form assumptions or on the reduced-form dynamics of the VAR.

\medskip
\noindent
\textbf{Orthogonality to U.S. monetary policy shocks and other financial shocks.}
Finally, we verify that our credit-supply shock is not contaminated by high-frequency U.S. monetary policy surprises. Figures~\ref{fig:Quarterly_CoMovement} and~\ref{fig:Quarterly_CoMovement_Lag} plot the credit-supply shock against the contemporaneous and one-quarter-lagged Bauer--Swanson (2023) shocks. In both cases, the fitted regression line is essentially flat and correlations are economically negligible: $-0.1463$ contemporaneously and $0.0625$ with a lag. This confirms that the credit-supply shock is orthogonal to monetary policy innovations and information effects, reinforcing our interpretation that it captures shifts in global banks’ balance-sheet capacity.

\medskip
\noindent
\textbf{Comparison with alternative financial-shock measures.}
To relate our identification strategy to prominent alternatives in the literature, we compare our purged credit-supply shock to two commonly used financial disturbances: the broker-dealer leverage
shock of \citet{cesa2018international} and the risk-appetite shock extracted from the VXO following \citet{di2022international}. For each case, we estimate a Bayesian VAR with Normal–Wishart priors
using quarterly U.S.\ data from 1986Q1 to 2019Q4—the longest sample for which the VXO is available and comparable to the coverage in \citet{cesa2018international}. The VAR includes, in addition to the
case-specific variable ordered first, the log dollar index, industrial production (log), CPI (log), and the federal funds rate. Implicitly, the recursive (Cholesky) structure allows the ordered-first variable to contemporaneously affect all others but not vice versa.

For each VAR, we recover the median structural shock implied by the Cholesky identification and compute its correlation with our baseline credit-supply shock. The correlation with the broker-dealer–first shock is modest (0.257), while the correlation with the VXO-first shock is negative and similarly small in magnitude (-0.322). The scatter plots in Figure~\ref{fig:CoMovement_Other_Financial_Shocks} illustrate the weak linear relationship in both cases. These results indicate that the disturbances extracted from recursive VARs using either broker-dealer leverage or the VXO capture different underlying forces than the high-frequency, net-worth–based credit-supply shocks that we employ.

This distinction is consistent with the interpretation that VAR innovations in leverage or the VXO reflect broader movements in global financial conditions, combining risk appetite, macroeconomic
expectations, and asset-demand shocks, while our purged credit-supply shock isolates high-frequency changes in intermediaries' balance-sheet capacity that move credit spreads and the excess bond 
premium in opposite directions. Accordingly, the two approaches  should be viewed as complementary, rather than substitutes, with ours focusing on a conceptually narrower source of variation that is more tightly linked to the mechanism in Section~\ref{subsec:stylized_model}.

\medskip
\noindent
\textbf{Summary.}
Across all identification strategies, shock definitions, econometric specifications, and sample adjustments considered in this subsection, the macroeconomic responses to global bank credit-supply shocks remain remarkably stable. The exchange rate appreciates, external borrowing costs fall, and investment and trade expand following a positive credit-supply shock. These robust patterns provide strong empirical support for the macroeconomic transmission mechanisms highlighted in the paper.

%%%%%%%%%%%%%%%%%%%%%%%%%%%%%%%%%%%%%%%%%%%%%%%%%%%%
\subsection{Micro Level Results \& Robustness } \label{subsec:additional_results_robustness_micro}

\subsubsection{Bank Level Results} 

\noindent
\textbf{Additional interactions at the bank level.}
We complement our benchmark bank-level analysis by examining how the transmission of global credit-supply shocks varies with additional dimensions of bank balance-sheet strength and liquidity. Specifically, we consider four indicators capturing distinct aspects of financial flexibility: (i) the ratio of liquid assets to total assets, (ii) the ratio of net worth to required regulatory capital, (iii) the share of demand deposits in total deposits, and (iv) the ratio of non-performing loans to total loans.

Figure~\ref{fig:Ratios_Other} presents the results from estimating Equation~\ref{eq:Regression_Bank_Level}, replacing the leverage interaction term with interactions between the credit-supply shock and each of the four alternative indicators. Panels (1) through (4) show a consistent pattern: banks with higher liquidity, stronger capitalization, a greater reliance on demand deposits, or higher non-performing loan ratios exhibit more pronounced lending responses to a positive credit-supply shock. These results reinforce the central role of financial heterogeneity in shaping the transmission of global financial conditions. Banks with greater balance-sheet flexibility respond more strongly to credit-supply expansions, whereas more indebted or financially constrained institutions display muted responses. This pattern aligns closely with our earlier findings on leverage and highlights the importance of financial strength in amplifying international spillovers.

\bigskip

\noindent
\textbf{Bank-level joint interaction effects.}
To assess the relative importance of leverage vis-à-vis these alternative balance-sheet dimensions, we estimate a richer specification that includes two interaction terms:
\begin{align}
    \Delta \ln \text{Loan}_{i,c,t+h} &= 
    \beta_h \times \text{Shock}^{\text{Supply}}_t 
    + \delta^{\text{Leverage}}_h \times \text{Shock}^{\text{Supply}}_t \times \text{Leverage}_{i,t-1} \nonumber \\
    &\quad + \delta^{\text{Other}}_h \times \text{Shock}^{\text{Supply}}_t \times \text{Other}_{i,t-1} 
    + \mu_{i,h} + \mu_{c,h} \nonumber \\
    &\quad + \Gamma^{\text{Bank}}_{i,t-1} + \Gamma^{\text{Macro}}_{i,t-1} + \epsilon_{i,c,t}, 
    \label{eq:Regression_Bank_Level_Double}
\end{align}
where leverage and each alternative variable are included simultaneously.

Figure~\ref{fig:Ratios_Other_Double} shows that across all specifications, the leverage interaction coefficient, $\delta^{\text{Leverage}}_h$, remains stable, negative, and statistically significant—closely resembling the magnitudes in Figures~\ref{fig:Ratios_Leverage_2by2} and~\ref{fig:Ratios_Leverage_TimeFE}. This confirms that leverage is a robust predictor of heterogeneous credit responses and a key margin through which global financial shocks influence domestic lending, even after controlling for other balance-sheet dimensions.

\bigskip

\noindent
\textbf{Domestic vs.\ foreign bank ownership.}
We next investigate whether the transmission of global credit-supply shocks differs systematically across domestic and foreign-owned banks. We interact the credit-supply shock with a dummy equal to one for banks with domestic capital and zero otherwise. Figure~\ref{fig:Ratios_Other_Nacional_Double} shows that domestically owned banks are less responsive to credit-supply expansions, particularly at short horizons. During the first six months after the shock, the interaction coefficients are negative and relatively large, indicating that domestic-capital banks expand lending by less than foreign-owned banks when global credit conditions improve. This suggests that foreign-owned banks—likely because of closer financial ties to global markets or balance-sheet structures more sensitive to global conditions—transmit global credit-supply shocks more forcefully.

Importantly, including this ownership interaction does not alter our main micro-level result. Even after controlling for systematic differences between domestic and foreign banks, the interaction between the credit-supply shock and firms’ leverage remains negative and statistically meaningful across horizons. Thus, the core finding—that more leveraged firms respond less to global credit-supply expansions—cannot be attributed to the ownership structure of the banks through which they borrow. Leverage remains a central and independent margin of heterogeneity in the micro-level transmission of global financial conditions.

\bigskip
\noindent
\textbf{Bank-level sign dependency.}
We also explore whether the transmission of credit-supply shocks is asymmetric across expansions and contractions. Specifically, we estimate:
\begin{align}
    \Delta \ln \text{Loan}_{i,c,t+h} &= 
    \beta^{Exp}_h \times \text{Shock}^{\text{Supply}}_t \times \mathbbm{1}[ \text{Shock}^{\text{Supply}}_t > 0 ] \nonumber \\
    &\quad + \beta^{Con}_h \times \text{Shock}^{\text{Supply}}_t \times \mathbbm{1}[ \text{Shock}^{\text{Supply}}_t < 0 ] \nonumber \\
    &\quad + \delta^{Exp}_h \times \text{Shock}^{\text{Supply}}_t \times \text{Leverage}_{i,t-1} \times \mathbbm{1}[ \text{Shock}^{\text{Supply}}_t > 0 ] \nonumber \\
    &\quad + \delta^{Con}_h \times \text{Shock}^{\text{Supply}}_t \times \text{Leverage}_{i,t-1} \times \mathbbm{1}[ \text{Shock}^{\text{Supply}}_t < 0 ] \nonumber \\
    &\quad + \mu_{i,h} + \mu_{c,h} + \Gamma^{\text{Bank}}_{i,t-1} + \Gamma^{\text{Macro}}_{i,t-1} + \epsilon_{i,c,t}, 
    \label{eq:Regression_Bank_Level_Sign}
\end{align}
allowing for separate responses depending on whether the shock is positive or negative.

Figures~\ref{fig:APA_Sign_Results} and~\ref{fig:PAS_Sign_Results} report the results for leverage measured as total liabilities to net worth and total liabilities to total assets. In both cases, expansionary shocks drive the heterogeneity: the interaction effects for expansions are positive and statistically significant at several horizons, whereas the interaction effects for contractions are small and statistically indistinguishable from zero. This indicates that global financial expansions, rather than contractions, generate most of the bank-level heterogeneity in credit supply.

\medskip
\noindent
We further assess robustness using a specification with time fixed effects:
\begin{align}
    \Delta \ln \text{Loan}_{i,c,t+h} &= 
    \tilde{\delta}^{Exp}_h \times \text{Shock}^{\text{Supply}}_t \times \text{Leverage}_{i,t-1} \times \mathbbm{1}[ \text{Shock}^{\text{Supply}}_t > 0 ] \nonumber \\
    &\quad + \tilde{\delta}^{Con}_h \times \text{Shock}^{\text{Supply}}_t \times \text{Leverage}_{i,t-1} \times \mathbbm{1}[ \text{Shock}^{\text{Supply}}_t < 0 ] \nonumber \\
    &\quad + \mu_{i,h} + \mu_{c,h} + \tau_t + \Gamma^{\text{Bank}}_{i,t-1} + \epsilon_{i,c,t}, 
    \label{eq:Regression_Bank_Level_Sign_Time_FE}
\end{align}
reported in Figures~\ref{fig:APA_Sign_Results_Time_FE} and~\ref{fig:PAS_Sign_Results_Time_FE}. The results are consistent: expansionary shocks generate strong and significant heterogeneity in lending responses, while contractionary shocks do not. Overall, the evidence supports the view that global financial expansions, rather than contractions, are the primary channel through which global bank net worth shocks are transmitted to domestic credit markets. Leverage critically determines which banks can take advantage of favorable global financial conditions and expand lending to firms.

%%%%%%%%%%%%%%%%%%%%%%%%%%%%%%%%%%%%%%%%%%%%%
\subsubsection{Firm Level Results} 

\noindent
\textbf{Role of credit and collateral dollarization at the firm--bank level.}
We next examine whether the heterogeneous responses documented above remain robust when allowing for additional sources of balance-sheet heterogeneity. To do so, we extend our benchmark specification by introducing a second interaction term that captures either the degree of credit dollarization, the degree of collateral dollarization, or the currency denomination of lending. Formally, we estimate:
\begin{align} \label{eq:Double_Interaction_LP_Firm}
    \Delta \ln \text{Loan}_{i,b,c,t+h} 
    &= \delta^{(1)}_h \cdot \left( \text{Shock}^{\text{Supply}}_t \times X^{(1)}_{i,b,c,t-1} \right) 
    + \delta^{(2)}_h \cdot \left( \text{Shock}^{\text{Supply}}_t \times X^{(2)}_{i,b,c,t-1} \right) \nonumber \\
    &\quad + \mu_c + \mu_{i,t} + \mu_{b,t} + \Gamma_{i,b,c,t-1} + \epsilon_{i,b,c,t+h},
\end{align}
where $X^{(1)}_{i,b,c,t-1}$ denotes the firm leverage ratio (total loans to total collateral), and $X^{(2)}_{i,b,c,t-1}$ represents one of the additional firm- or loan-level characteristics described above. The coefficients $\delta^{(1)}_h$ and $\delta^{(2)}_h$ capture heterogeneous responses to the global credit-supply shock along these two dimensions. All controls, fixed effects, and clustering structures are identical to those in the benchmark specification. This specification allows us to assess whether the leverage-based heterogeneity remains robust after accounting for other firm- or contract-level features that may also mediate the transmission of global financial conditions.

Figures~\ref{fig:ShareDoll_Double} and \ref{fig:ShareAssets_Double} report the results for the share of dollar debt and share of dollar-priced collateral, respectively. We first interact the credit-supply shock with the share of firms' dollar-denominated debt. Firms with a larger share of USD liabilities exhibit a significantly stronger expansion in borrowing following a positive shock. This pattern is consistent with a valuation channel: because the domestic currency appreciates in response to the shock (see Section~\ref{subsec:macro_main_results}), the domestic-currency value of USD debt falls, thereby relaxing firms' balance-sheet constraints. The effect remains strong and statistically meaningful even after jointly controlling for leverage, indicating that currency-of-liability exposure captures an independent dimension of balance-sheet sensitivity to global financial shocks.

In contrast, firms pledging a greater share of collateral valued in dollars respond less to the same positive shock. This negative interaction is also consistent with a valuation effect. The appreciation of the domestic currency reduces the domestic-currency value of USD-priced collateral, tightening collateral-based borrowing capacity and leading banks to remain more conservative toward these firms even as global financial conditions ease. This pattern persists in the joint specification with leverage, confirming that the result is not driven by indebtedness alone.

Across all double-interaction specifications, the leverage interaction remains stable, negative, and economically meaningful, reinforcing our main finding that more indebted firms benefit less from improvements in global credit supply. Taken together, these results show that the balance-sheet implications of global shocks operate through multiple channels—liability-side dollarization, collateral-side dollarization, and firm leverage—with the latter emerging as the most robust and quantitatively important source of heterogeneity in firm-level responses.

\noindent
\textbf{Results after the Great Financial Crisis.} 
As an additional robustness check, we examine whether the leverage-dependent transmission mechanism identified in our main results remains present in the more recent post-crisis period. To do so, we re-estimate the benchmark specification in Equation~\ref{eq:Benchmark_LP_Firm} restricting the sample to start in January 2010, after the end of the Great Financial Crisis (GFC).

Figure~\ref{fig:Comparison_After2010} presents the results. The left panel reproduces our baseline estimates using the full sample, while the right panel shows the results for the post-GFC subsample. The estimated interaction coefficient remains negative and statistically significant across both samples, confirming the robustness of our main finding that more indebted firms are less able to expand borrowing in response to positive global credit supply shocks. Importantly, the estimated effects appear somewhat larger in magnitude during the post-GFC period, with wider confidence intervals reflecting the shorter sample size. This suggests that the relevance of firm balance sheet heterogeneity may have become even more pronounced in the years following the GFC, potentially reflecting changes in global banking regulation, cross-border lending practices, and the composition of international capital flows over the past decade. In summary, these results reinforce the conclusion that firm leverage constitutes a persistent and quantitatively significant margin of heterogeneity in the transmission of global credit supply shocks to emerging market borrowers.

\bigskip
\noindent
\textbf{Heterogeneity across firm size categories.}
We also examine whether the leverage-based heterogeneity varies across firm size classes using the self-reported MiPyME size categories (Micro, Small, Medium, Large). Appendix Figure~\ref{fig:Total_Leverage_By_Size} reports sector-specific responses for the subsample of firms with valid size identifiers. The leverage interaction is negative for all size groups except micro firms, for which the estimates are noisier and close to zero, consistent with the notion that micro firms face non-collateral lending frictions that limit their ability to expand borrowing even when leverage is low. The negative differential effect of leverage is strongest for small and medium-sized firms—segments that rely heavily on bank credit and for which collateral constraints bind more tightly. Large firms display a milder and more volatile response, reflecting their greater access to alternative funding sources. Overall, the pattern across size groups confirms that leverage remains a robust predictor of firms' sensitivity to global credit-supply shocks.

\bigskip
\noindent
\textbf{Heterogeneity across sectors of economic activity.}
We further assess whether the heterogeneous responses documented above vary across firms' main sectors of activity. Because sector information is self-reported and not available for all borrowers, we estimate the leverage interaction separately for the subsample of firms with valid sector identifiers. Appendix Figure~\ref{fig:Total_Leverage_By_Sector} reports sector-specific impulse responses for Agriculture, Manufacturing, Services, Construction, and Retail, along with the pooled results for all firms reporting a sector. The leverage interaction remains negative across all sectors, confirming that the main result---that more indebted firms benefit less from improvements in global credit supply---is not driven by sectoral composition. While the magnitude of the effect varies somewhat across sectors, the overall pattern is remarkably consistent, providing additional evidence that leverage is a robust predictor of firm-level sensitivity to global financial conditions.

\bigskip
\noindent\textbf{Role of Age-Adjusted Leverage.}  
A growing body of work documents a strong empirical relationship between firm leverage and firm age, and shows that younger, more indebted firms tend to respond more strongly to financial and monetary policy shocks \citep{cloyne2018monetary,ottonello2020financial,jeenas2023firm}. Because leverage in our data naturally declines with firm--bank relationship age, we assess whether our heterogeneous effects are simply capturing this mechanical age--leverage gradient. To do so, we detrend firm--bank leverage using either a linear or a linear-plus-quadratic function of relationship age and re-estimate our benchmark specification with these detrended leverage measures. Appendix Figure~\ref{fig:Total_Leverage_Age_Adjusted} presents the resulting heterogeneous impulse responses. The magnitude, timing, and persistence of the estimates are remarkably similar to the baseline: firms with higher (age-adjusted) leverage continue to exhibit significantly weaker borrowing responses to a positive global credit supply shock. Taken together, this exercise shows that our main heterogeneous result is not driven by mechanical age patterns. Instead, it reflects economically meaningful differences in firms’ balance-sheet sensitivity to changes in global bank net worth.

\section{Concluding Remarks} \label{sec:concluding_remarks}

This paper has examined how shocks to global banks' net worth spill over into Emerging Market Economies, and through which domestic channels these shocks propagate. By combining high-frequency identification of global bank balance-sheet shocks with a macro-to-micro empirical framework, we provide evidence that improvements in global intermediaries' financial conditions ease external financing premia, stimulate cross-border bank flows, and generate broad expansions in economic activity across EMEs. These effects operate primarily through banking sectors: global banks expand their foreign exposures, domestic intermediaries accumulate external liabilities, and domestic credit grows steadily in the quarters following the shock.

A central contribution of the paper is to show that these aggregate dynamics mask substantial heterogeneity at both the bank and firm levels. Banks with stronger balance sheets, those with lower leverage, higher liquidity, or more stable funding—transmit improvements in global financial conditions more forcefully. On the borrower side, firms’ responses vary sharply with leverage, currency composition of liabilities, maturity structures, and the pledgeability of collateral. Firms with high leverage, elevated foreign-currency exposure, reliance on medium-term debt, or collateral that is not priced to market exhibit markedly weaker expansions in borrowing. These patterns are consistent with a balance-sheet mechanism in which the sensitivity of repayment thresholds and collateral valuations shapes the equity value of new borrowing when global bank net worth rises.

Taken together, the evidence points to a two-sided amplification mechanism: global shocks interact with the financial positions of both intermediaries and borrowers. Stronger banks and financially unconstrained firms benefit disproportionately from global credit easing, while weaker banks and more indebted firms transmit and absorb these shocks only partially. This mechanism highlights the importance of domestic balance-sheet heterogeneity in shaping the cross-country and cross-firm incidence of the global financial cycle. Future work could explore how macroprudential policies, foreign-currency regulation, or collateral requirements interact with global bank balance-sheet shocks, and whether similar mechanisms operate in other emerging markets where detailed micro data are available.

\bigskip

\bigskip

\bigskip

\bigskip

\noindent
\textbf{Acknowledgments.} We thank participants at the McGill University brown bag seminar, the McMaster University seminar series, the Universidad de San Andres departmental seminar, the Dalhousie University departmental seminar, the IIEP seminar series, the Banco de México Sixth Biennial Conference on Financial Stability, the Banco Central del Uruguay Jornadas Anuales de Economía, the Banco Central de la República Argentina seminar series, the BAM Workshop, and the Canadian Economic Association annual meeting for valuable comments and suggestions. We are especially grateful to Joao Santos, Javier Garcia Cicco, Damian Pierri, Juan Carlos Hallak, Lorenzo Menna (discussant), Martin Tobal, and Hernan Seoane for insightful discussions and feedback. We thank the Central Bank of Uruguay for granting secure access to the confidential credit registry data used in this study and for their assistance during the data review and disclosure process. All remaining errors are our own. The authors declare that they have no financial relationships, research support, or potential conflicts of interest related to this research. The empirical analysis relies on confidential administrative microdata accessed under strict confidentiality agreements with the corresponding authorities. All results have been reviewed to ensure that no identifying or confidential information is disclosed. Aside from the restricted access nature of the microdata, no other disclosures apply.

%%%%%%%%%%%%%%%%%%%%%%%%%%%%%%%%%%%%%%%%%%%%%%%%%%%%%%%%%%%%%%%%%%%%%%%%%%%%%%%%%%%%%%
\newpage
\bibliography{main.bib}

@article{jarocinski2020deconstructing,
  title={Deconstructing monetary policy surprises—the role of information shocks},
  author={Jaroci{\'n}ski, Marek and Karadi, Peter},
  journal={American Economic Journal: Macroeconomics},
  volume={12},
  number={2},
  pages={1--43},
  year={2020}
}

@techreport{rey2015dilemma,
  title={Dilemma not trilemma: the global financial cycle and monetary policy independence},
  author={Rey, H{\'e}l{\`e}ne},
  year={2015},
  institution={National Bureau of Economic Research}
}

@article{miranda2021transmission,
  title={The transmission of monetary policy shocks},
  author={Miranda-Agrippino, Silvia and Ricco, Giovanni},
  journal={American Economic Journal: Macroeconomics},
  volume={13},
  number={3},
  pages={74--107},
  year={2021}
}

@article{ottonello2020financial,
  title={Financial heterogeneity and the investment channel of monetary policy},
  author={Ottonello, Pablo and Winberry, Thomas},
  journal={Econometrica},
  volume={88},
  number={6},
  pages={2473--2502},
  year={2020},
  publisher={Wiley Online Library}
}

@article{gertler2015monetary,
  title={Monetary policy surprises, credit costs, and economic activity},
  author={Gertler, Mark and Karadi, Peter},
  journal={American Economic Journal: Macroeconomics},
  volume={7},
  number={1},
  pages={44--76},
  year={2015}
}

@book{canova2013panel,
  title={Panel Vector Autoregressive Models: A Survey},
  author={Canova, Fabio and Ciccarelli, Matteo},
  year={2013},
  publisher={Emerald Group Publishing Limited}
}

@article{miranda2020us,
  title={US monetary policy and the global financial cycle},
  author={Miranda-Agrippino, Silvia and Rey, H{\'e}lene},
  journal={The Review of Economic Studies},
  volume={87},
  number={6},
  pages={2754--2776},
  year={2020},
  publisher={Oxford University Press}
}

@article{camararamirez2022,
  title={The Transmission of US Monetary Policy Shocks: The Role of Investment \& Financial Heterogeneity},
  author={Camara, Santiago and Ramirez-Venegas, Sebastian},
  journal={Working Paper},
  year={2022},
  publisher={Northwestern University}
}

@techreport{ottonello2022financial,
  title={Financial intermediaries and the macroeconomy: Evidence from a high-frequency identification},
  author={Ottonello, Pablo and Song, Wenting},
  year={2022},
  institution={National Bureau of Economic Research}
}

@article{morelli2022global,
  title={Global banks and systemic debt crises},
  author={Morelli, Juan M and Ottonello, Pablo and Perez, Diego J},
  journal={Econometrica},
  volume={90},
  number={2},
  pages={749--798},
  year={2022},
  publisher={Wiley Online Library}
}

@article{di2022international,
  title={International spillovers and local credit cycles},
  author={Di Giovanni, Julian and Kalemli-{\"O}zcan, {\c{S}}ebnem and Ulu, Mehmet Fatih and Baskaya, Yusuf Soner},
  journal={The Review of Economic Studies},
  volume={89},
  number={2},
  pages={733--773},
  year={2022},
  publisher={Oxford University Press}
}

@techreport{avdjiev2017gross,
  title={Gross capital flows by banks, corporates and sovereigns},
  author={Avdjiev, Stefan and Hardy, Bryan and Kalemli-{\"O}zcan, Sebnem and Serv{\'e}n, Luis},
  year={2017},
  institution={National Bureau of Economic Research}
}

@article{bruno2015cross,
  title={Cross-border banking and global liquidity},
  author={Bruno, Valentina and Shin, Hyun Song},
  journal={The Review of Economic Studies},
  volume={82},
  number={2},
  pages={535--564},
  year={2015},
  publisher={Oxford University Press}
}

@article{camara2025international,
  title={The international monetary transmission mechanism},
  author={Camara, Santiago and Christiano, Lawrence and Dalg{\i}c, H{\"u}sn{\"u}},
  journal={NBER Macroeconomics Annual},
  volume={39},
  number={1},
  pages={65--140},
  year={2025},
  publisher={The University of Chicago Press Chicago, IL}
}

@article{cesa2018international,
  title={International credit supply shocks},
  author={Cesa-Bianchi, Ambrogio and Ferrero, Andrea and Rebucci, Alessandro},
  journal={Journal of International Economics},
  volume={112},
  pages={219--237},
  year={2018},
  publisher={Elsevier}
}

@article{baskaya2017capital,
  title={Capital flows and the international credit channel},
  author={Baskaya, Yusuf Soner and Di Giovanni, Julian and Kalemli-{\"O}zcan, {\c{S}}ebnem and Peydr{\'o}, Jos{\'e}-Luis and Ulu, Mehmet Fatih},
  journal={Journal of International Economics},
  volume={108},
  pages={S15--S22},
  year={2017},
  publisher={Elsevier}
}

@article{khwaja2008tracing,
  title={Tracing the impact of bank liquidity shocks: Evidence from an emerging market},
  author={Khwaja, Asim Ijaz and Mian, Atif},
  journal={American Economic Review},
  volume={98},
  number={4},
  pages={1413--1442},
  year={2008},
  publisher={American Economic Association}
}

@article{schnabl2012international,
  title={The international transmission of bank liquidity shocks: Evidence from an emerging market},
  author={Schnabl, Philipp},
  journal={The Journal of Finance},
  volume={67},
  number={3},
  pages={897--932},
  year={2012},
  publisher={Wiley Online Library}
}

@article{morais2019international,
  title={The international bank lending channel of monetary policy rates and QE: Credit supply, reach-for-yield, and real effects},
  author={Morais, Bernardo and Peydr{\'o}, Jos{\'e}-Luis and Rold{\'a}n-Pe{\~n}a, Jessica and Ruiz-Ortega, Claudia},
  journal={The Journal of Finance},
  volume={74},
  number={1},
  pages={55--90},
  year={2019},
  publisher={Wiley Online Library}
}

@article{jeenas2023firm,
  title={Firm balance sheet liquidity, monetary policy shocks, and investment dynamics},
  author={Jeenas, Priit},
  journal={Working paper},
  year={2023}
}

@techreport{cloyne2018monetary,
  title={Monetary policy, corporate finance and investment},
  author={Cloyne, James and Ferreira, Clodomiro and Froemel, Maren and Surico, Paolo},
  year={2018},
  institution={National Bureau of Economic Research}
}

@article{paravisini2015dissecting,
  title={Dissecting the effect of credit supply on trade: Evidence from matched credit-export data},
  author={Paravisini, Daniel and Rappoport, Veronica and Schnabl, Philipp and Wolfenzon, Daniel},
  journal={The review of economic studies},
  volume={82},
  number={1},
  pages={333--359},
  year={2015},
  publisher={Oxford University Press}
}

@article{camara2025spillovers,
  title={Spillovers of us interest rates: Monetary policy \& information effects},
  author={Camara, Santiago},
  journal={Journal of International Economics},
  volume={154},
  pages={104059},
  year={2025},
  publisher={Elsevier}
}

\newpage
\appendix

\section{Stylized Model: Full Description and Derivations}
\label{sec:appendix_stylized_model}

This appendix presents the full version of the stylized model that underpins the main text. The model draws on the structure of \cite{morelli2022global} and on the global bank credit supply interpretation of \cite{ottonello2022financial}. Its purpose is to establish two theoretical results used in our empirical analysis, namely, an increase in global bank net worth reduces emerging market loan spreads, and the relative response of firms to such a shock is not monotone in either leverage or collateral quality.

\subsection{Environment}

There are two periods, \(t=0,1\). A global bank intermediates funds between a representative United States household sector and a continuum of emerging market firms indexed by \(i\in[0,1]\). The bank enters with net worth \(n_0\) and raises deposits from United States households at the gross return \(R^d\).

\paragraph{United States households and ownership structure.}
United States households are risk neutral and own the global bank. They supply deposits and value a stable return on these deposits. The bank distributes its profits back to households as dividends. Households do not freely recapitalize the bank. Instead, equity issuance is costly in the short run.

\paragraph{Equity adjustment costs.}
Following \cite{ottonello2022financial}, issuing new equity at \(t=0\) incurs a quadratic adjustment cost
\[
\Phi(e_0) = \frac{\phi}{2} e_0^2,
\]
where \(e_0\) denotes equity issuance and \(\phi>0\). These costs imply that the bank net worth relevant for lending decisions is effectively predetermined, and unexpected movements in net worth have real consequences.

\paragraph{Risk neutrality of the global bank.}
The global bank is risk neutral and maximizes expected profits subject to the leverage constraint and the cost of adjusting equity. Since both the bank and households are risk neutral, asset pricing reflects expected repayment values and the only frictions affecting lending allocations are the leverage constraint and equity adjustment costs.

\subsection{EME Firms}

Each emerging market firm chooses capital \(k_i\) at \(t=0\) and produces
\[
y_i = \varepsilon_i A_i k_i^\alpha,
\qquad \varepsilon_i\sim U[0,1], \quad 0<\alpha<1.
\]
The firm carries legacy debt \(D_{0i}\ge 0\) and borrows \(d_i\) at price \(q_i\). Funds borrowed equal capital investment, \(k_i=d_i\). The firm repays in full if
\[
\varepsilon_i A_i k_i^\alpha \ge \frac{d_i}{q_i}+D_{0i}.
\]
Define the default threshold
\[
\varepsilon_i^* = 
\frac{(d_i/q_i)+D_{0i}}{A_i k_i^\alpha}.
\]

\paragraph{Interpretation of legacy debt.}
Although the model is written in a single good and a single currency, the term \(D_{0i}\) should be interpreted broadly as the effective repayment burden faced by firm \(i\). In our credit registry, this burden reflects several dimensions of firm balance sheet risk. Firms with a larger share of liabilities denominated in foreign currency face higher effective repayment volatility because domestic revenues are exposed to exchange rate movements. Firms with short maturity debt or large rollover needs face higher short term repayment pressures. Firms with high overall leverage have larger debt overhang. Treating \(D_{0i}\) as a sufficient statistic for these repayment risks allows the model to accommodate the empirical heterogeneity we observe without introducing additional goods, currencies, or dynamic structures.

\paragraph{Pledgeability and collateral quality.}
If firm \(i\) defaults, the bank recovers a fraction \(\theta_i\in(0,1)\) of output,
\[
\text{Recovery}_i = \theta_i \varepsilon_i A_i k_i^\alpha.
\]
The parameter \(\theta_i\) indexes collateral quality and pledgeability. Firms with more liquid collateral, longer maturity liabilities, lower leverage ratios, or collateral priced to market have higher \(\theta_i\). Firms with poor collateral or significant rollover needs have lower pledgeability. These dimensions map directly to the micro variables in our credit registry.

\subsection{United States Firm}

The bank can also lend to a representative United States borrower with high pledgeability \(\theta_{US}>\theta_i\). This captures the safer nature of advanced economy borrowers and helps rationalize why some earnings announcement shocks may contain information about non financial borrowers, which is precisely the component isolated and removed in the credit supply decomposition of \cite{ottonello2022financial}.

\subsection{Global Bank}

At \(t=0\), the global bank chooses United States loans \(b^{US}\), emerging market loans \(b(i)\), deposits \(d_0\), and equity issuance \(e_0\). The balance sheet satisfies
\[
q^{US}b^{US} + \int_0^1 q_i b(i)\,di = n_0 + d_0 + e_0 - \Phi(e_0),
\]
where \(\Phi(e_0)\) is the equity adjustment cost defined earlier. Because equity issuance is costly, the bank behaves as if its lending capacity is tightly linked to its initial net worth \(n_0\).

\subsubsection{Leverage constraint}

The bank faces a leverage or capital constraint
\[
R^d d_0 \le 
\theta_{US}q^{US}b^{US} 
+ \int_0^1 \theta_i q_i b(i)\,di
+ \lambda n_0,
\]
where \(\lambda>0\). Let \(\mu\ge0\) be the multiplier. When \(\mu>0\), the constraint binds.

\subsubsection{Loan Pricing}

Profit maximization yields the pricing condition
\[
\frac{1}{q_i} = R^d + \mu(1-\theta_i).
\]
Thus
\begin{equation}
\label{eq:spread_app}
\frac{1}{q_i} - R^d = \mu(1-\theta_i).
\end{equation}

\subsection{Net Worth and Spread Sensitivity}

Because equity issuance is costly and net worth enters the leverage constraint, increases in \(n_0\) relax the constraint and reduce \(\mu\). Formally, if the constraint binds then \(\mu'(n_0)<0\). Differentiating \eqref{eq:spread_app},
\[
\frac{\partial}{\partial n_0}
\left(\frac{1}{q_i}-R^d\right)
=
(1-\theta_i)\mu'(n_0)
<0.
\]

Thus, positive net worth shocks reduce emerging market loan spreads for all \(i\). This justifies isolating the component of net worth surprises that moves spreads and the excess bond premium in opposite directions.

\subsection{Firm Optimization}

Firm \(i\) chooses \(k_i=d_i\) to maximize expected equity
\[
\Pi_i(d_i;q_i) = 
\int_{\varepsilon_i\ge\varepsilon_i^*}
\left[
\varepsilon_iA_i k_i^\alpha - \frac{d_i}{q_i} - D_{0i}
\right]d\varepsilon_i.
\]
Let \(k_i^*(n_0)\) be the optimum. Then
\[
\frac{\partial k_i^*}{\partial n_0} > 0,
\]
and therefore aggregate emerging market investment rises with \(n_0\).

\subsection{Ambiguity in the Cross Firm Response}

Totally differentiating the first order condition gives
\[
\frac{\partial k_i^*}{\partial n_0}
=
-\,\frac{(1-\theta_i)\mu'(n_0)\,H_i(k_i^*;D_{0i},\theta_i)}{\Pi_{ii}},
\]
where \(\Pi_{ii}<0\). The term \(H_i\) captures how leverage, debt maturity, collateral quality, and rollover exposure shift the default threshold. Lower pledgeability increases the sensitivity of spreads to \(n_0\), while higher effective debt burdens reduce the marginal value of new borrowing. Depending on which effect dominates, either high leverage or low leverage firms may respond more strongly.

\subsection{Summary}

The model delivers two implications that guide our empirical analysis. First, increases in global banks net worth reduce emerging market loan spreads and expand aggregate credit. Second, the relative response of firms depends on leverage, currency mismatch, debt maturity, and collateral quality. These are precisely the dimensions we exploit in our micro level regressions.

%%%%%%%%%%%%%%%%%%%%%%%%%%%%%%%%%%%%%%%%%%%%
\newpage
\section{Additional Data Details} \label{sec:appendix_data_details}

\noindent
\textbf{Country list.} Below, Table \ref{tab:eme_countries_macro} which presents the Emerging Market Economies that are part of our benchmark sample. 
\begin{table}[htbp]
    \centering
    \caption{Emerging Market Economies in the Macro Panel}
    \label{tab:eme_countries_macro}
    \begin{tabular}{l c}
        \hline\hline
        \textbf{Country Name} & \textbf{Quarters in Sample} \\ \hline
        Argentina & 64 \\
        \textbf{Bosnia and Herzegovina} & \textbf{70} \\
        Botswana & 56 \\
        \textbf{Brazil} & \textbf{70} \\
        \textbf{Bulgaria} & \textbf{70} \\
        \textbf{Chile} & \textbf{70} \\
        \textbf{China, P.R.: Hong Kong} & \textbf{70} \\
        Colombia & 60 \\
        \textbf{Hungary} & \textbf{70} \\
        \textbf{India} & \textbf{70} \\
        \textbf{Indonesia} & \textbf{70} \\
        \textbf{Mexico} & \textbf{70} \\
        Montenegro & 56 \\
        Peru & 52 \\
        \textbf{Philippines} & \textbf{70} \\
        \textbf{South Africa} & \textbf{70} \\
        Thailand & 68 \\
        \textbf{Türkiye, Rep. of} & \textbf{70} \\ \hline\hline
    \end{tabular}
    \vspace{0.25em}
    \begin{minipage}{0.85\linewidth}
        \footnotesize \textbf{Note:} The table lists the 18 Emerging Market Economies included in the macroeconomic panel. Countries shown in bold constitute the balanced benchmark sample, available for all 70 quarters (2002Q3–2019Q4). The benchmark unbalanced panel used in the main analysis includes only countries observed for at least 40 quarters, ensuring coverage of major global financial events, including the 2008–2009 Global Financial Crisis.
    \end{minipage}
\end{table}

\bigskip

\begin{figure}[ht]
    \centering
    \caption{Evolution of Firm Bank Loans in Uruguay (2003–2019)}
    \label{fig:Aggregate_Evolution}
    \begin{subfigure}[b]{0.45\textwidth}
        \centering
        \includegraphics[width=\textwidth]{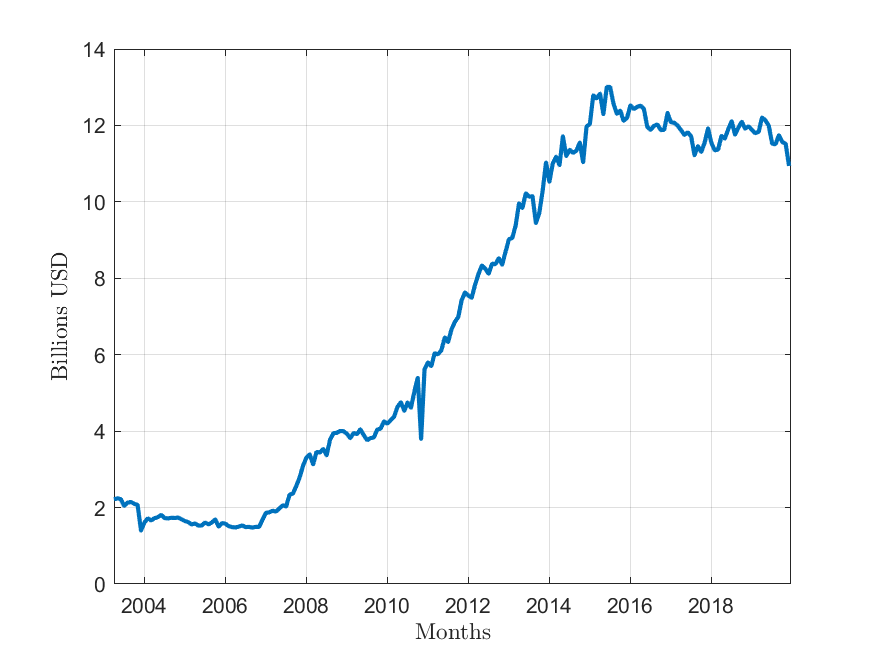}
        \caption{Foreign-Currency Loans}
        \label{fig:Aggregate_Debt_FC}
    \end{subfigure}
    \begin{subfigure}[b]{0.45\textwidth}
        \centering
        \includegraphics[width=\textwidth]{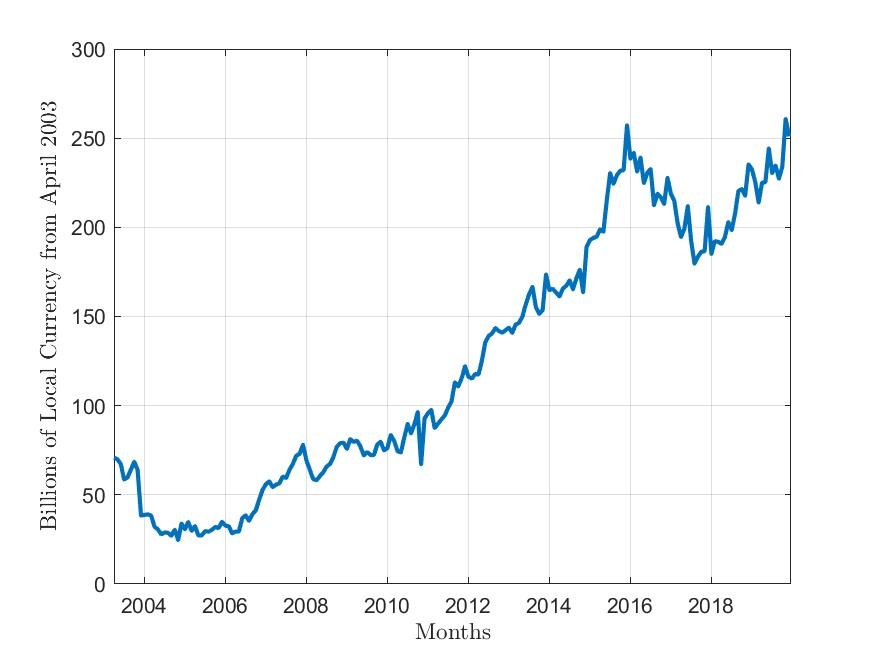}
        \caption{Local-Currency Loans}
        \label{fig:Aggregate_Debt_LC}
    \end{subfigure}
    \floatfoot{\textbf{Note:} The left panel reports foreign-currency loans, converted to U.S. dollars using the monthly exchange rate. The right panel shows local-currency loans deflated by a domestic price index and expressed in December 2019 purchasing power.}
\end{figure}

\begin{figure}
    \centering
    \includegraphics[width=0.75\linewidth]{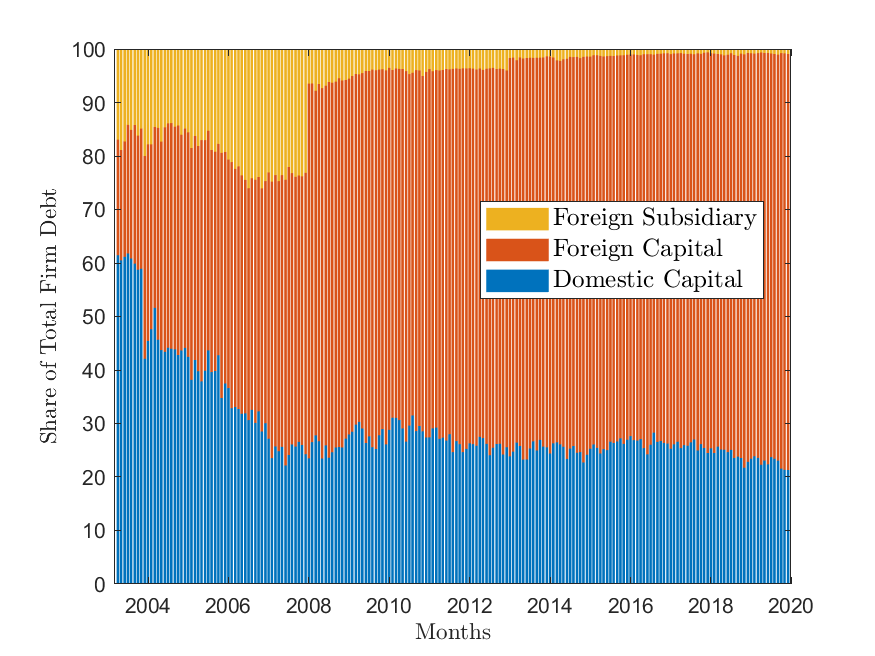}
    \caption{Share of Firm Debt by Type of Bank}
    \label{fig:Bank_Shares}
\end{figure}

%%%%%%%%%%%%%%%%%%%%%%%%%%%%%%%%%%%%%%%%%%%%
\newpage
\section{Macroeconomic Impact - Additional Results} \label{sec:appendix_macro_results}

%%%% LP Benchmark results
\begin{figure*}[p]
    \centering
    \includegraphics[width=0.95\textwidth]{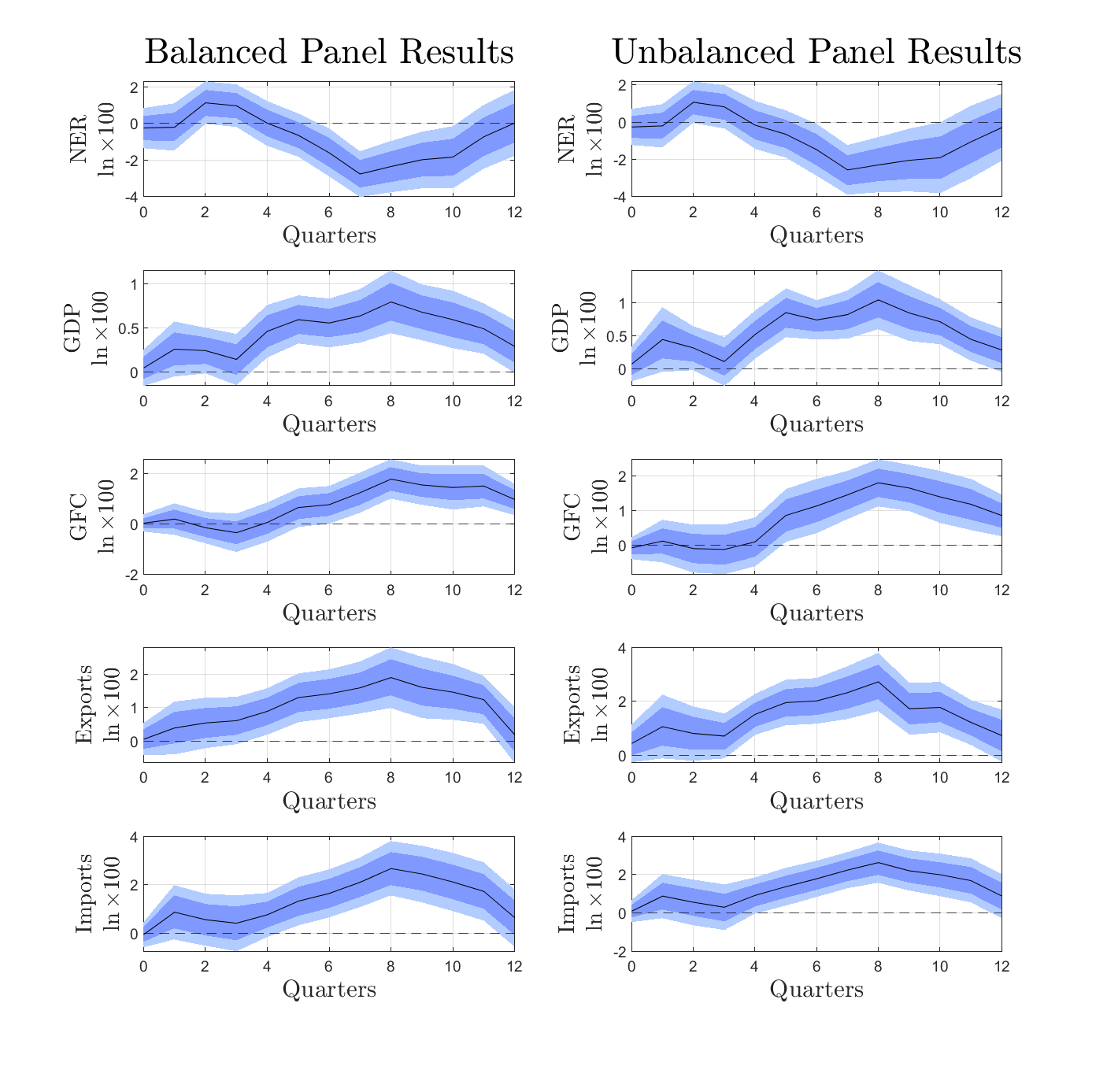}
    \caption{Macroeconomic Responses of EMEs to a Global Bank Net-Worth Shock \\ Local Projection Estimates }
    \label{fig:Benchmark_Regressions_Comparison}
    \floatfoot{\textbf{Note:} The figure displays impulse responses of five key macroeconomic variables to a one–standard–deviation increase in global banks’ net worth, estimated using panel Local Projections. Each row corresponds to a different variable, while the left and right columns show results for the balanced and unbalanced panels of EMEs, respectively. The black line shows point estimates, while the dark and light blue shaded areas denote 68\% and 90\% confidence intervals based on standard errors clustered at the date level. In the text, each panel is referenced as Panel~(i,j), where $i$ denotes the row and $j$ the column position in the figure.}
\end{figure*}

%%%% LP Financial Prices and Quantities
\begin{figure*}[p]
    \centering
    \includegraphics[width=0.80\textwidth]{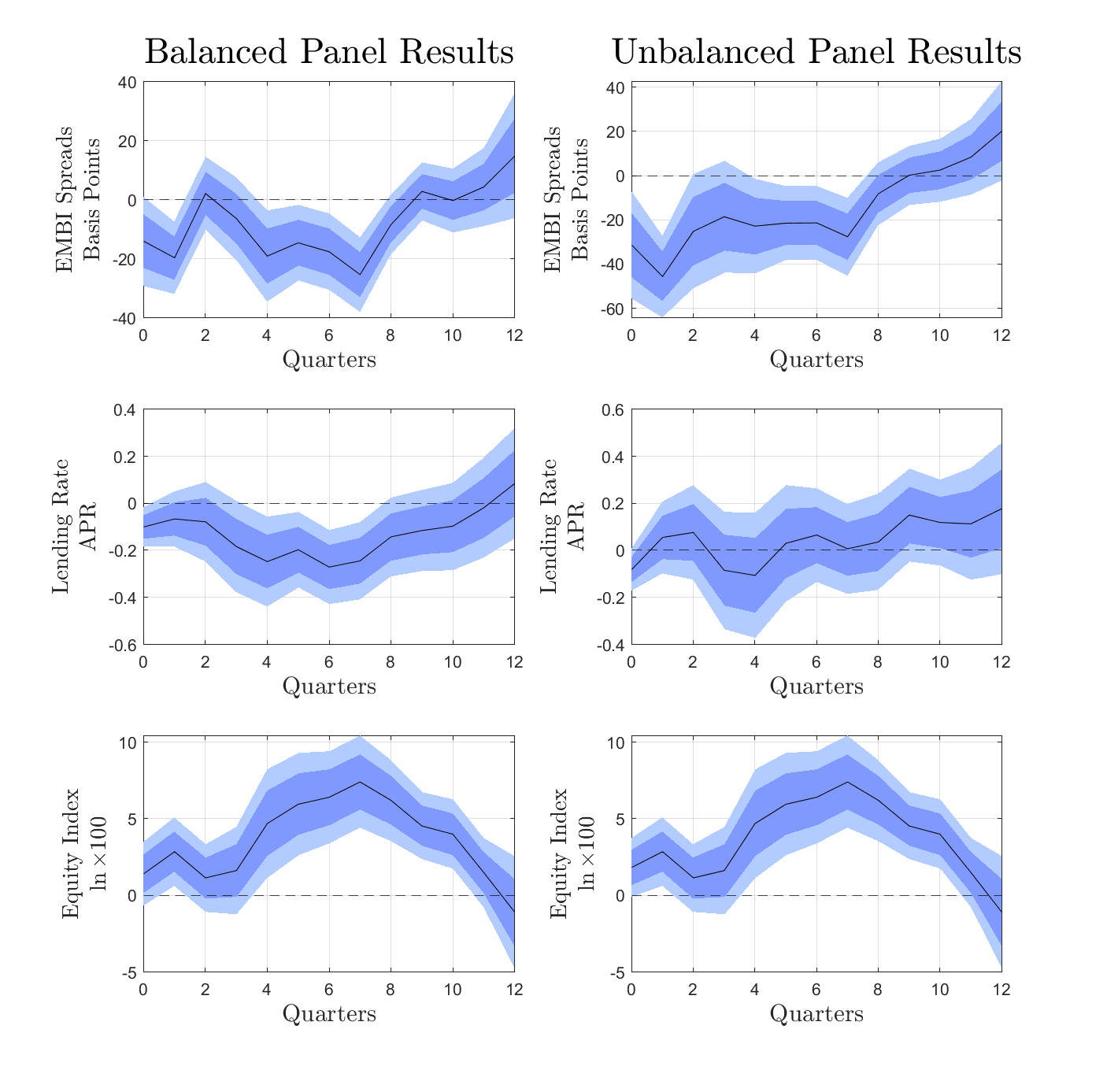}
    \caption{Responses of Financial Prices in EMEs to a Global Bank Net-Worth Shock \\ Local Projection Estimates}
    \label{fig:LP_Financial_Prices}
    \floatfoot{\textbf{Note:} The figure displays impulse responses of three financial price variables—EMBI spreads, an equity index, and a domestic lending rate—to a one–standard–deviation improvement in global banks’ net worth, estimated using panel Local Projections. Each row corresponds to a different variable, and each column reports results for the balanced and unbalanced samples, respectively. The black line shows point estimates, while the dark and light blue shaded areas denote 68\% and 90\% confidence intervals based on standard errors clustered at the date level. In the text, each panel is referenced as Panel~(i,j), where $i$ denotes the row and $j$ the column position in the figure.}
\end{figure*}

\begin{figure*}[p]
    \centering
    \includegraphics[width=0.80\textwidth]{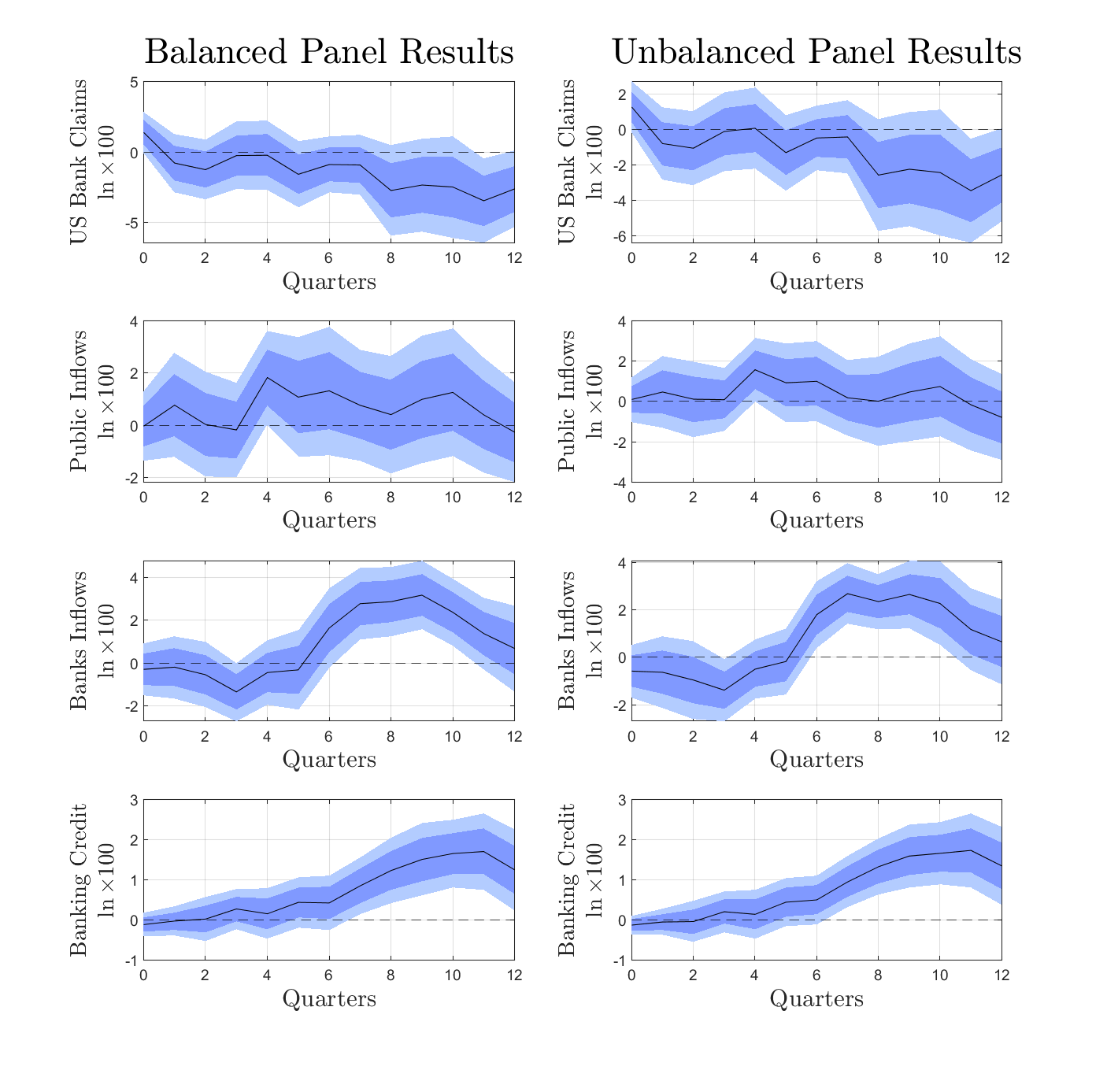}
    \caption{Responses of Financial Quantities in EMEs to a Global Bank Net-Worth Shock \\ Local Projection Estimates}
    \label{fig:LP_Financial_Quantities}
    \floatfoot{\textbf{Note:} The figure displays impulse responses of four financial quantity variables—foreign liabilities of the public sector (public inflows), foreign liabilities of the banking sector (bank inflows), domestic bank credit to the non-financial private sector (banking credit), and U.S. banks’ consolidated claims on EMEs (U.S. bank claims)—to a one–standard–deviation improvement in global banks’ net worth, estimated using panel Local Projections. Each row corresponds to a different variable, and each column reports results for the balanced and unbalanced samples, respectively. The black line shows point estimates, while the dark and light blue shaded areas denote 68\% and 90\% confidence intervals based on standard errors clustered at the date level. In the text, each panel is referenced as Panel~(i,j), where $i$ denotes the row and $j$ the column position in the figure.}
\end{figure*}

%%%%%%%%%% Uruguay
\begin{figure}[H]
    \centering
    \includegraphics[width=0.8\textwidth]{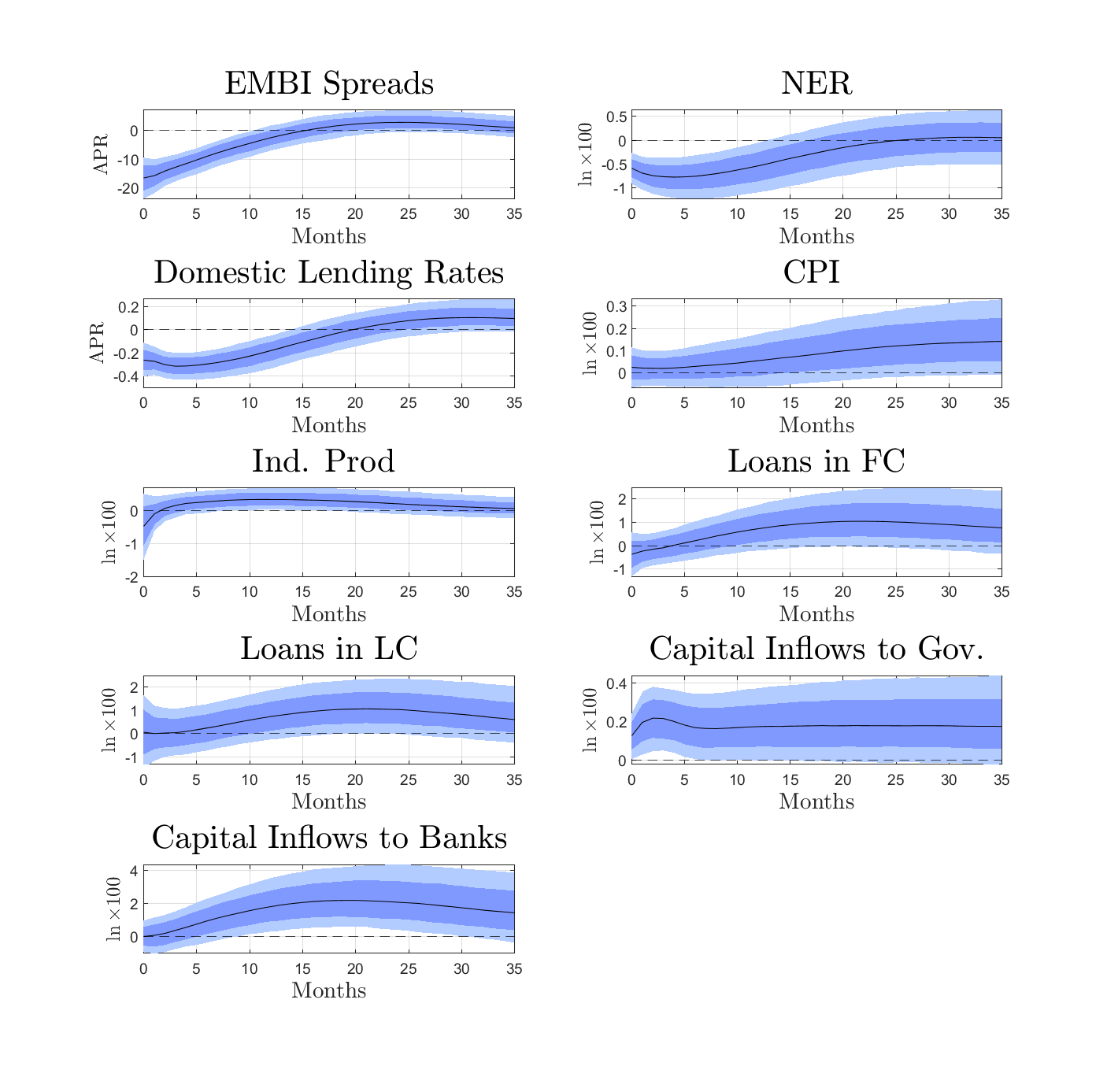}
    \caption{Macroeconomic Responses of Uruguay to a Global Bank Net-Worth Shock}
    \label{fig:Uruguay_Macro_Responses}
    \floatfoot{\textbf{Note:} The figure shows impulse responses of key Uruguayan macro and financial variables to a one–standard–deviation improvement in global banks’ net worth, estimated using a Bayesian VAR on monthly data (2003–2019). The black line shows the median impulse response, and the shaded areas denote 68\% and 90\% credible intervals. The estimated model is a standard Bayesian VAR with Minnesota priors and six monthly lags, corresponding to two quarters as in the benchmark quarterly VAR. Capital inflow series were interpolated to a monthly frequency using cubic spline interpolation, consistent with methodologies adopted in related macro–financial studies.}
\end{figure}

\begin{figure}[H]
    \centering
    \includegraphics[width=0.9\textwidth]{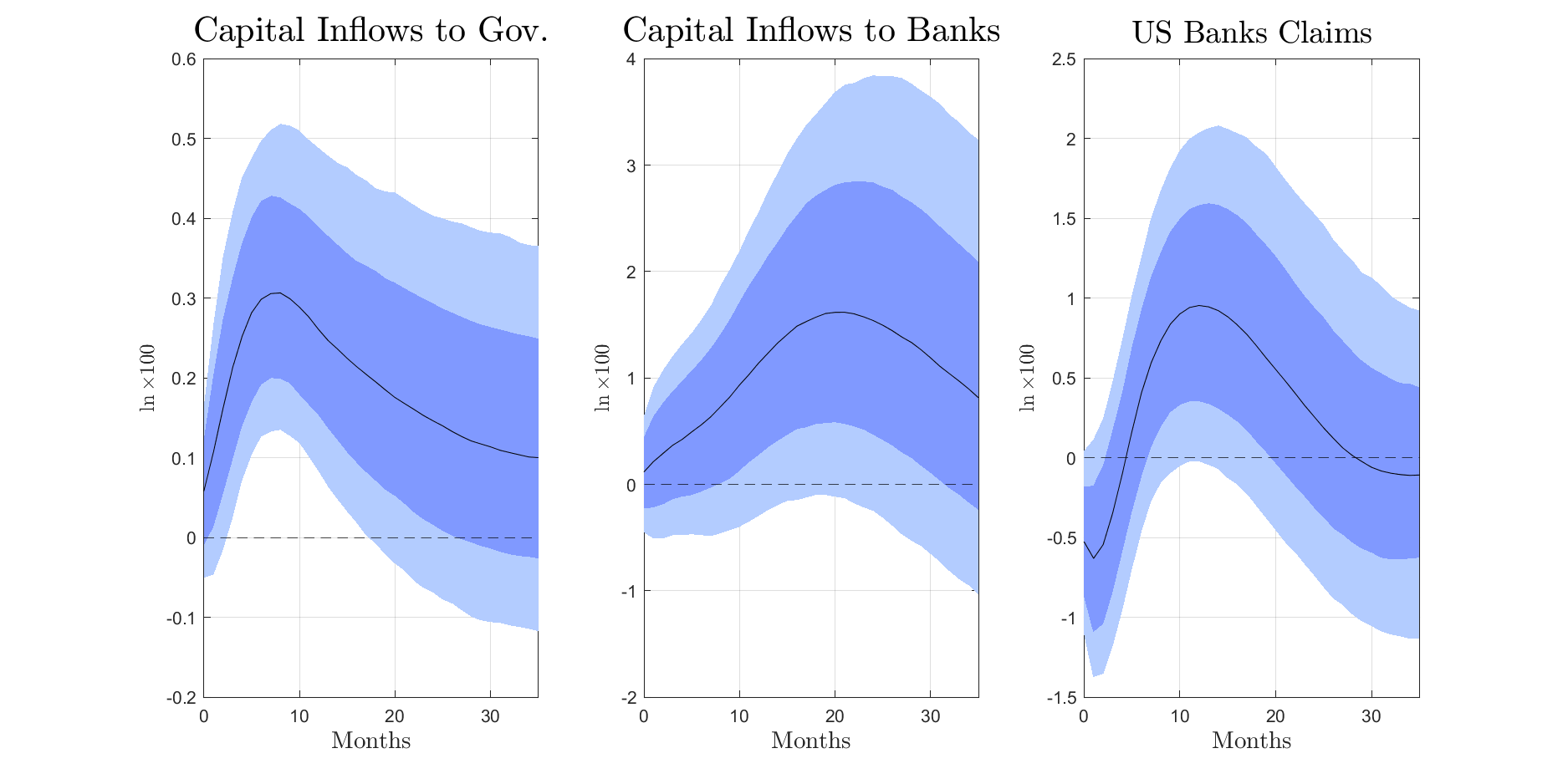}
    \caption{Financial Flow Response of Uruguay to a Global Bank Net-Worth Shock}
    \label{fig:Uruguay_Macro_Responses_Flows}
    \floatfoot{\textbf{Note:} The figure shows impulse responses of capital inflows to the government and banking sectors, and of U.S. banks’ consolidated claims on Uruguay, to a one–standard–deviation improvement in global banks’ net worth, estimated using a Bayesian VAR on monthly data (2006–2019). The black line shows the median impulse response, and the shaded areas denote 68\% and 90\% credible intervals. The estimated model is a standard Bayesian VAR with Minnesota priors and six monthly lags, corresponding to two quarters as in the benchmark quarterly VAR. Capital inflow series were interpolated to a monthly frequency using cubic spline interpolation, consistent with methodologies adopted in related macro–financial studies.}
\end{figure}

%%%%%%%%%%%%%%%%%%%%%%%%%%%%%%%%%%
\begin{figure*}
    \centering
    \includegraphics[width=0.95\textwidth]{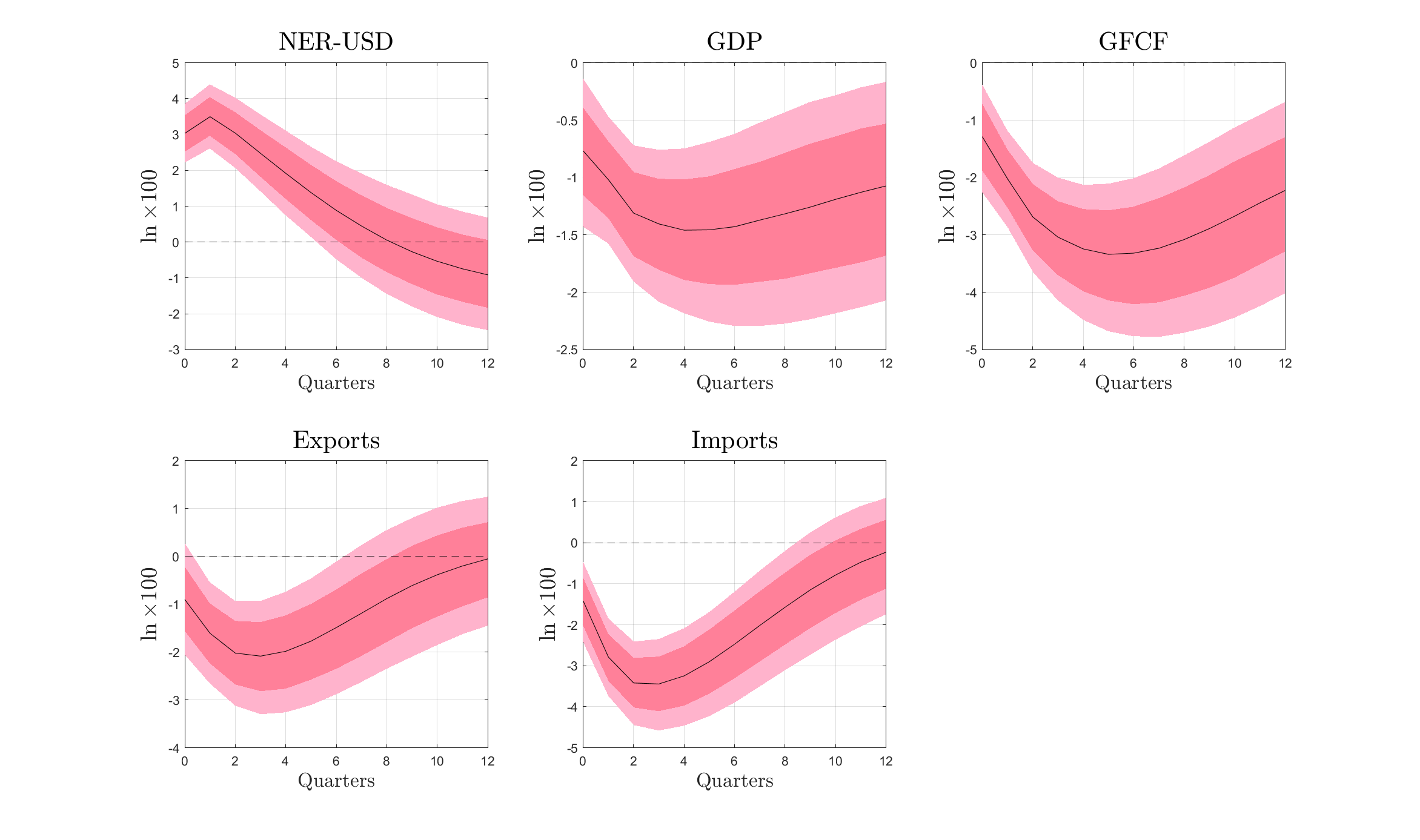}
    \caption{Macroeconomic Responses of EMEs to a Global Bank Net-Worth Shock}
    \label{fig:Pooled_EME_CD}
    \floatfoot{\textbf{Note:} The figure displays impulse responses of five key macroeconomic variables to a one–standard–deviation increase in global banks’ net worth, estimated using the pooled panel Bayesian VAR associated with a positive co-movement with the EBP. Each row corresponds to a different variable, and the left and right columns report results for the balanced and unbalanced samples, respectively. The black line shows the median impulse response function (IRF), while the dark and light blue shaded areas denote the 68\% and 90\% credible intervals, respectively. Panels are referenced as Panel (i,j), where $i$ denotes the row and $j$ the column position.}
\end{figure*}

\newpage
\begin{figure*}
    \centering
    \includegraphics[width=0.95\textwidth]{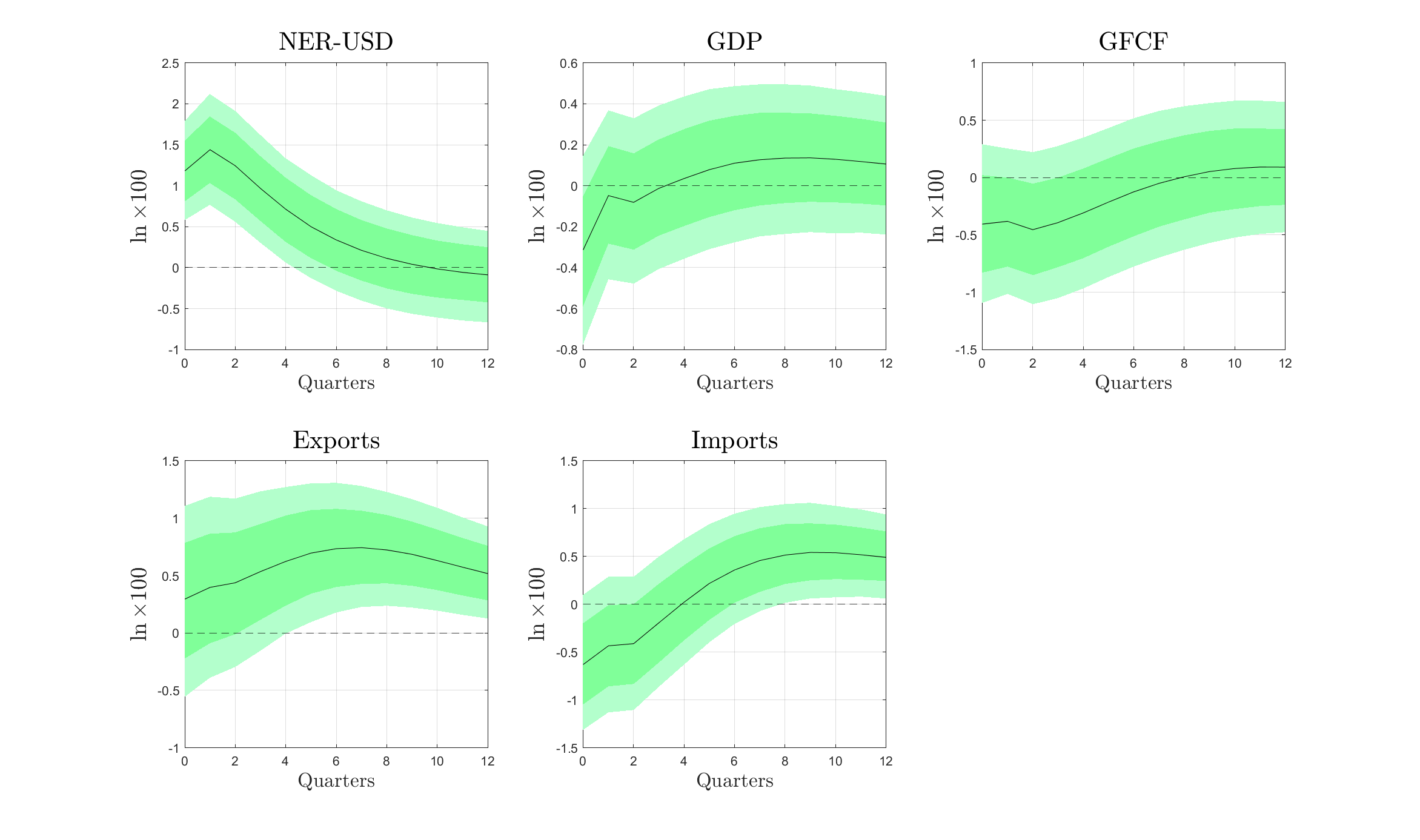}
    \caption{Macroeconomic Responses of EMEs to the Total Net-Worth Surprise}
    \label{fig:Pooled_EME_Outside}
    \floatfoot{\textbf{Note:} The figure displays impulse responses of five key macroeconomic variables to a one–standard–deviation innovation in the total net-worth surprise, estimated using the pooled panel Bayesian VAR. Each row corresponds to a different variable, and the left and right columns report results for the balanced and unbalanced samples, respectively. The black line shows the median impulse response function (IRF), while the dark and light blue shaded areas denote the 68\% and 90\% credible intervals, respectively.}
\end{figure*}

\newpage
\begin{figure*}
    \centering
    \includegraphics[width=0.95\textwidth]{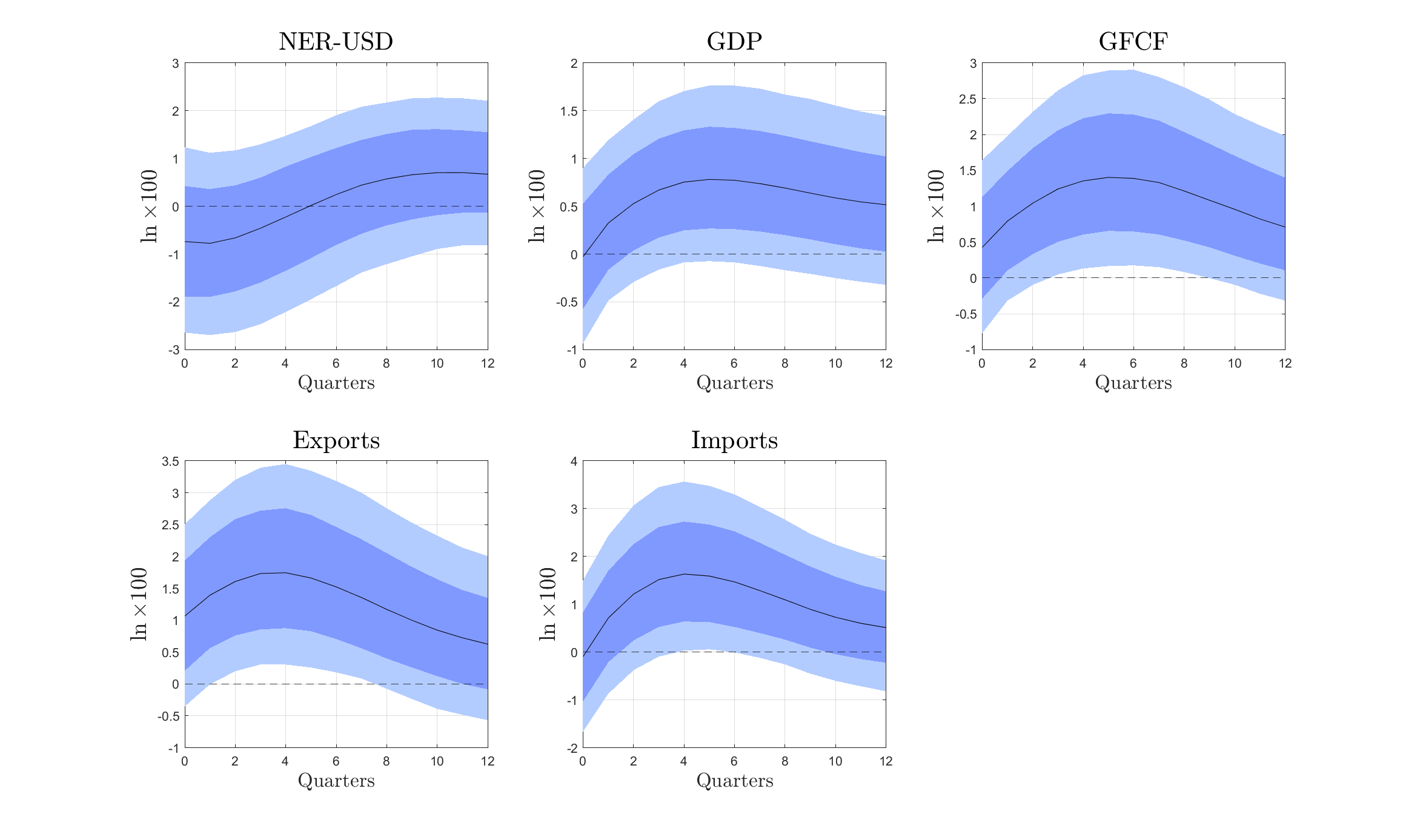}
    \caption{Macroeconomic Responses of EMEs to a Global Bank Net-Worth Shock \\ Average VAR Specification}
    \label{fig:Average_EME}
    \floatfoot{\textbf{Note:} The figure displays impulse responses of five key macroeconomic variables to a one–standard–deviation increase in global banks’ net worth, estimated using the average VAR specification with Minnesota Priors. Each row corresponds to a different variable, and the left and right columns report results for the balanced and unbalanced samples, respectively. The black line shows the median impulse response function (IRF), while the dark and light blue shaded areas denote the 68\% and 90\% credible intervals, respectively. In the text, each panel is referenced as Panel (i,j), where $i$ denotes the row and $j$ the column position in the figure.}
\end{figure*}

\newpage
\begin{figure}
     \centering
     \begin{subfigure}[b]{0.7\textwidth}
         \centering
         \includegraphics[width=\textwidth]{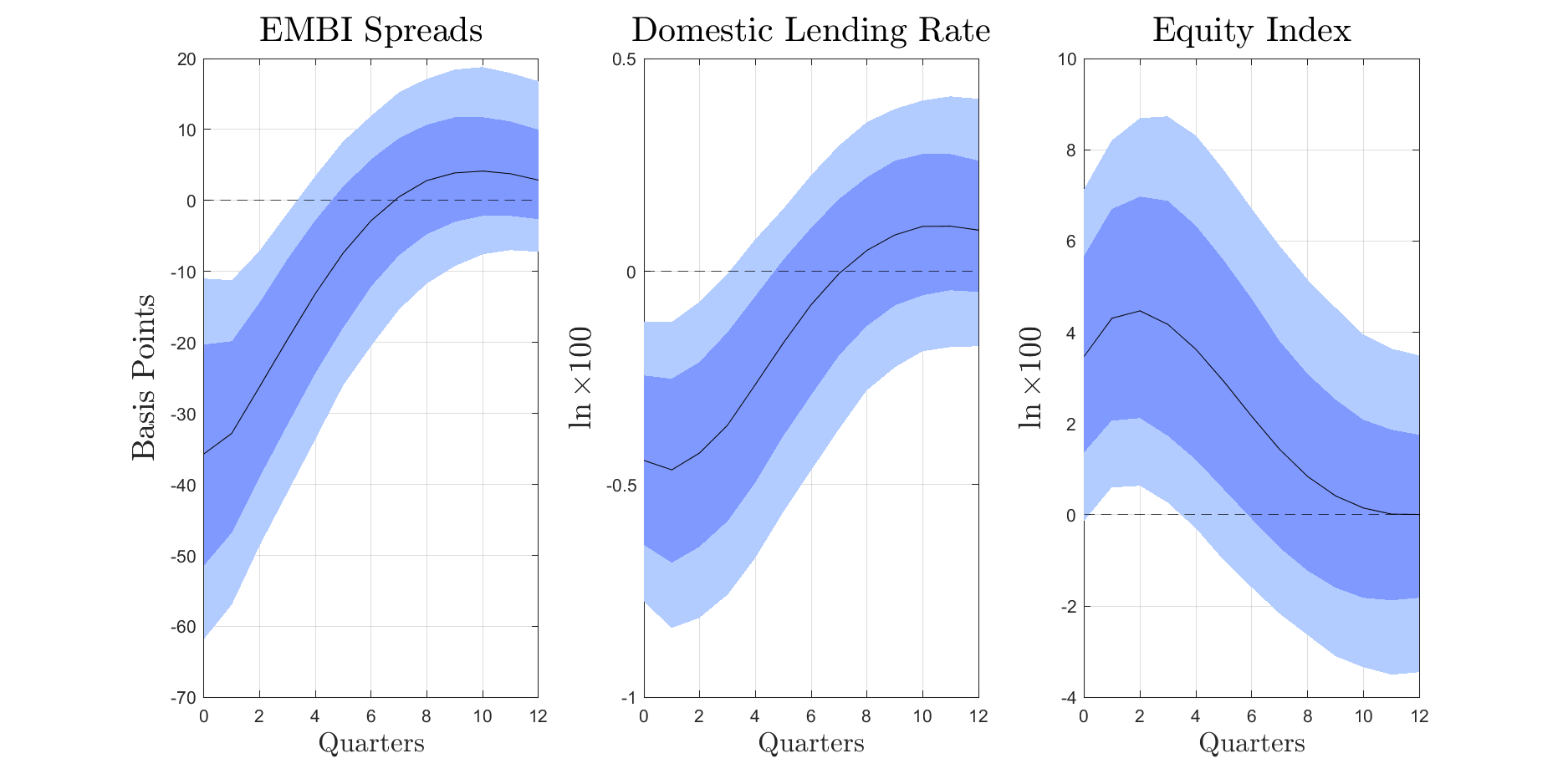}
         \caption{Financial Prices}
         \label{fig:Average_EME_Financial_Prices}
     \end{subfigure}
     \hfill
     \begin{subfigure}[b]{0.7\textwidth}
         \centering
         \includegraphics[width=\textwidth]{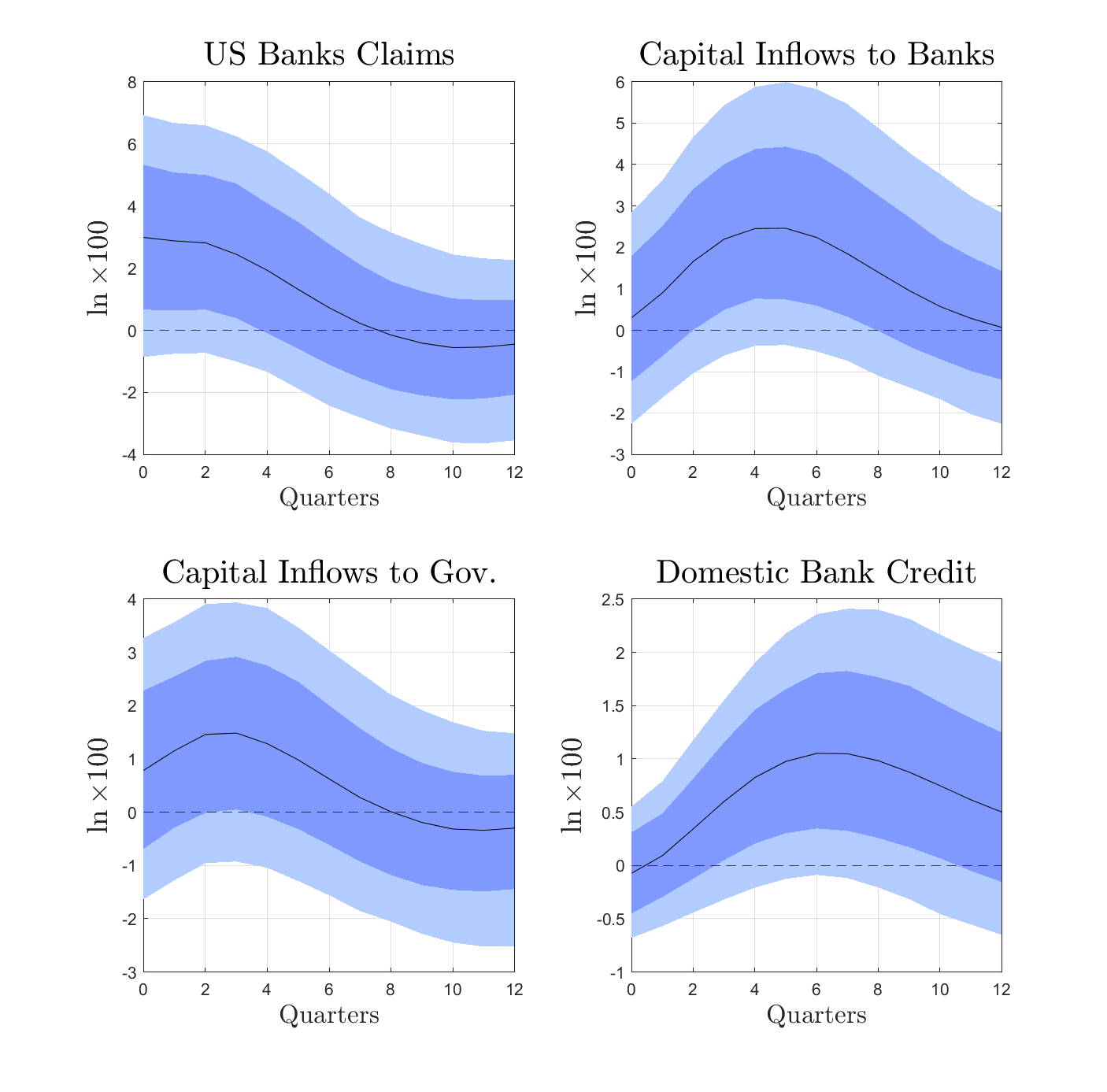}
         \caption{Financial Quantities}
         \label{fig:Average_EME_Financial_Quantities}
     \end{subfigure}
    \caption{Macroeconomic Responses of EMEs to a Global Bank Net-Worth Shock \\ Average VAR Specification – Prices and Quantities}
    \label{fig:Average_Financial_Prices_Quantities}
    \floatfoot{\textbf{Note:} The figures display impulse responses of three key financial price variables and four key financial quantity variables to a one–standard–deviation increase in global banks’ net worth, estimated using the average VAR specification with Minnesota Priors. Each row corresponds to a different variable, and the left and right columns report results for the balanced and unbalanced samples, respectively. The black line shows the median impulse response function (IRF), while the dark and light blue shaded areas denote the 68\% and 90\% credible intervals, respectively. In the text, each panel is referenced as Panel (i,j), where $i$ denotes the row and $j$ the column position in the figure.}
\end{figure}

\newpage
\begin{figure*}[p]
    \centering
    \includegraphics[width=0.95\textwidth]{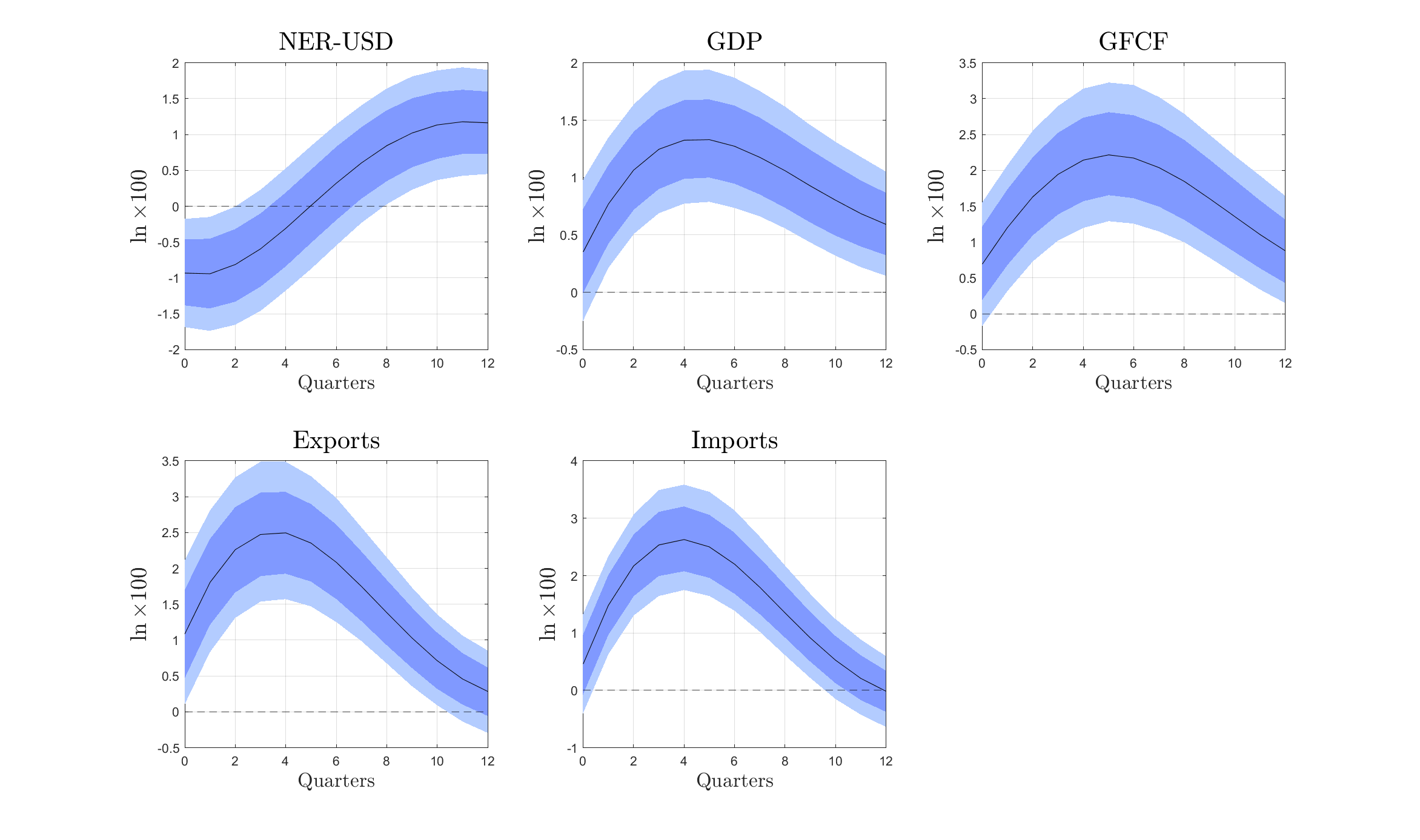}
    \caption{Macroeconomic Responses of EMEs to a Global Bank Net-Worth Shock \\ 1 - Lag}
    \label{fig:Pooled_EME_Lags1}
    \floatfoot{\textbf{Note:} The figure displays impulse responses of five key macroeconomic variables to a one–standard–deviation increase in global banks’ net worth, estimated using the pooled panel Bayesian VAR. Each row corresponds to a different variable, and the left and right columns report results for the balanced and unbalanced samples, respectively. The black line shows the median impulse response function (IRF), while the dark and light blue shaded areas denote the 68\% and 90\% credible intervals, respectively. In the text, each panel is referenced as Panel~(i,j), where $i$ denotes the row and $j$ the column position in the figure.}
\end{figure*}

\newpage
\begin{figure*}[p]
    \centering
    \includegraphics[width=0.95\textwidth]{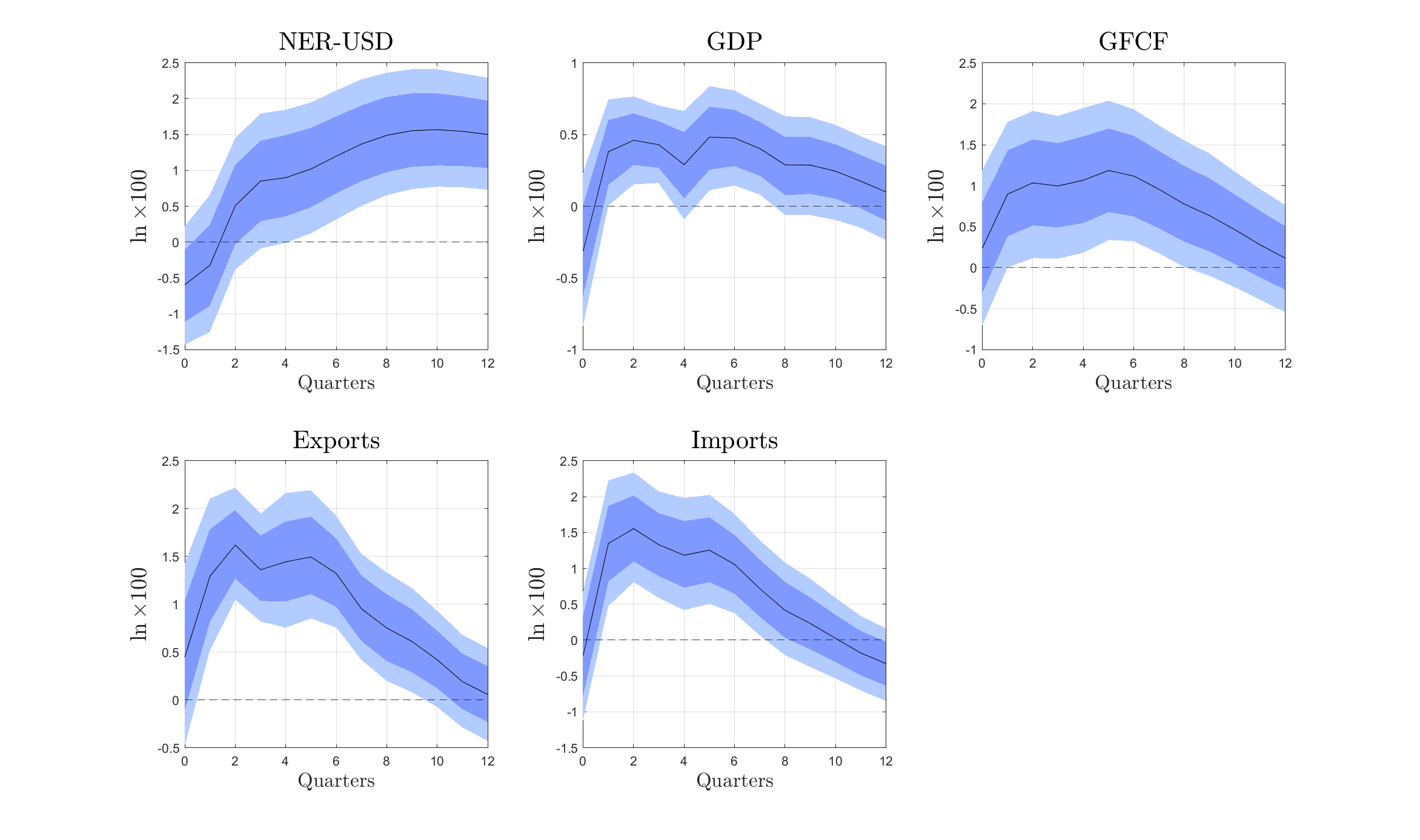}
    \caption{Macroeconomic Responses of EMEs to a Global Bank Net-Worth Shock \\ 4 - Lag}
    \label{fig:Pooled_EME_Lags4}
    \floatfoot{\textbf{Note:} The figure displays impulse responses of five key macroeconomic variables to a one–standard–deviation increase in global banks’ net worth, estimated using the pooled panel Bayesian VAR. Each row corresponds to a different variable, and the left and right columns report results for the balanced and unbalanced samples, respectively. The black line shows the median impulse response function (IRF), while the dark and light blue shaded areas denote the 68\% and 90\% credible intervals, respectively. In the text, each panel is referenced as Panel~(i,j), where $i$ denotes the row and $j$ the column position in the figure.}
\end{figure*}

%%%%%%%%%%%%%%%%%%%%%%%%%%%%%%%%%%%%%%
%% LP Robustness
\newpage
\begin{figure*}[p]
    \centering
    \includegraphics[width=0.95\textwidth]{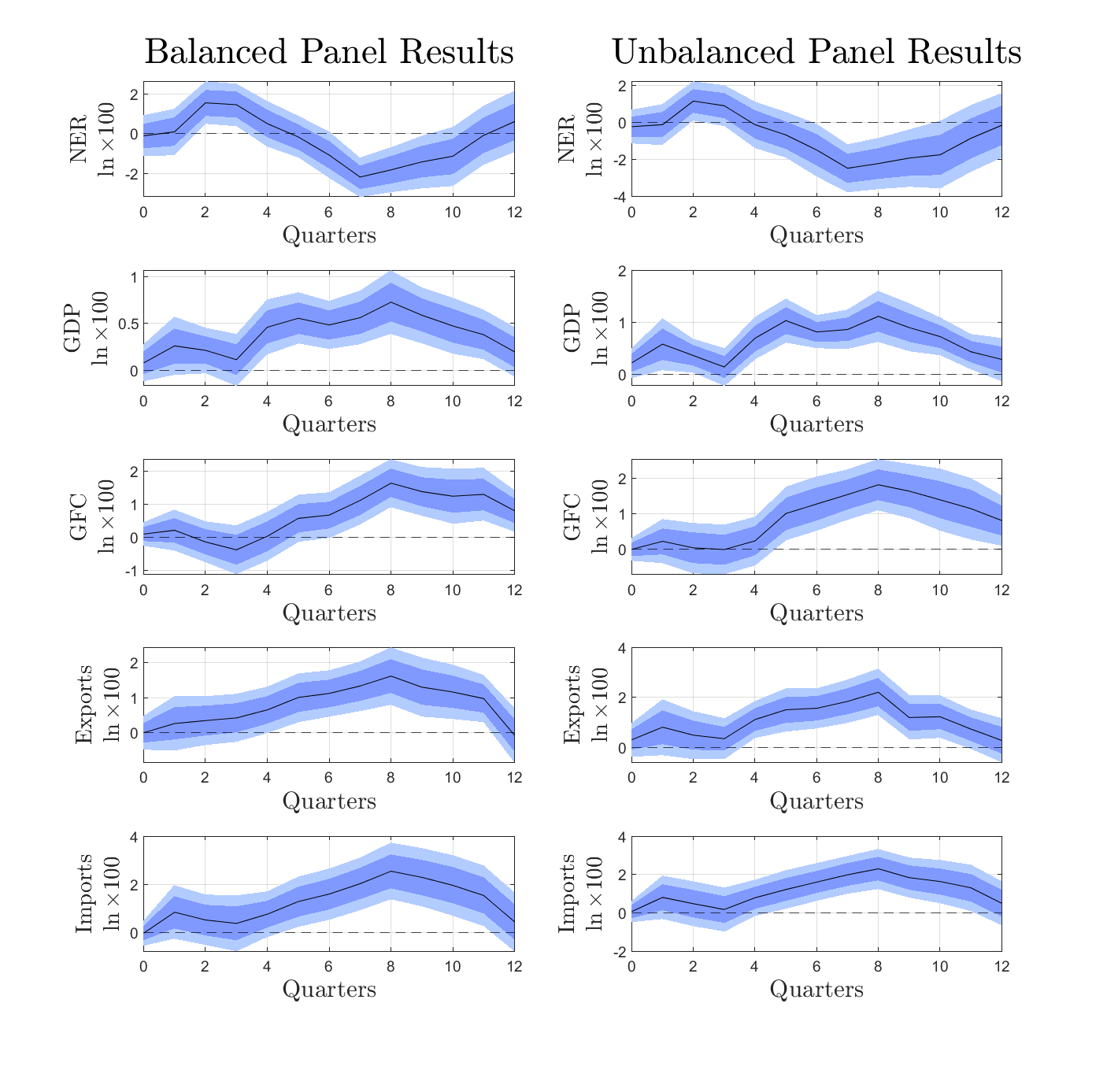}
    \caption{Macroeconomic Responses of EMEs to a Global Bank Net-Worth Shock \\ Local Projection Estimates with FE }
    \label{fig:CountryFE_Regressions_Comparison}
    \floatfoot{\textbf{Note:} The figure displays impulse responses of five key macroeconomic variables to a one–standard–deviation increase in global banks’ net worth, estimated using panel Local Projections with country fixed effects. Each row corresponds to a different variable, while the left and right columns show results for the balanced and unbalanced panels of EMEs, respectively. The black line shows point estimates, while the dark and light blue shaded areas denote 68\% and 90\% confidence intervals based on standard errors clustered at the date level. In the text, each panel is referenced as Panel~(i,j), where $i$ denotes the row and $j$ the column position in the figure.}
\end{figure*}

%% With Linear and Quadratic Trends
\newpage
\begin{figure*}[p]
    \centering
    \includegraphics[width=0.95\textwidth]{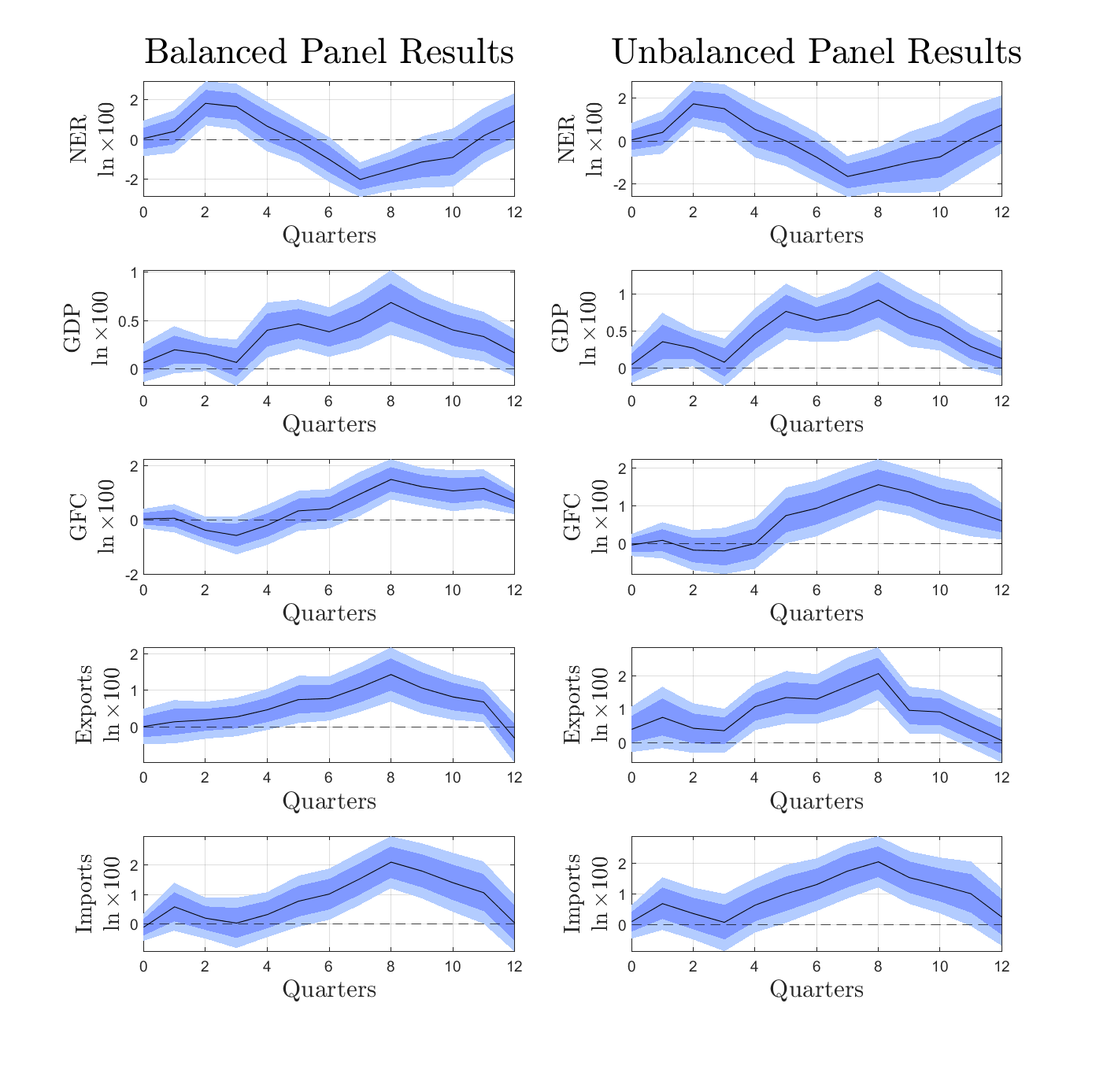}
    \caption{Macroeconomic Responses of EMEs to a Global Bank Net-Worth Shock \\ Local Projection Estimates with Linear \& Quadratic Trend }
    \label{fig:Trends_Regressions_Comparison}
    \floatfoot{\textbf{Note:} The figure displays impulse responses of five key macroeconomic variables to a one–standard–deviation increase in global banks’ net worth, estimated using panel Local Projections with linear and quadratic trends. Each row corresponds to a different variable, while the left and right columns show results for the balanced and unbalanced panels of EMEs, respectively. The black line shows point estimates, while the dark and light blue shaded areas denote 68\% and 90\% confidence intervals based on standard errors clustered at the date level. In the text, each panel is referenced as Panel~(i,j), where $i$ denotes the row and $j$ the column position in the figure.}
\end{figure*}

%% Not Controlling for Lagged Values of Credit Supply Shock
\newpage
\begin{figure*}[p]
    \centering
    \includegraphics[width=0.95\textwidth]{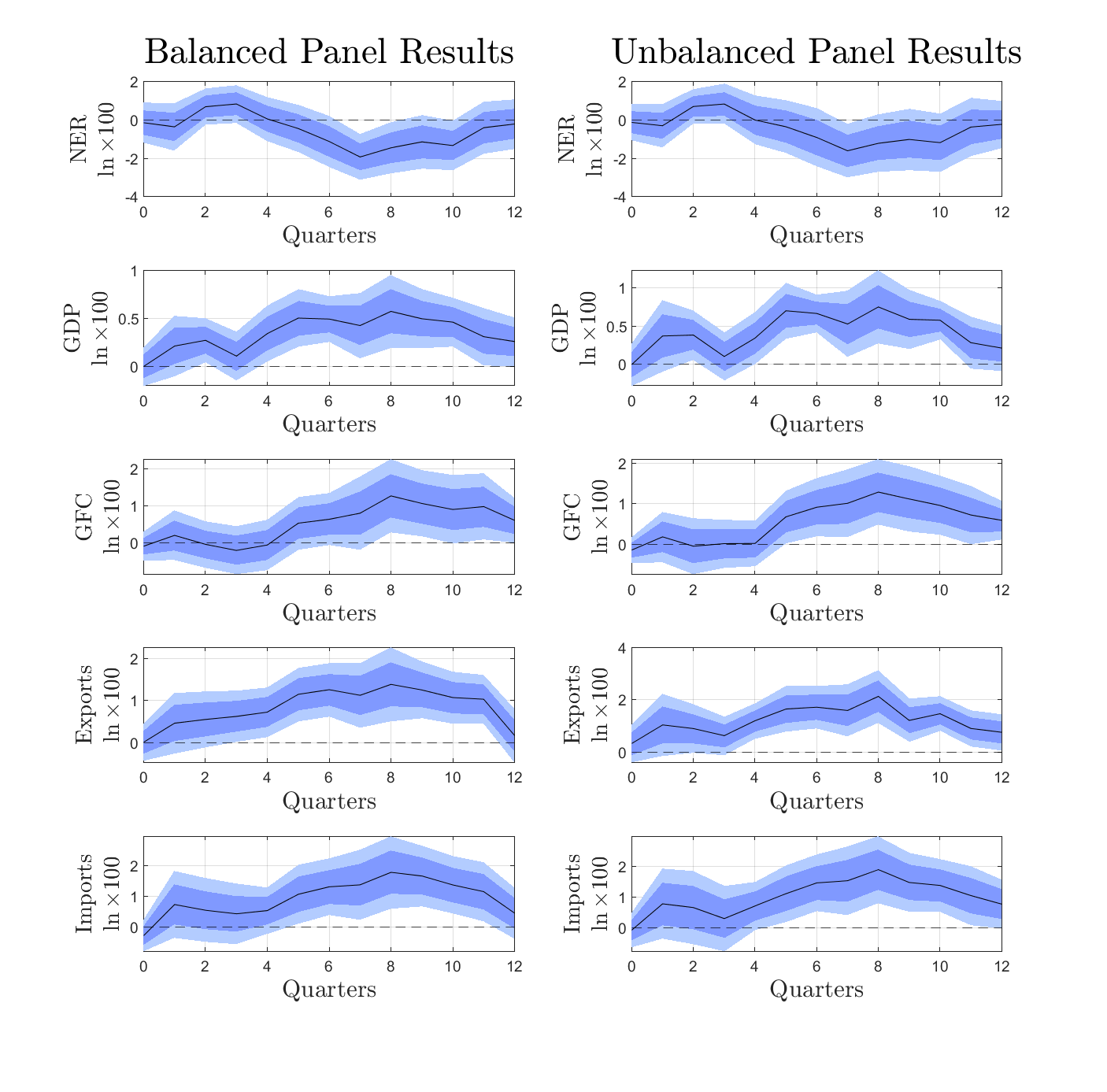}
    \caption{Macroeconomic Responses of EMEs to a Global Bank Net-Worth Shock \\ Local Projection Estimates with No Lags of Shocks}
    \label{fig:NoShockLags_Regressions_Comparison}
    \floatfoot{\textbf{Note:} The figure displays impulse responses of five key macroeconomic variables to a one–standard–deviation increase in global banks’ net worth, estimated using panel Local Projections, not controlling for lagged values of shocks. Each row corresponds to a different variable, while the left and right columns show results for the balanced and unbalanced panels of EMEs, respectively. The black line shows point estimates, while the dark and light blue shaded areas denote 68\% and 90\% confidence intervals based on standard errors clustered at the date level. In the text, each panel is referenced as Panel~(i,j), where $i$ denotes the row and $j$ the column position in the figure.}
\end{figure*}

%%%%%%%%%%%%%%%%%%%%%%%%%%%%%%%%%%%%%%
%% IV Specifications
\newpage
\begin{figure*}[p]
    \centering
    \includegraphics[width=0.80\textwidth]{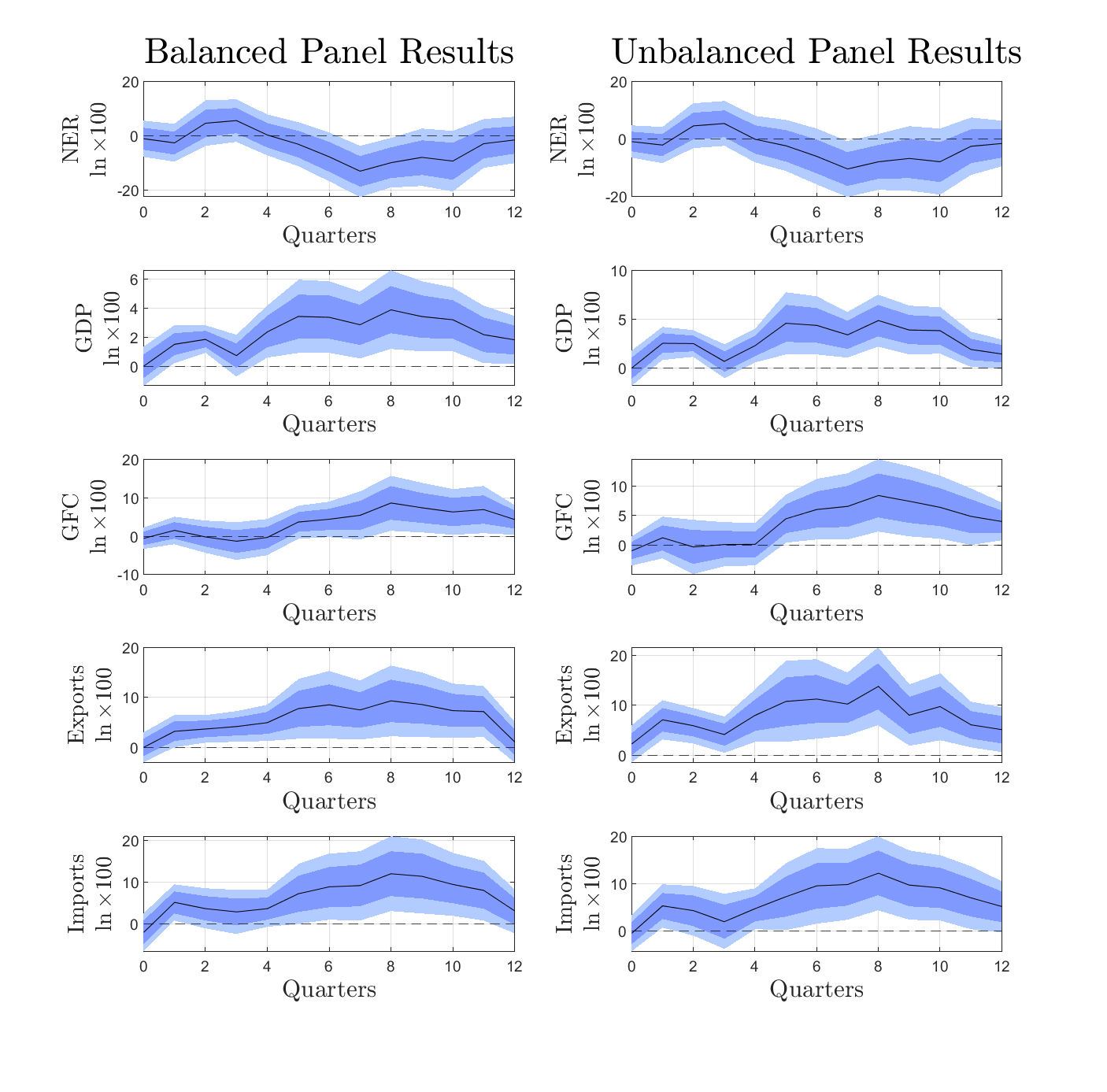}
    \caption{Responses of Benchmark Variables in EMEs to a Global Bank Net-Worth Shock \\ IV Local Projection Estimates}
    \label{fig:IV_Regressions_Comparison}
    \floatfoot{\textbf{Note:} The figure displays impulse responses of five key macroeconomic variables to a one–standard–deviation increase in global banks’ net worth, estimated using a panel IV Local Projection specification. We use the credit supply shock as an instrument for the EBP. We normalize the shock to a 100-basis-point decrease in the EBP. Each row corresponds to a different variable, while the left and right columns show results for the balanced and unbalanced panels of EMEs, respectively. The black line shows point estimates, while the dark and light blue shaded areas denote 68\% and 90\% confidence intervals based on standard errors clustered at the date level. In the text, each panel is referenced as Panel~(i,j), where $i$ denotes the row and $j$ the column position in the figure.}
\end{figure*}

\newpage
\begin{figure*}[p]
    \centering
    \includegraphics[width=0.80\textwidth]{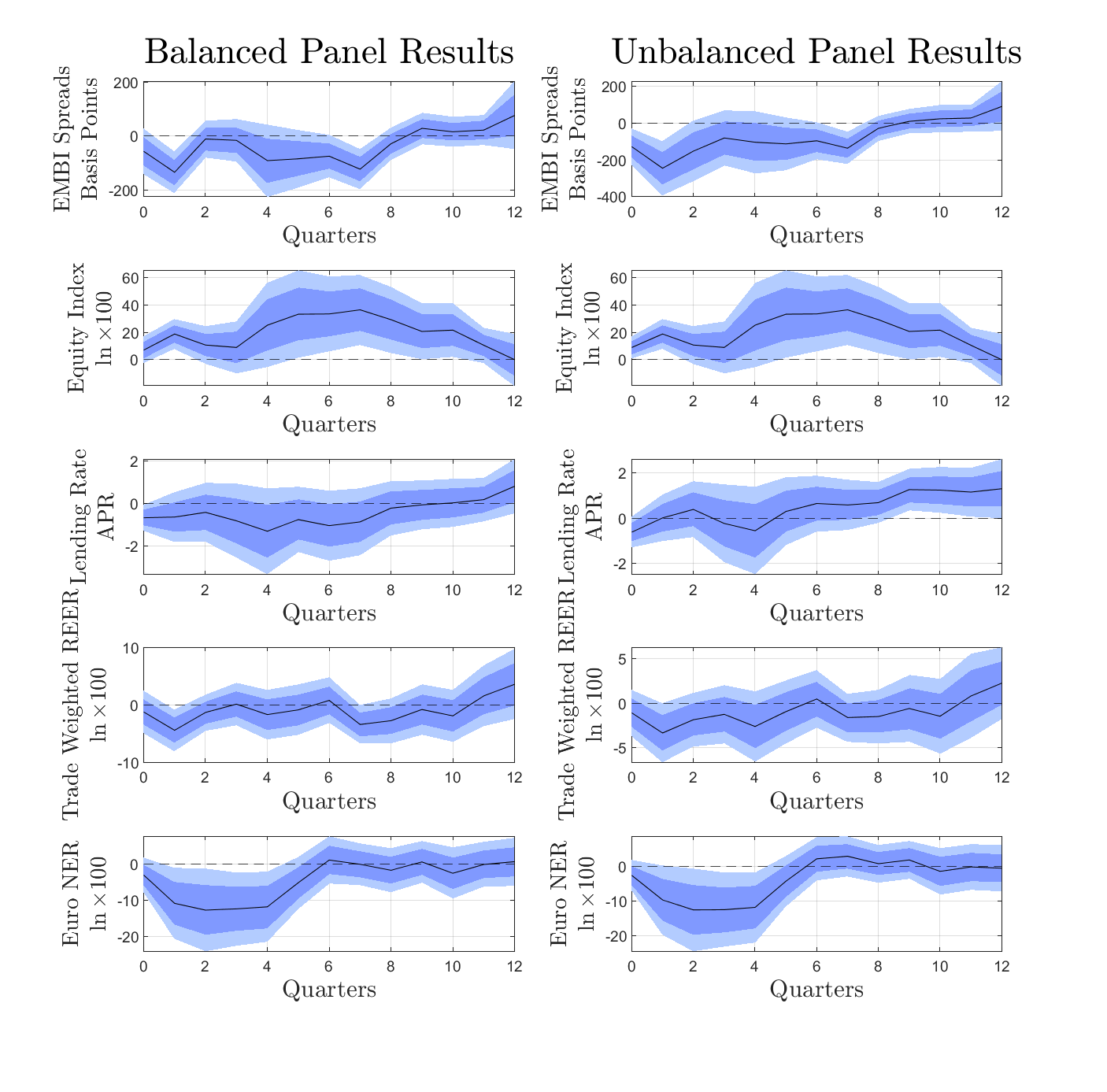}
    \caption{Responses of Financial Prices in EMEs to a Global Bank Net-Worth Shock \\ IV Local Projection Estimates}
    \label{fig:IV_Additional_Variables_Prices_Comparison}
    \floatfoot{\textbf{Note:} The figure displays impulse responses of three financial price variables—EMBI spreads, an equity index, and a domestic lending rate—to a one–standard–deviation improvement in global banks’ net worth, estimated using panel IV Local Projections. Each row corresponds to a different variable, and each column reports results for the balanced and unbalanced samples, respectively. The black line shows point estimates, while the dark and light blue shaded areas denote 68\% and 90\% confidence intervals based on standard errors clustered at the date level. In the text, each panel is referenced as Panel~(i,j), where $i$ denotes the row and $j$ the column position in the figure.}
\end{figure*}

\newpage
\begin{figure*}[p]
    \centering
    \includegraphics[width=0.80\textwidth]{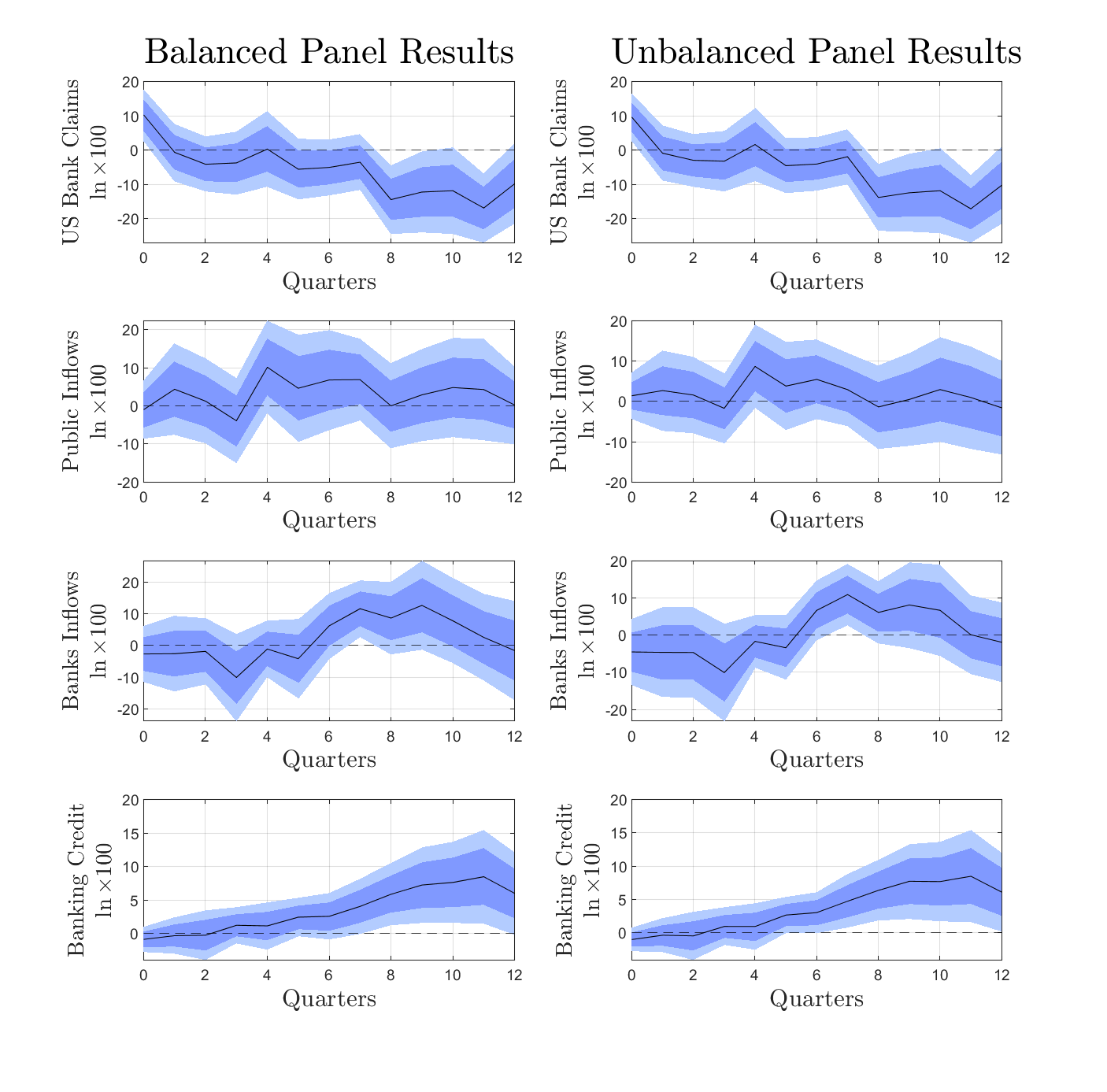}
    \caption{Responses of Financial Quantities in EMEs to a Global Bank Net-Worth Shock \\ IV Local Projection Estimates}
    \label{fig:IV_Additional_Variables_Quantities_Comparison}
    \floatfoot{\textbf{Note:} The figure displays impulse responses of four financial quantity variables—foreign liabilities of the public sector (public inflows), foreign liabilities of the banking sector (bank inflows), domestic bank credit to the non-financial private sector (banking credit), and U.S. banks’ consolidated claims on EMEs (U.S. bank claims)—to a one–standard–deviation improvement in global banks’ net worth, estimated using panel IV Local Projections. Each row corresponds to a different variable, and each column reports results for the balanced and unbalanced samples, respectively. The black line shows point estimates, while the dark and light blue shaded areas denote 68\% and 90\% confidence intervals based on standard errors clustered at the date level. In the text, each panel is referenced as Panel~(i,j), where $i$ denotes the row and $j$ the column position in the figure.}
\end{figure*}

\newpage
\begin{figure}
    \centering
    \begin{subfigure}[b]{0.48\textwidth}
        \centering
        \includegraphics[width=\textwidth]{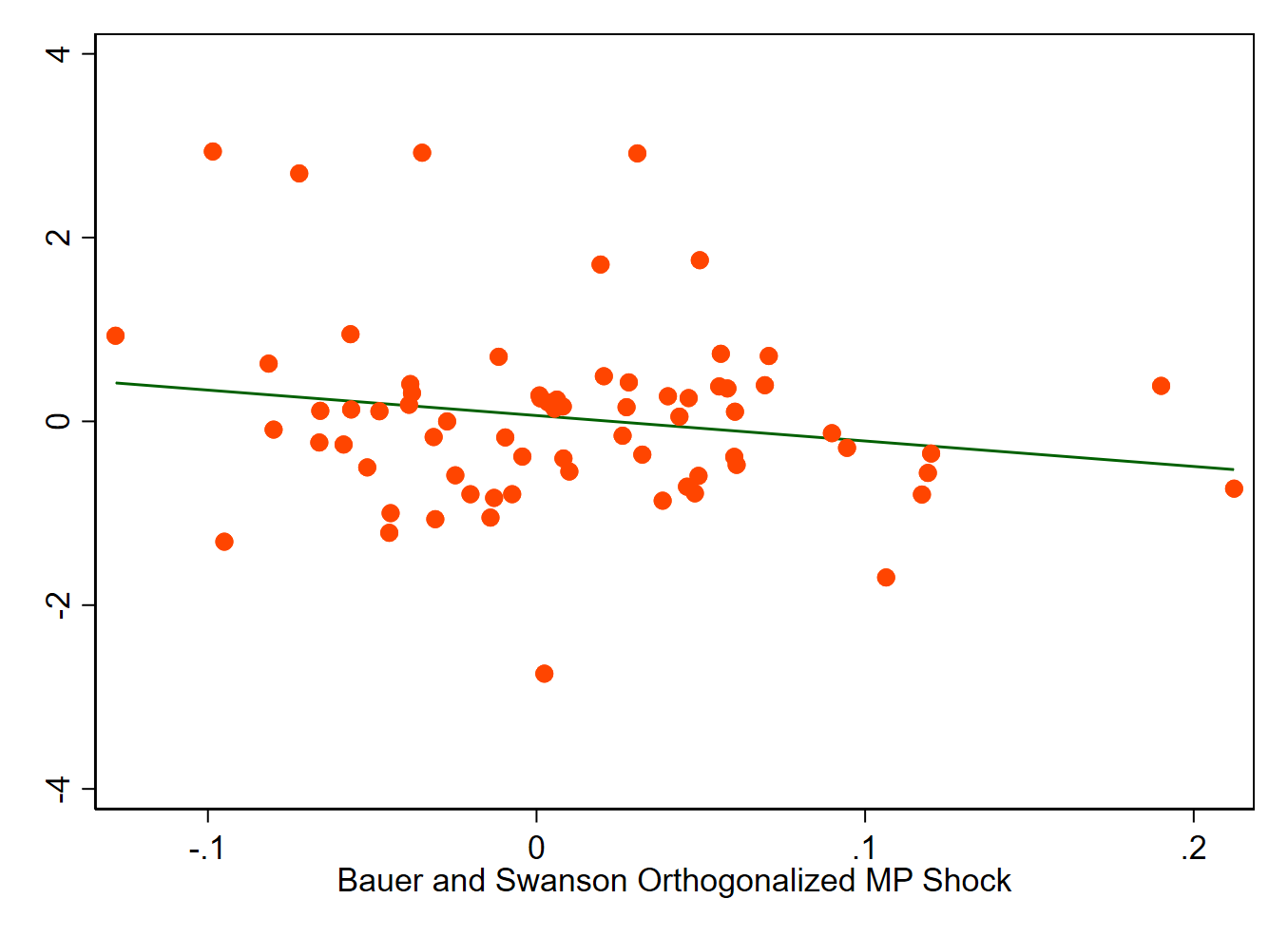}
        \caption{Contemporaneous correlation with Bauer--Swanson shock}
        \label{fig:Quarterly_CoMovement}
    \end{subfigure}
    \hfill
    \begin{subfigure}[b]{0.48\textwidth}
        \centering
        \includegraphics[width=\textwidth]{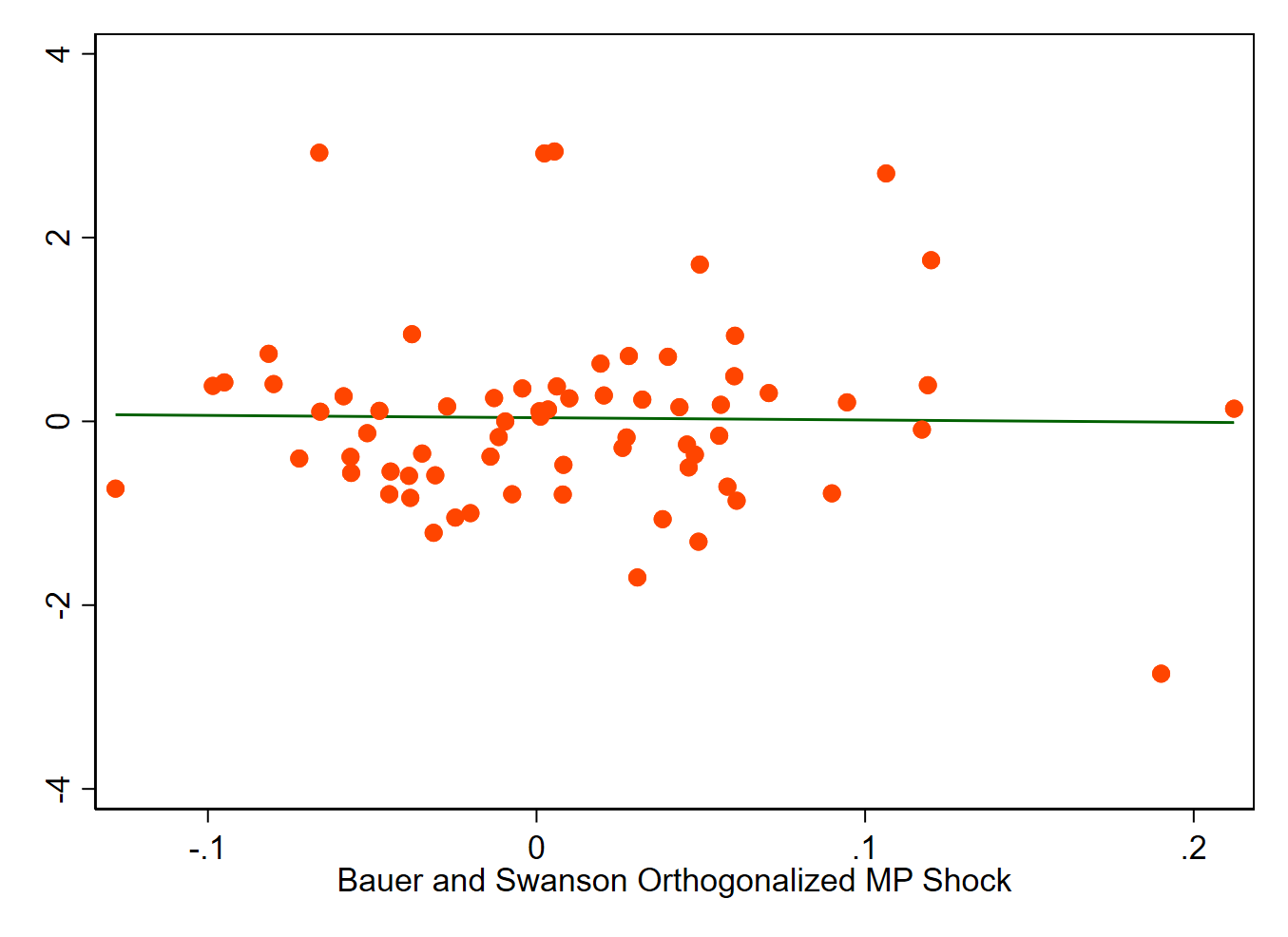}
        \caption{Lagged (one-quarter) correlation with Bauer--Swanson shock}
        \label{fig:Quarterly_CoMovement_Lag}
    \end{subfigure}
    
    \caption{Scatter plots of the global credit-supply shock against orthogonalized U.S. monetary policy shocks from Bauer and Swanson (2023).}
    \label{fig:CoMovement_BS}
\end{figure}

\newpage
\begin{figure}[ht]
    \centering
    \begin{subfigure}[b]{0.48\linewidth}
        \centering
        \includegraphics[width=\linewidth]{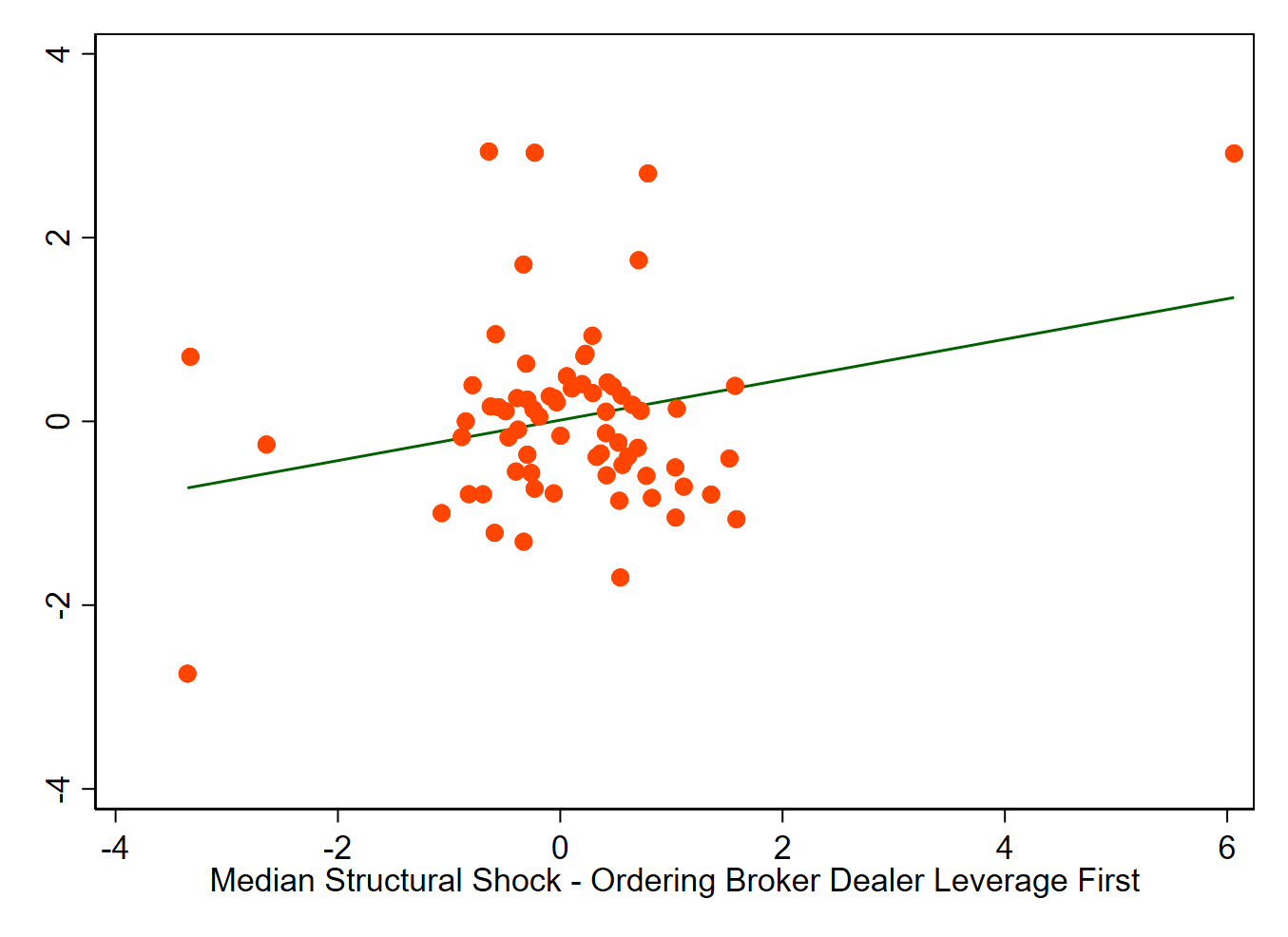}
        \caption{Broker-Dealer Leverage Ordered First \\ (Correlation = 0.257)}
        \label{fig:Quarterly_CoMovement_Leverage_First}
    \end{subfigure}
    \hfill
    \begin{subfigure}[b]{0.48\linewidth}
        \centering
        \includegraphics[width=\linewidth]{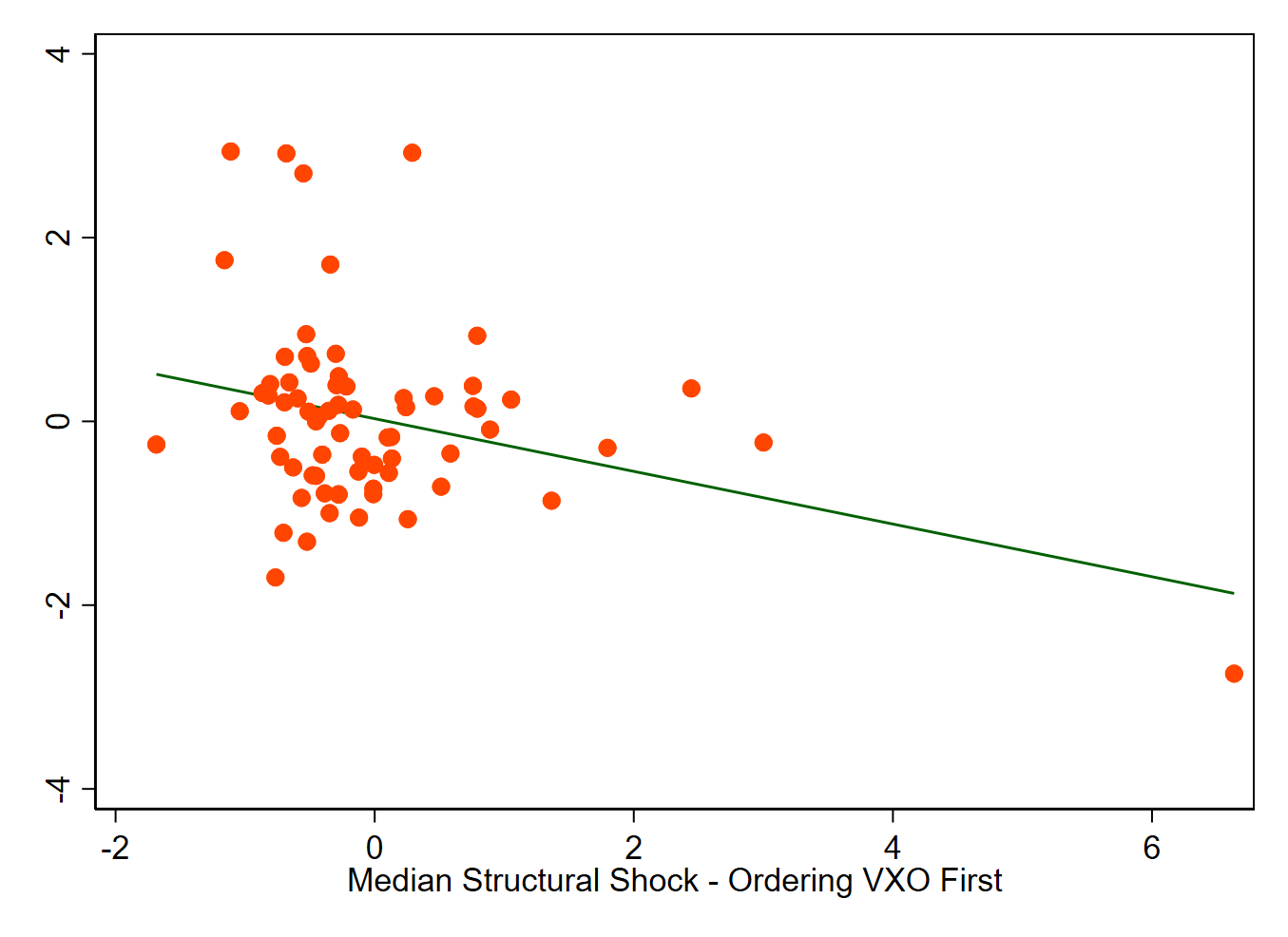}
        \caption{VXO Ordered First \\ (Correlation = -0.322)}
        \label{fig:Quarterly_CoMovement_VXO_First}
    \end{subfigure}
    \caption{Correlation of the global credit-supply shock with alternative financial shocks.}
    \floatfoot{Each panel plots the purged credit-supply shock against the median structural shock implied     by a Cholesky-identified VAR in which broker-dealer leverage or the VXO is ordered first.}
    \label{fig:CoMovement_Other_Financial_Shocks}
\end{figure}

%%%%%%%%%%%%%%%%%%%%%%%%%%%%%%%%%%%%%%%%%%%%
\newpage
\section{Microeconomic Transmission - Additional Figures} \label{sec:appendix_micro_results}

\subsection{Bank Level Results  - Additional Figures}

\noindent
\begin{figure}[ht]
\centering
\includegraphics[width=0.85\linewidth]{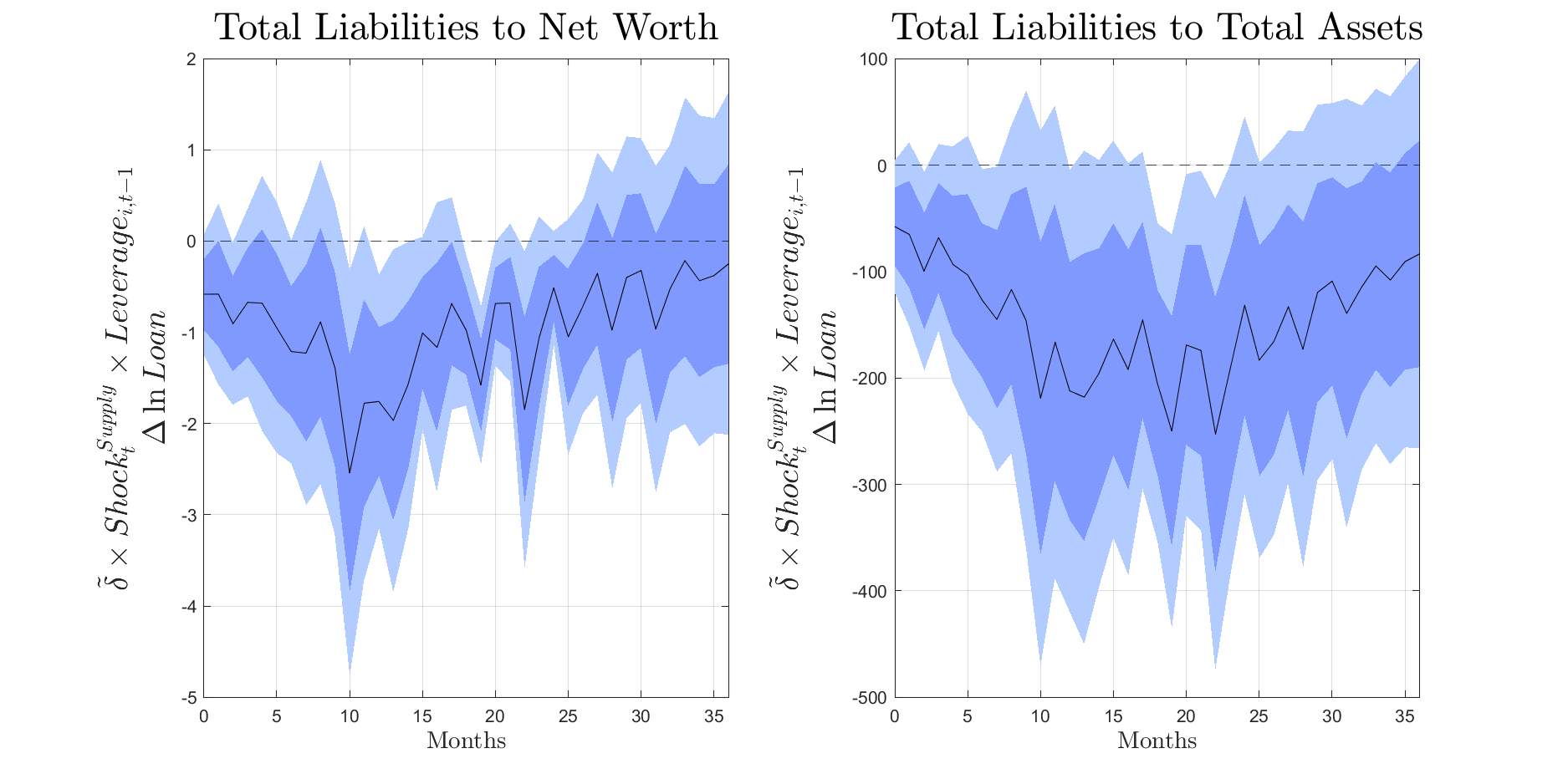}
\caption{Bank-Level Responses to Global Credit Supply Shocks \\ Time Fixed Effects Specification}
\label{fig:Ratios_Leverage_TimeFE}
\floatfoot{
\textbf{Note:} The figure reports the interaction coefficients $\tilde{\delta}_h$ from the time fixed effects specification: $
\Delta \ln(\text{Loan}_{i,c,t+h})
= \tilde{\delta}_h\, \text{Shock}^{\text{Supply}}_t \times \text{Leverage}_{i,t-1}
+ \mu_{i,h} + \mu_{c,h} + \tau_t
+ \Gamma^{\text{Bank}}_{i,t-1}
+ \epsilon_{i,c,t+h},$ where $\tau_t$ denotes time fixed effects absorbing all contemporaneous domestic and global shocks. 
The left panel uses leverage as liabilities-to-net-worth; the right panel uses liabilities-to-assets. Solid lines denote point estimates; shaded areas represent 68\% and 90\% confidence intervals.}
\end{figure}

\begin{figure}[ht]
\centering
\includegraphics[width=0.95\linewidth]{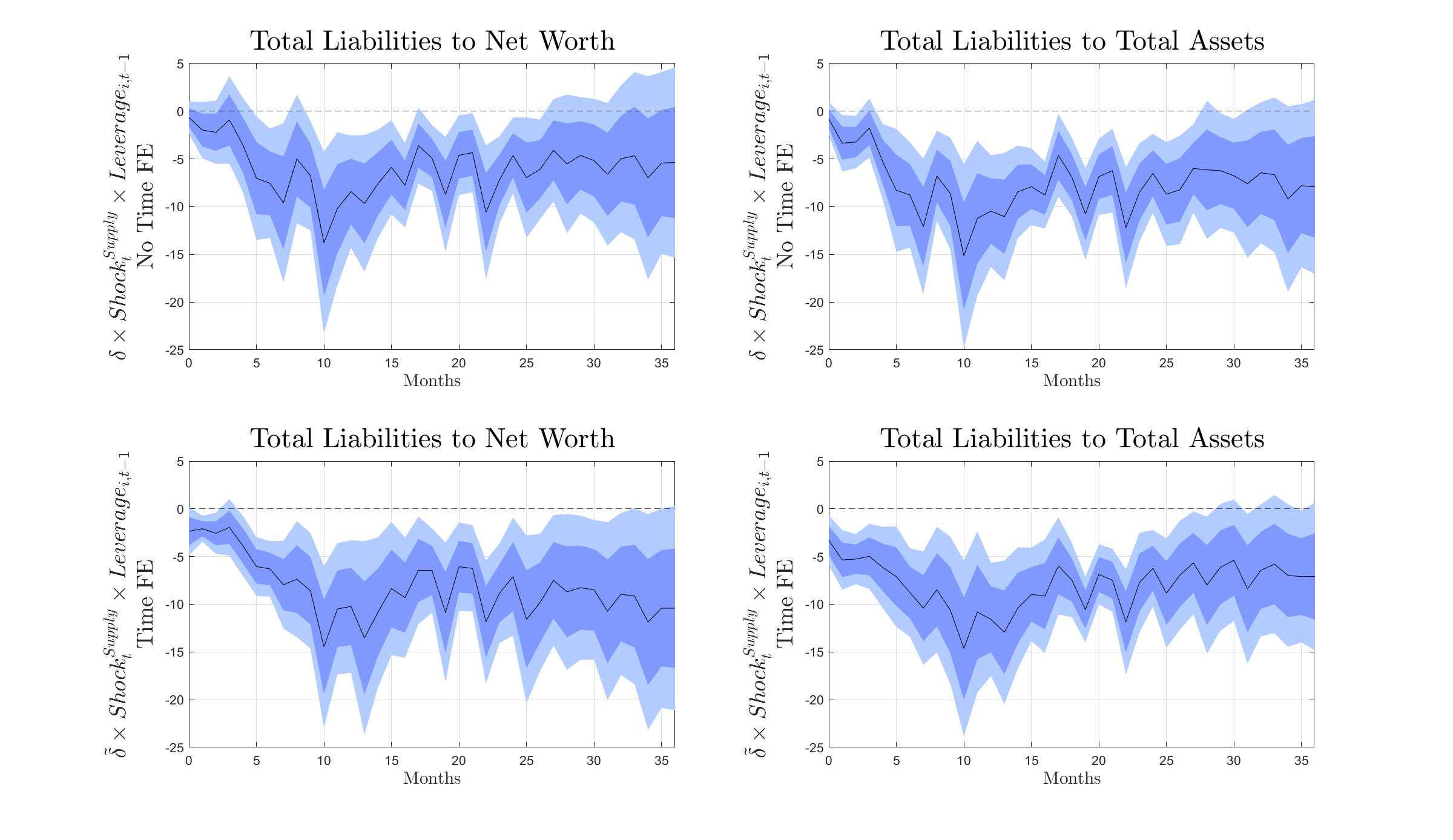}
\caption{Bank-Level Responses Using Within-Bank Normalized Leverage}
\label{fig:Ratios_Leverage_2by2_Norm}
\floatfoot{\textbf{Note:} Estimates of $\delta_h$ using leverage standardized at the bank level. Solid lines denote point estimates; shaded areas represent 68\% and 90\% confidence intervals.}
\end{figure}

\noindent
\textbf{Additional interactions at the bank level.} Figures that present the interaction coefficients between the negative credit supply shock and four variables that reflect banks' financial liquidity: (i) ratio of liquid assets to total assets, (ii) ratio of net worth capitalization to required regulatory capital, (iii) ratio of demand deposits to total deposits, and, (iv) share of non-performing loans. 
\begin{figure}[ht]
    \centering
    \includegraphics[width=0.85\linewidth]{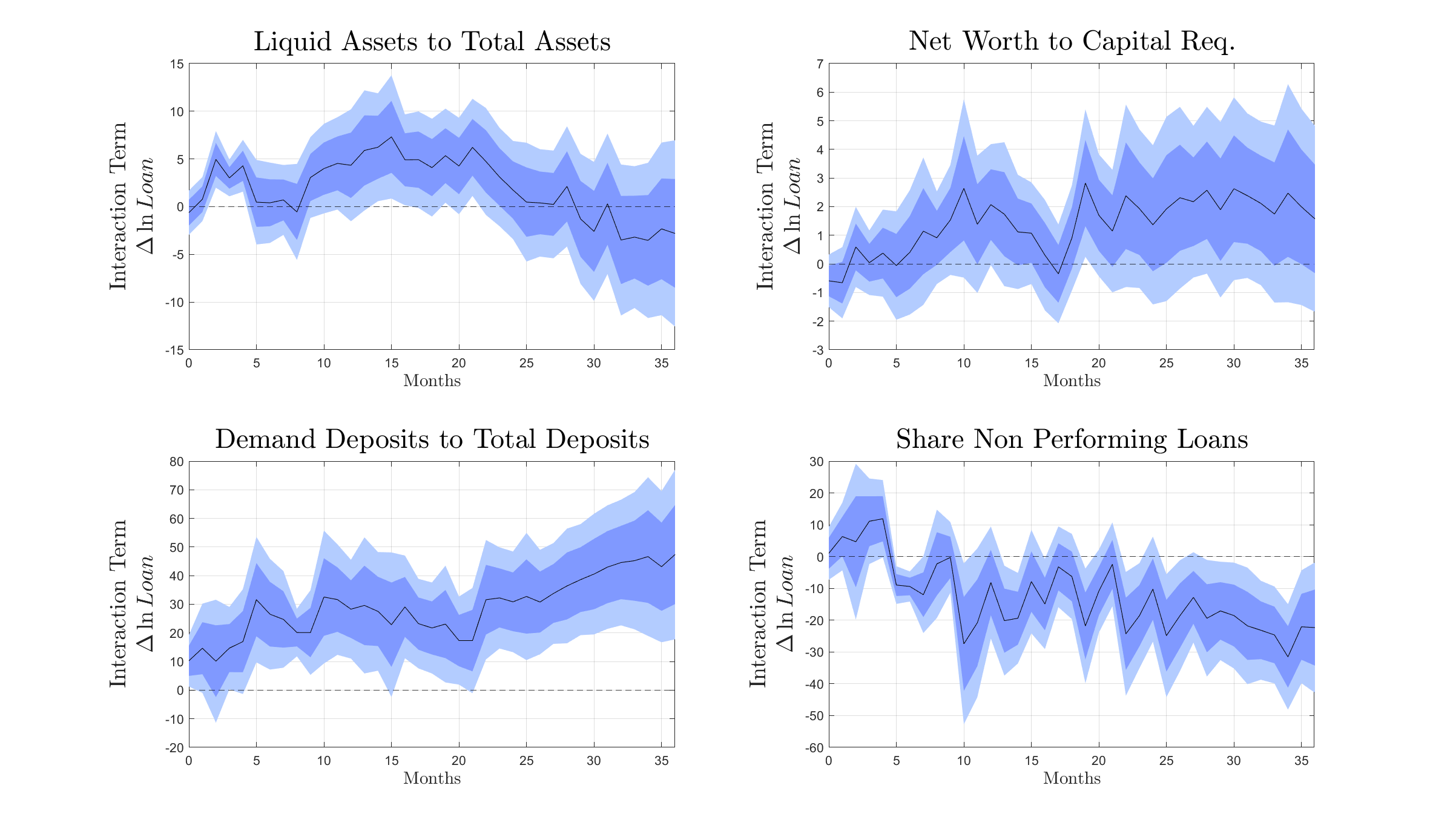}
    \caption{Bank Level Results - Additional Interactions}
    \label{fig:Ratios_Other}
    \floatfoot{\textbf{Note:} The panel is comprised of four panels ordered in two rows and two columns, which present the estimated coefficient $\delta_h$ from estimating Equation \ref{eq:Regression_Bank_Level}, replacing leverage with each of the following variables: share of liquid assets under 30 day maturity to total assets (top-left panel); net worth to regulatory capital requirements (top-right panel); share of demand deposits to total deposits (bottom-left panel); share of non performing loans to total loans (bottom-right panel). The solid black line presents the point estimate, the dark blue shaded area represents the 68\% confidence interval and the light blue shaded area represents the 90\% confidence interval.}
\end{figure}

\begin{figure}[ht]
    \centering
    \includegraphics[width=0.95\linewidth]{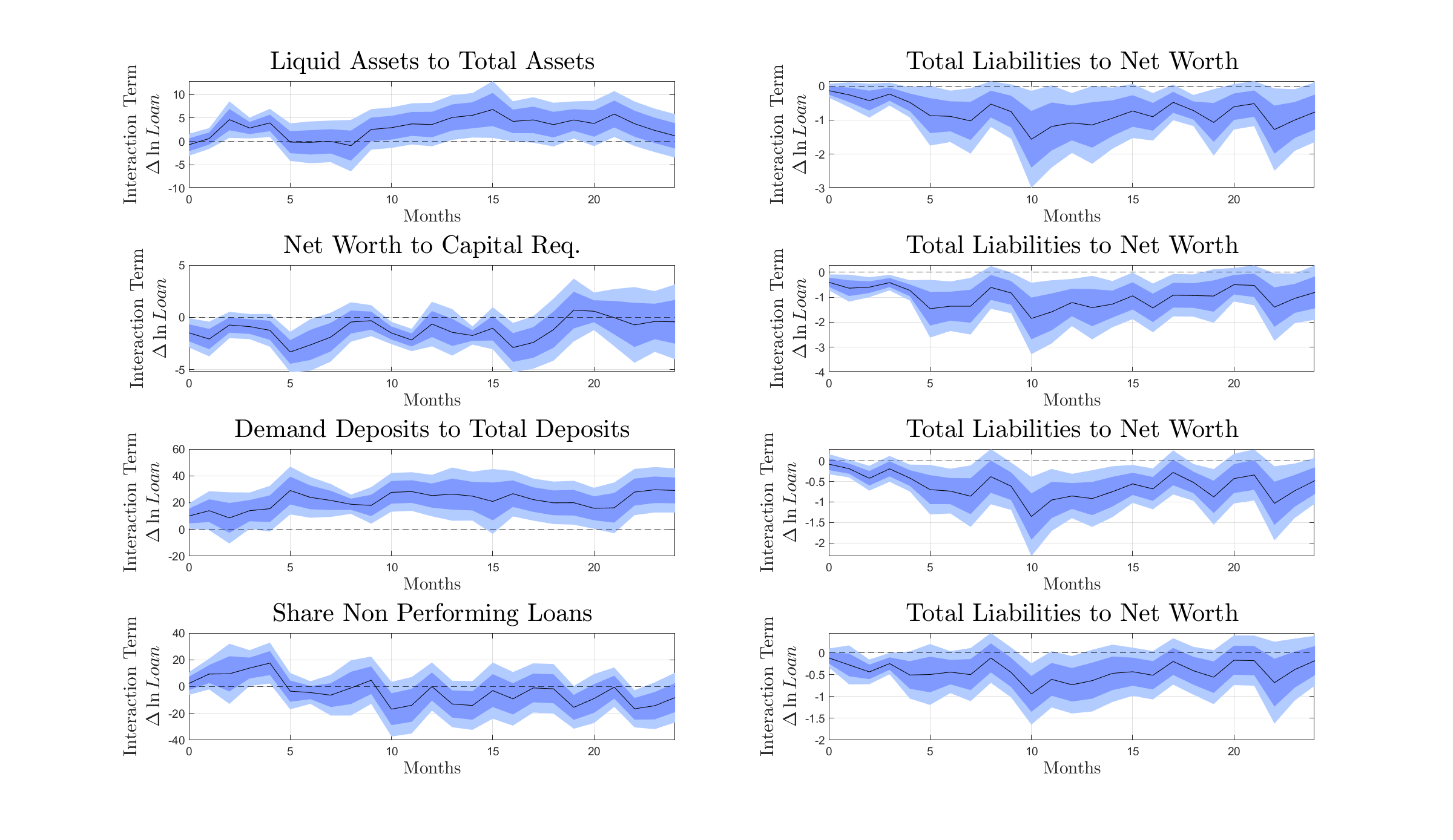}
    \caption{Bank Level Results - Leverage \& Additional Interactions}
    \label{fig:Ratios_Other_Double}
    \floatfoot{\textbf{Note:} The panel is comprised of eight panels ordered in four rows and two columns. which present the estimated coefficient $\delta_h$ from estimating Equation \ref{eq:Regression_Bank_Level_Double} with the double interaction terms. Apart from including leverage we include the following four variables: share of liquid assets under 30 day maturity to total assets; net worth to regulatory capital requirements; share of demand deposits to total deposits; share of non performing loans to total loans. The solid black line presents the point estimate, the dark blue shaded area represents the 68\% confidence interval and the light blue shaded area represents the 90\% confidence interval.} 
\end{figure}

%%%%%%%%%%%%%%%%%%%%%%%%%%%%%%%%%%%%%%%%%%%%%%%%%%%%%%%%%%%%%%%%%
%%% Bank Level Domestic Capital Bank
\newpage
\begin{figure}[ht]
    \centering
    \includegraphics[width=0.9\linewidth]{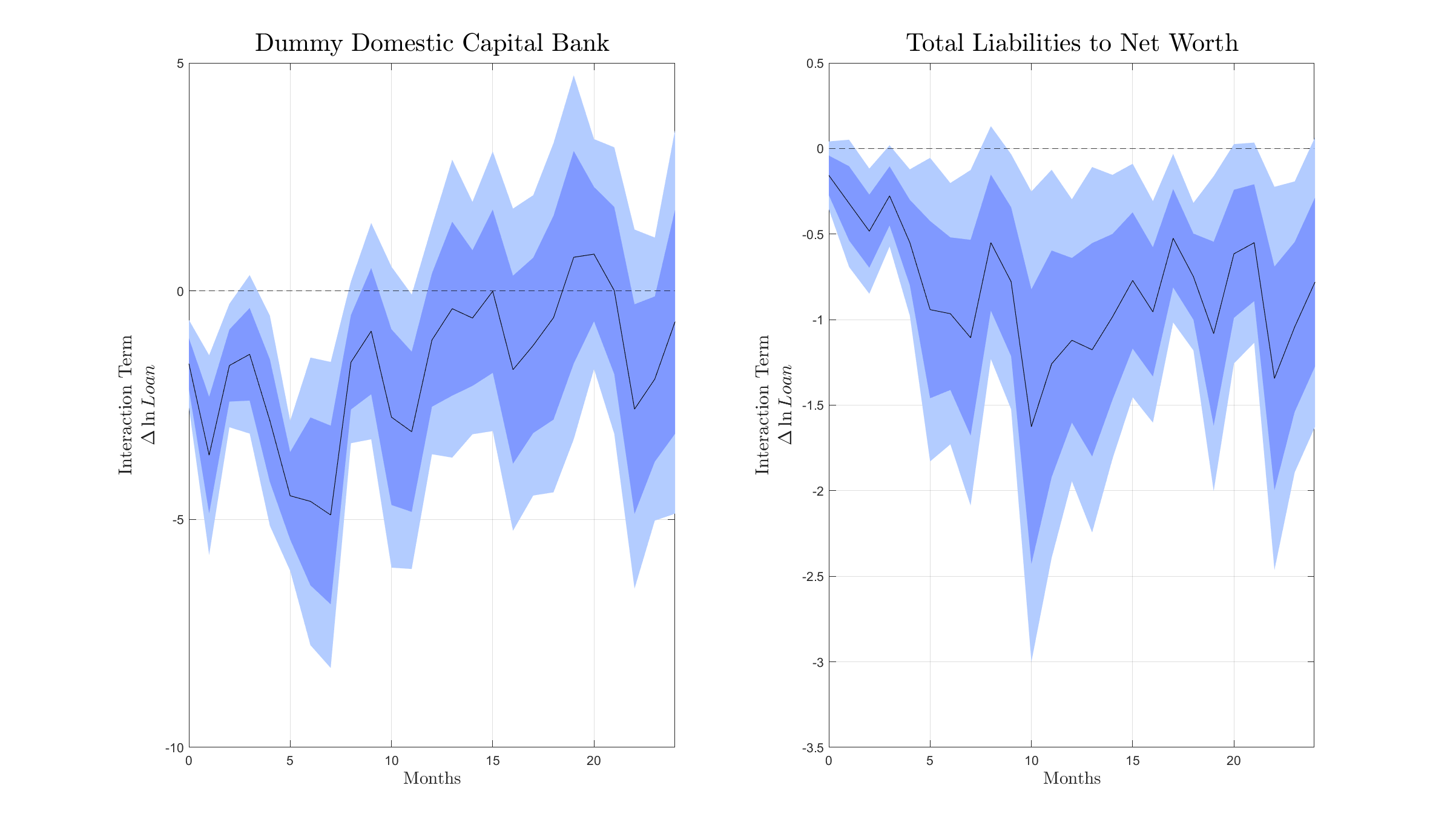}
    \caption{Bank Level Results - Role of Domestic Banks}
    \label{fig:Ratios_Other_Nacional_Double}
    \floatfoot{\textbf{Note:} The panel is comprised of two panels ordered in one row and two columns, which present the estimated coefficient $\delta_h$ from estimating Equation \ref{eq:Regression_Bank_Level_Double} with the double interaction terms. Apart from including leverage we include the interaction between the shock and a dummy variable that takes the value of 1 if the bank is of domestic capital and zero otherwise. The solid black line presents the point estimate, the dark blue shaded area represents the 68\% confidence interval and the light blue shaded area represents the 90\% confidence interval.}

\end{figure}

%%%%%%%%%%%%%%%%%%%%%%%%%%%%%%%%%%%%%%%%%%%%%%%%%%%%%%%%%%%%%%%%%
%%% Bank Level Sign-Dependency
\newpage

\noindent
\textbf{Bank-level sign dependency results.} We present the results for the sign-dependent bank-level results. We argue that the resulting responses are greater in magnitude and with relatively tighter confidence intervals for expansionary shocks. This is particularly evident for the more stringent econometric regression with time fixed effects, in Equation \ref{eq:Regression_Bank_Level_Sign_Time_FE}
\begin{figure}[ht]
    \centering
    \includegraphics[width=0.75\linewidth]{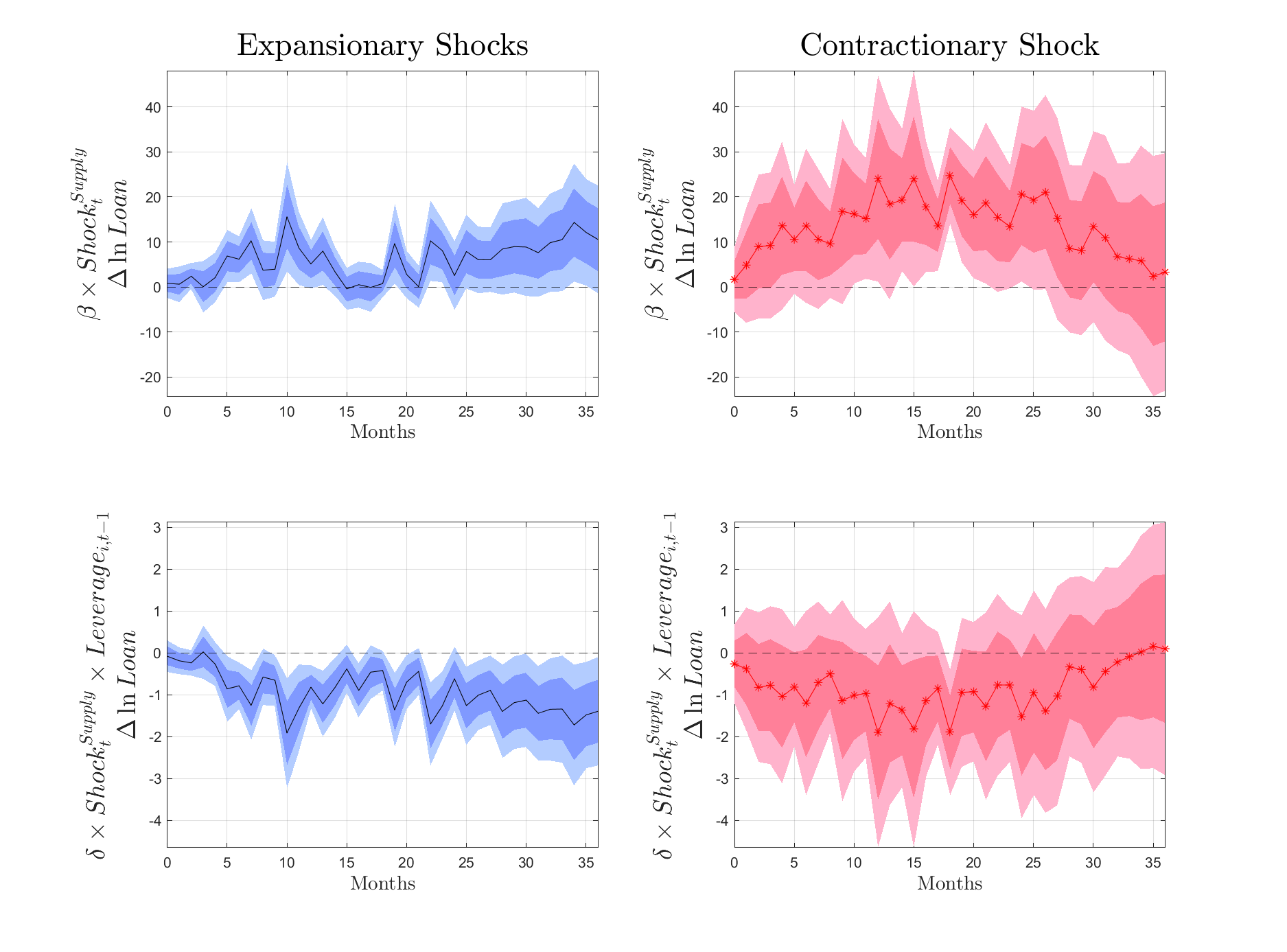}
    \caption{Bank Level Results - Sign Dependency \\ Total Liabilities to Net Worth}
    \label{fig:APA_Sign_Results}
    \floatfoot{\textbf{Note:} The panel is comprised of four panels ordered in two rows and two columns, which present the estimated coefficient $\delta_h$ from estimating Equation \ref{eq:Regression_Bank_Level_Sign} using the measure of leverage as the ratio of total liabilities to net worth. The left column presents the results for a negative or contractionary credit supply shock. The solid black line presents the point estimate, the dark blue shaded area represents the 68\% confidence interval and the light blue shaded area represents the 90\% confidence interval. The right column presents the results for a positive or expansionary credit supply shock. The red-starred line presents the point estimate, the dark red shaded area represents the 68\% confidence interval and the light red shaded area represents the 90\% confidence interval.}
\end{figure}
\begin{figure}[ht]
    \centering
    \includegraphics[width=0.75\linewidth]{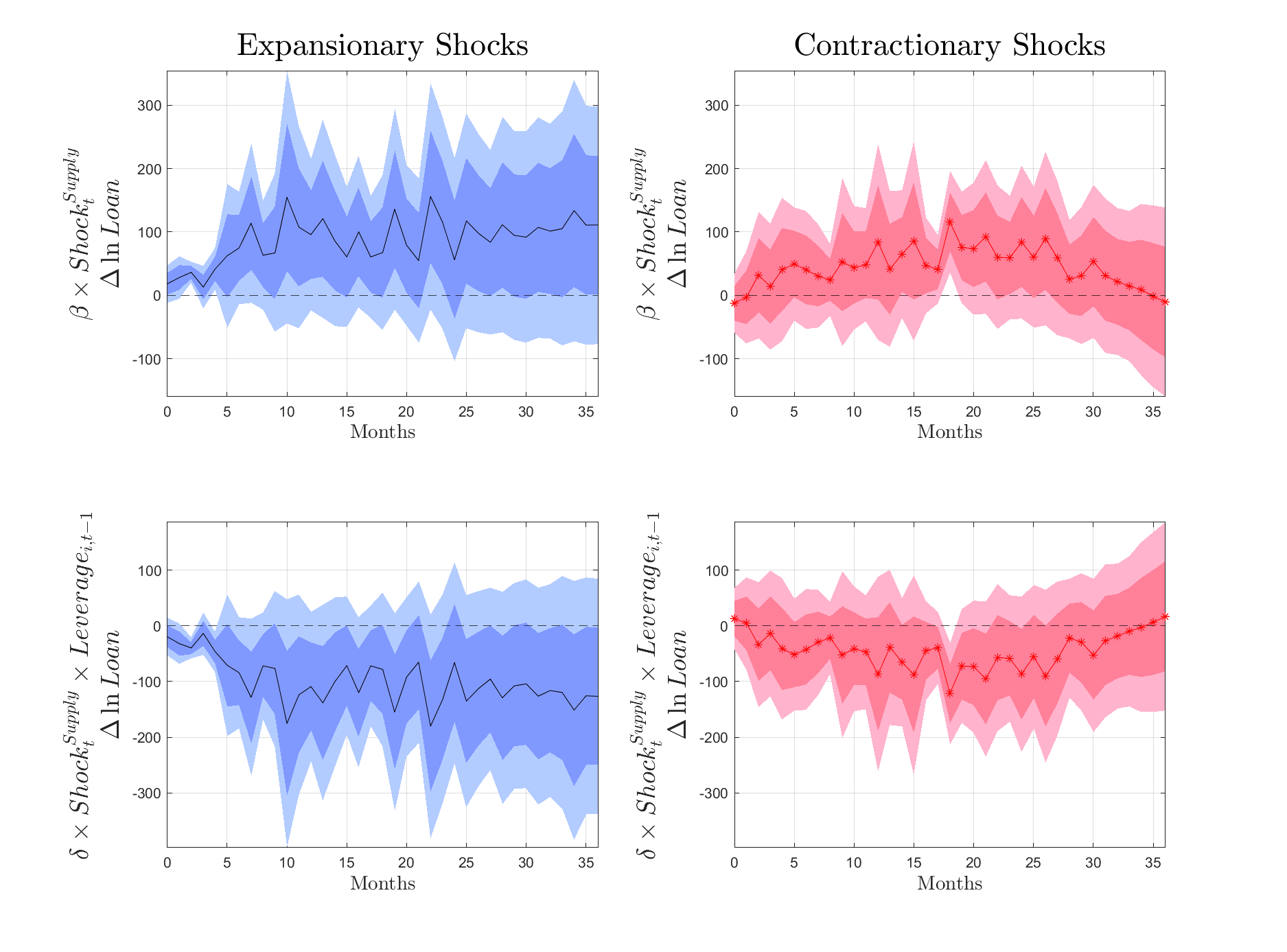}
    \caption{Bank Level Results - Sign Dependency \\ Total Liabilities to Total Assets}
    \label{fig:PAS_Sign_Results}
    \floatfoot{\textbf{Note:} The panel is comprised of four panels ordered in two rows and two columns, which present the estimated coefficient $\delta_h$ from estimating Equation \ref{eq:Regression_Bank_Level_Sign} using the measure of leverage as the ratio of total liabilities to net worth. The left column presents the results for a negative or contractionary credit supply shock. The solid black line presents the point estimate, the dark blue shaded area represents the 68\% confidence interval and the light blue shaded area represents the 90\% confidence interval. The right column presents the results for a positive or expansionary credit supply shock. The red-starred line presents the point estimate, the dark red shaded area represents the 68\% confidence interval and the light red shaded area represents the 90\% confidence interval.}
\end{figure}
% TIME FE
\begin{figure}[ht]
    \centering
    \includegraphics[width=0.75\linewidth]{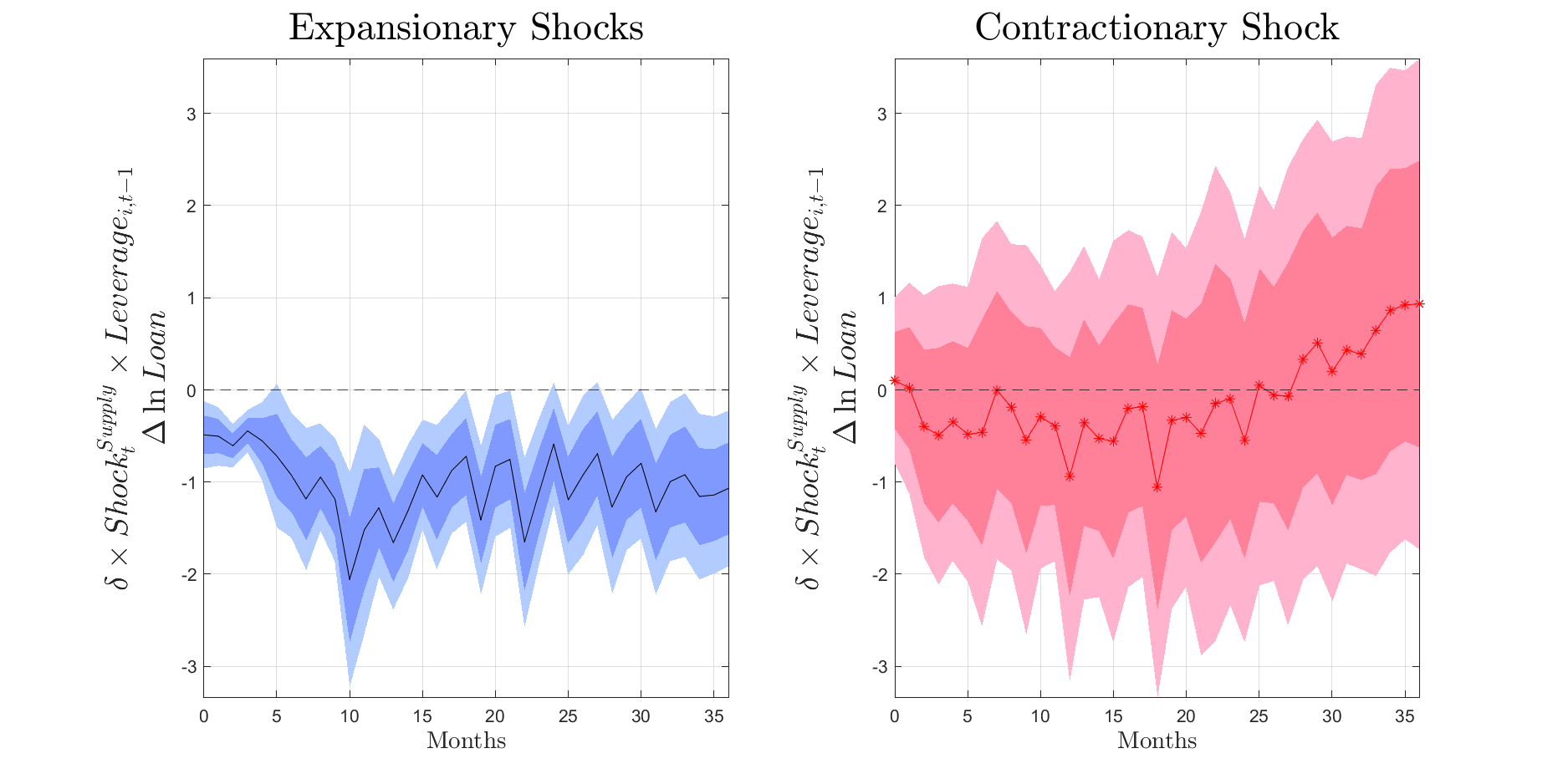}
    \caption{Bank Level Results - Sign Dependency \\  Total Liabilities to Net Worth}
    \label{fig:APA_Sign_Results_Time_FE}
    \floatfoot{\textbf{Note:} The panel is comprised of two panels ordered in one row and two columns, which present the estimated coefficient $\delta_h$ from estimating Equation \ref{eq:Regression_Bank_Level_Sign_Time_FE} using the measure of leverage as the ratio of total liabilities to net worth. The left column presents the results for a negative or contractionary credit supply shock. The solid black line presents the point estimate, the dark blue shaded area represents the 68\% confidence interval and the light blue shaded area represents the 90\% confidence interval. The right column presents the results for a positive or expansionary credit supply shock. The red-starred line presents the point estimate, the dark red shaded area represents the 68\% confidence interval and the light red shaded area represents the 90\% confidence interval.}
\end{figure}

\begin{figure}[ht]
    \centering
    \includegraphics[width=0.75\linewidth]{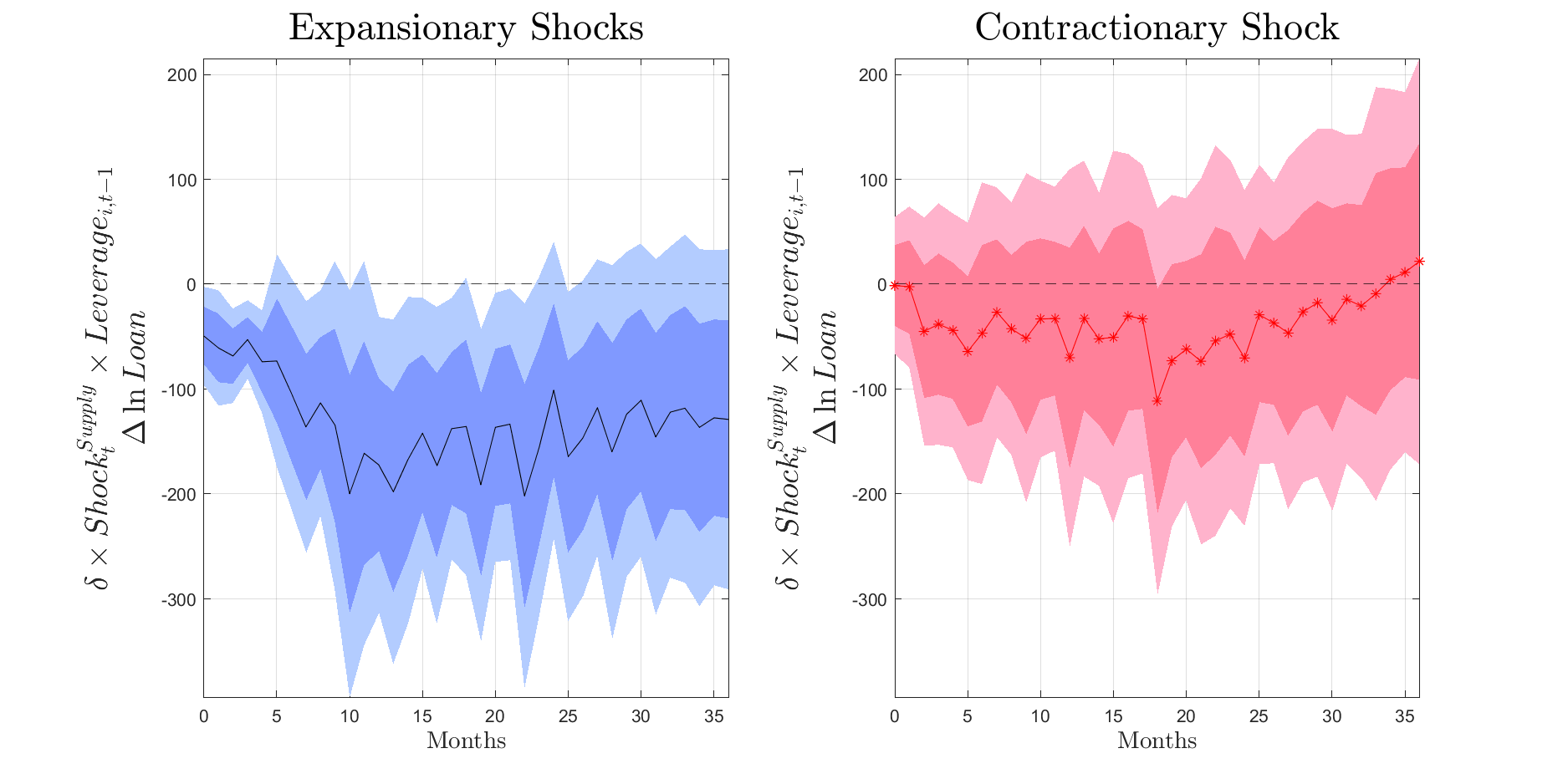}
    \caption{Bank Level Results - Sign Dependency \\ Total Liabilities to Total Assets}
    \label{fig:PAS_Sign_Results_Time_FE}
    \floatfoot{\textbf{Note:} The panel is comprised of two panels ordered in one row and two columns, which present the estimated coefficient $\delta_h$ from estimating Equation \ref{eq:Regression_Bank_Level_Sign_Time_FE} using the measure of leverage as the ratio of total liabilities to net worth. The left column presents the results for a negative or contractionary credit supply shock. The solid black line presents the point estimate, the dark blue shaded area represents the 68\% confidence interval and the light blue shaded area represents the 90\% confidence interval. The right column presents the results for a positive or expansionary credit supply shock. The red-starred line presents the point estimate, the dark red shaded area represents the 68\% confidence interval and the light red shaded area represents the 90\% confidence interval.}
\end{figure}

%%%%%%%%%%%%%%%%%%%%%%%%%%%%%%%%%%%%%%%%%%%%%%%%%%%%%
\subsection{Firm Level Results  - Additional Figures}

\begin{figure}[ht]
    \centering
    \includegraphics[width=0.75\linewidth]{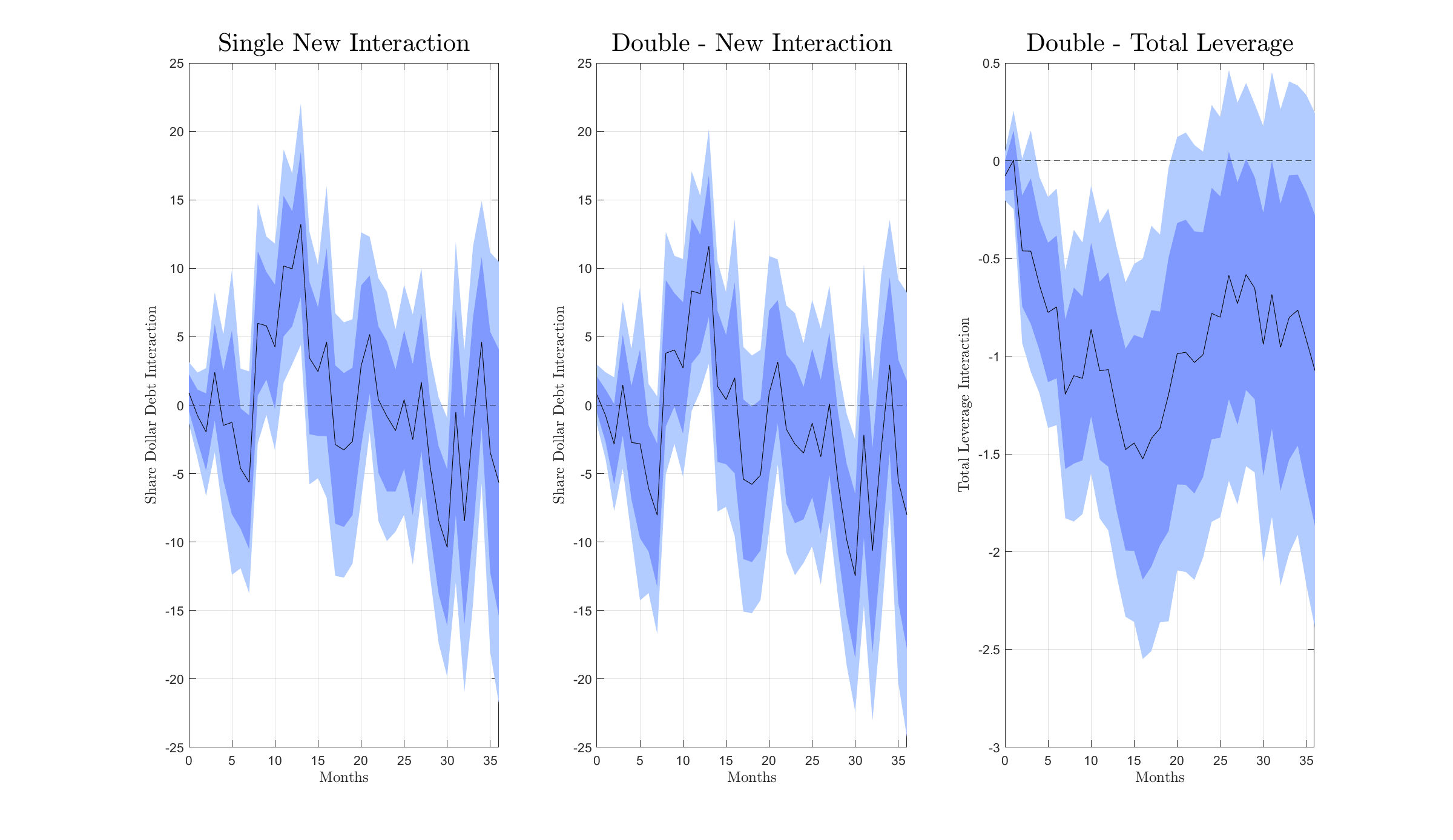}
    \caption{Firm Level Results - Share of Dollar Debt}
    \label{fig:ShareDoll_Double}
    \floatfoot{\textbf{Note:} The panel is comprised of three panels, organized in one row and three columns. The left column presents the estimated interaction coefficient of estimating Equation \ref{eq:Benchmark_LP_Firm} replacing total leverage. The middle and right columns present the results for the new interaction and total leverage, as estimated by Equation \ref{eq:Double_Interaction_LP_Firm}. The solid black line presents the point estimate, the dark blue shaded area represents the 68\% confidence interval and the light blue shaded area represents the 90\% confidence interval. The right column presents the results for a positive or expansionary credit supply shock. The red-starred line presents the point estimate, the dark red shaded area represents the 68\% confidence interval and the light red shaded area represents the 90\% confidence interval.}
\end{figure}

\begin{figure}[ht]
    \centering
    \includegraphics[width=0.75\linewidth]{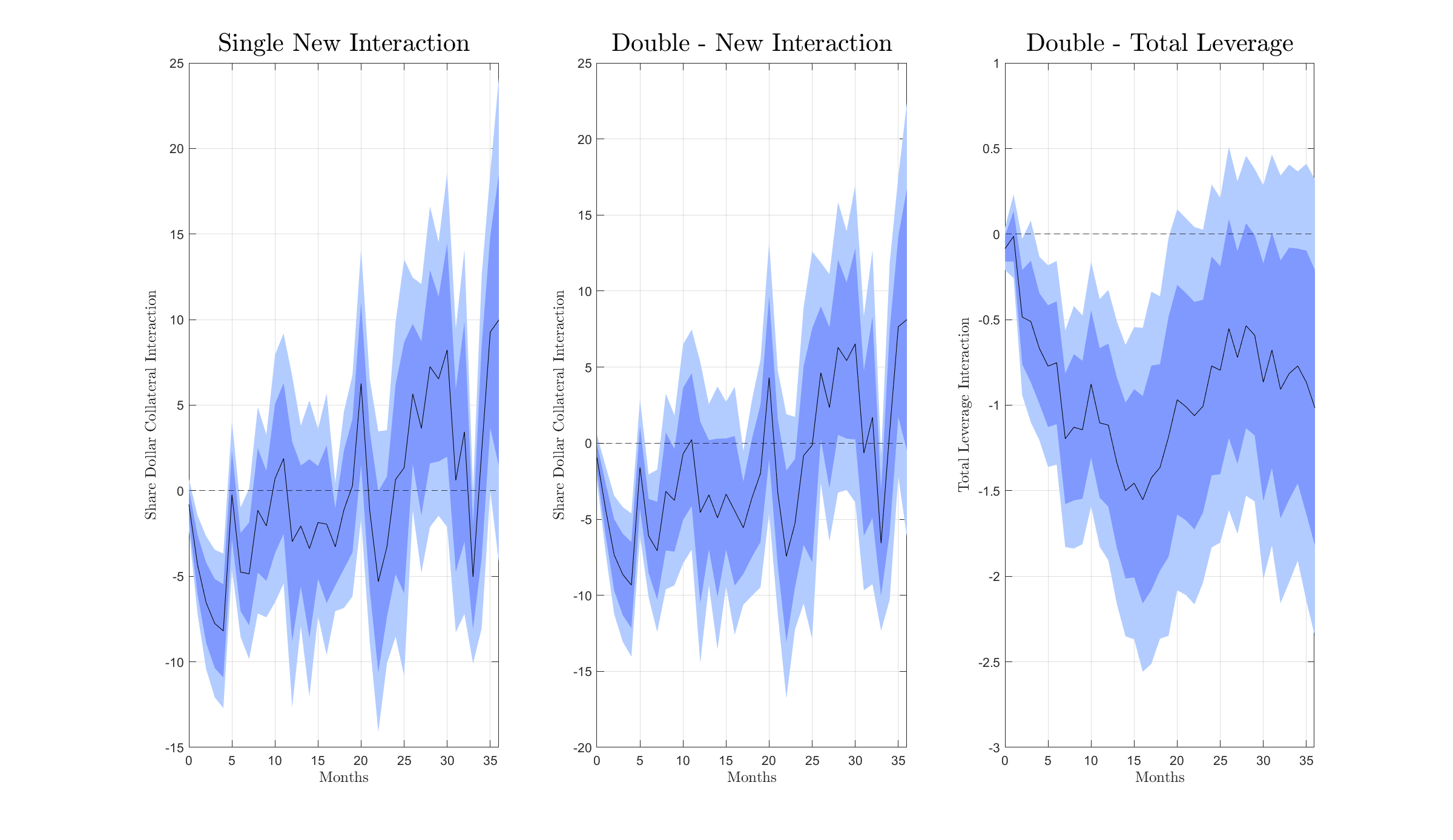}
    \caption{Firm Level Results - Share of Dollar Priced Collateral}
    \label{fig:ShareAssets_Double}
    \floatfoot{\textbf{Note:} The panel is comprised of three panels, organized in one row and three columns. The left column presents the estimated interaction coefficient of estimating Equation \ref{eq:Benchmark_LP_Firm} replacing total leverage. The middle and right columns present the results for the new interaction and total leverage, as estimated by Equation \ref{eq:Double_Interaction_LP_Firm}. The solid black line presents the point estimate, the dark blue shaded area represents the 68\% confidence interval and the light blue shaded area represents the 90\% confidence interval. The right column presents the results for a positive or expansionary credit supply shock. The red-starred line presents the point estimate, the dark red shaded area represents the 68\% confidence interval and the light red shaded area represents the 90\% confidence interval.}
\end{figure}

%%%%%%%%%%%%%%%%% Results across Sectors
\newpage
\begin{figure}[ht]
    \centering
    \includegraphics[width=0.75\linewidth]{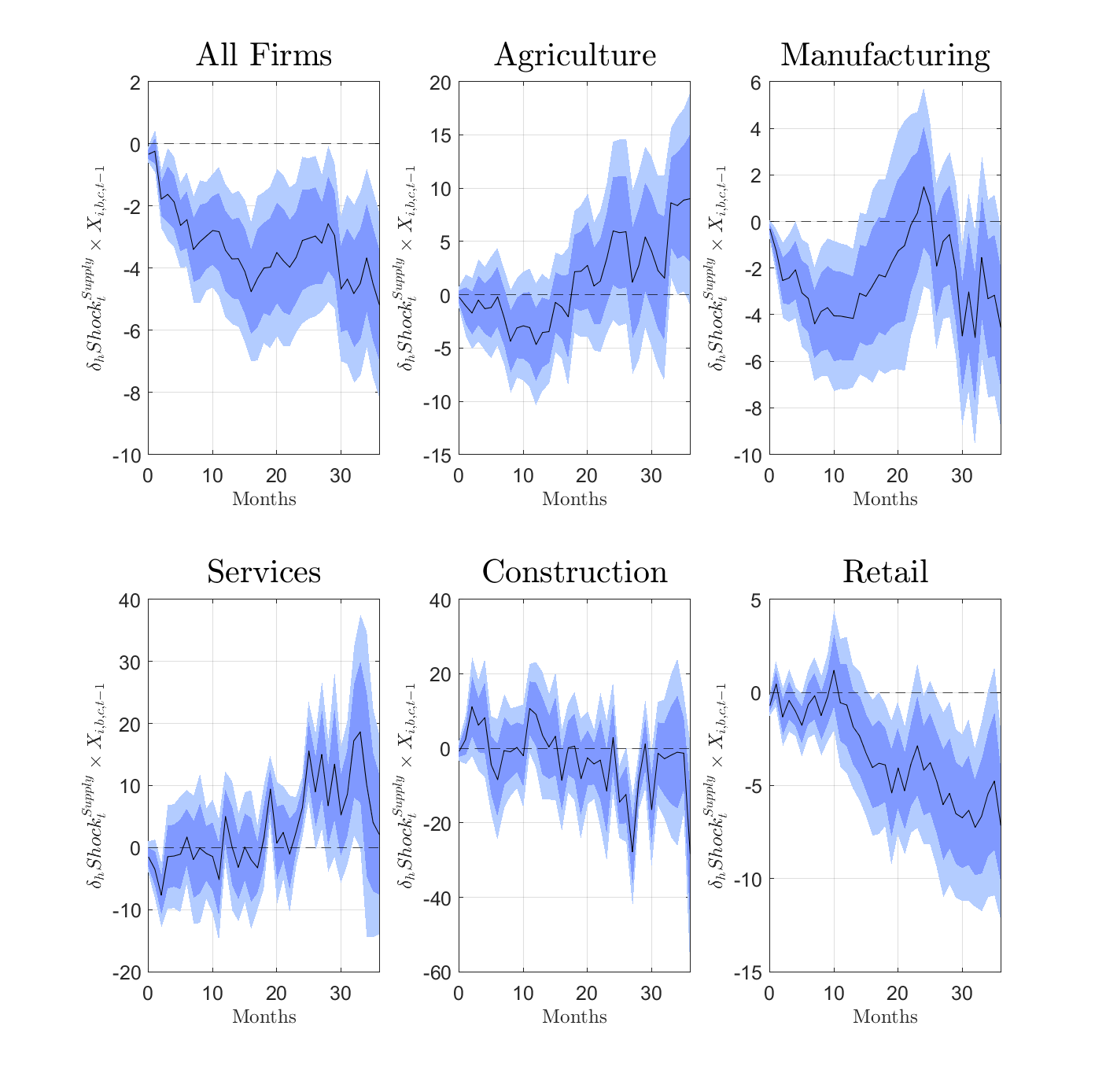}
    \caption{Firm-Level Heterogeneity in Leverage Effects Across Sectors}
    \label{fig:Total_Leverage_By_Sector}
    \floatfoot{\textbf{Note:} 
    The figure consists of six panels arranged in two rows and three columns, reporting the estimated heterogeneous response coefficient $\delta_h$ from Equation~\ref{eq:Benchmark_LP_Firm}. Each panel corresponds to a different self-reported main sector of economic activity: Agriculture, Manufacturing, Services, Construction, and Retail. Because sector information is not available for all borrowers, the top-left panel reports results for the subsample of all firms that do report a sector. The solid black line shows the point estimate, while the dark and light blue shaded areas represent the 68\% and 90\% confidence intervals, respectively.}
\end{figure}

\begin{figure}[ht]
    \centering
    \includegraphics[width=0.75\linewidth]{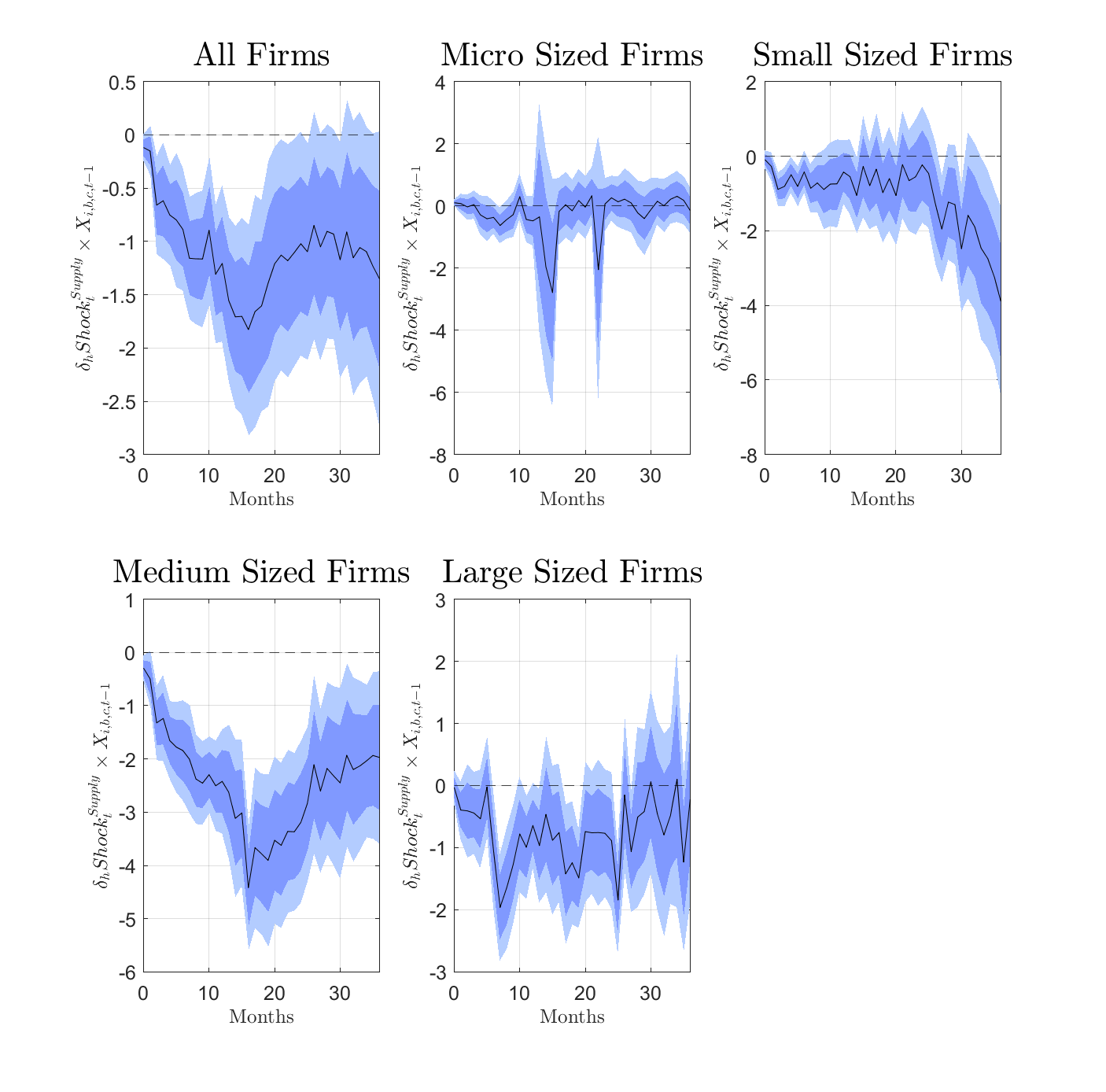}
    \caption{Firm-Level Heterogeneity in Leverage Effects Across Size Categories}
    \label{fig:Total_Leverage_By_Size}
    \floatfoot{\textbf{Note:}
    The figure reports the estimated heterogeneous response coefficient $\delta_h$ from Equation~\ref{eq:Benchmark_LP_Firm} separately for firms in each MiPyME size category: Micro, Small, Medium, and Large. Because size is self-reported and not available for all borrowers, the top-left panel shows results for the subsample of all firms that do report size information. The solid black line shows the point estimate; dark and light blue shaded areas represent the 68\% and 90\% confidence intervals, respectively.}
\end{figure}

%%%%%%%%%%%%%%%%% Results After GFC
\newpage
\noindent
\begin{figure}[ht]
    \centering
    \includegraphics[width=0.75\linewidth]{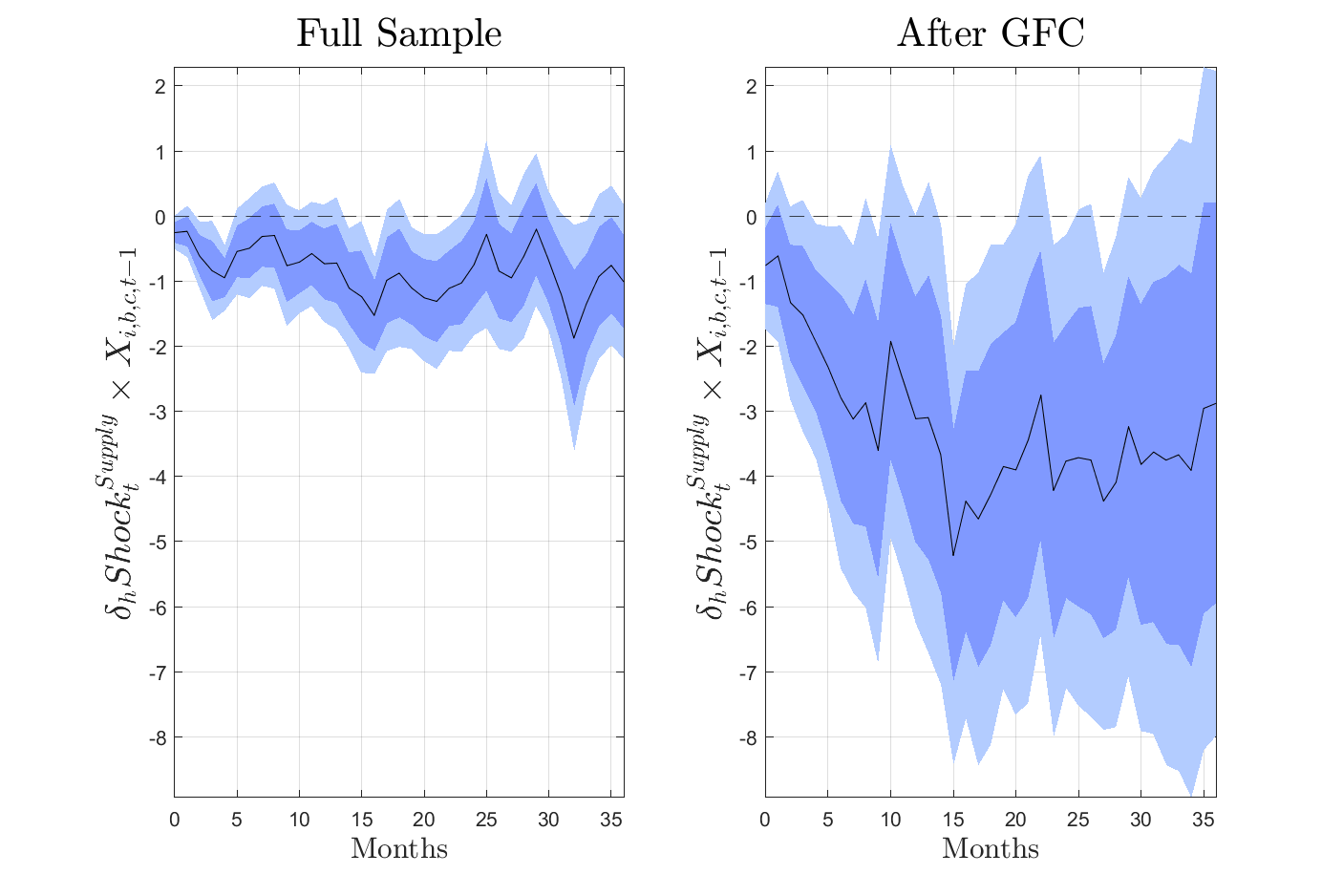}
    \caption{Firm Level Results - After the GFC}
    \label{fig:Comparison_After2010}
    \floatfoot{\textbf{Note:} The panel is comprised of two panels ordered in one row and two columns, which present the estimated coefficient $\delta_h$ from estimating Equation \ref{eq:Benchmark_LP_Firm}. The left column presents our benchmark results for the full sample. The right column presents the results for a sample starting after the Great Financial Crisis, in January 2010. The solid black line presents the point estimate, the dark blue shaded area represents the 68\% confidence interval and the light blue shaded area represents the 90\% confidence interval.}
\end{figure}

%%%%%%%%%%%%%%%%% 
\newpage
\noindent 
\begin{figure}[ht]
    \centering
    \includegraphics[width=0.85\linewidth]{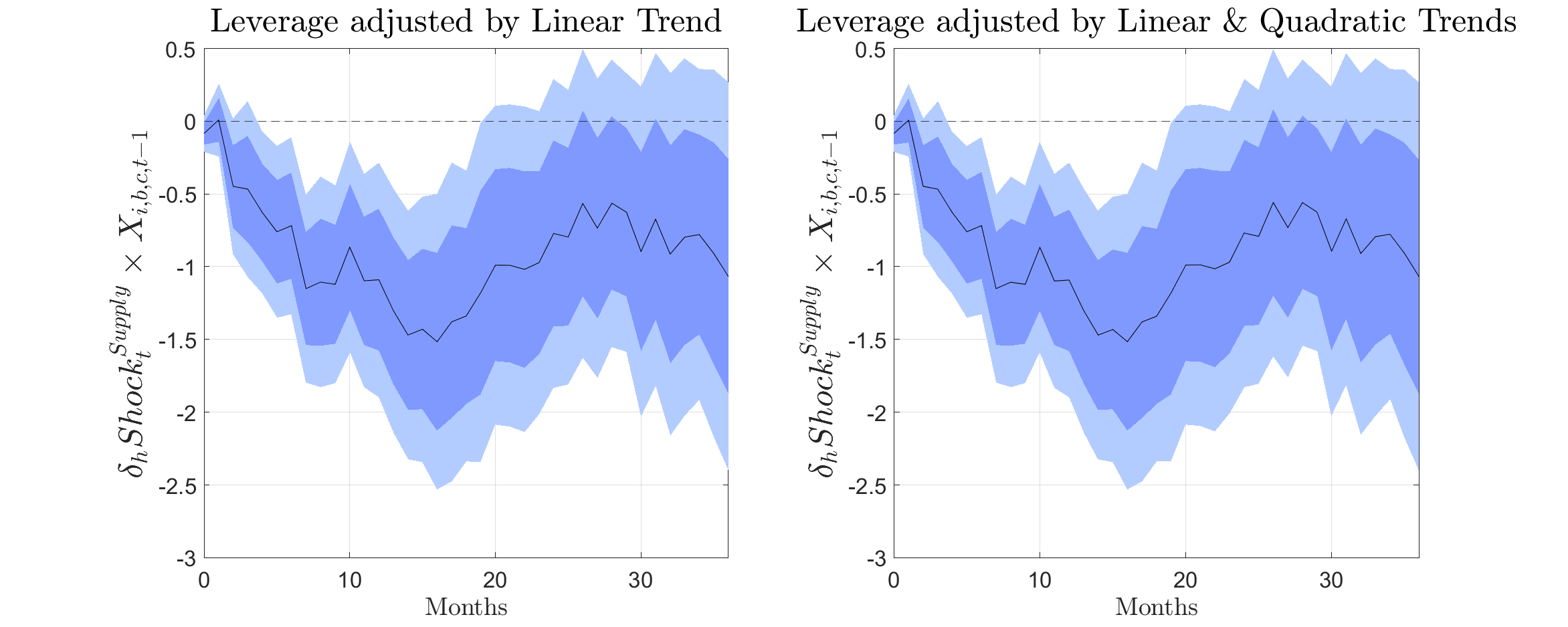}
    \caption{Firm Level Results - Leverage Adjusted by Age}
    \label{fig:Total_Leverage_Age_Adjusted}
    \floatfoot{\textbf{Note:} The panel is comprised of two panels ordered in one row and two columns, which present the estimated coefficient $\delta_h$ from estimating Equation \ref{eq:Benchmark_LP_Firm}. The left column presents our results when using a measure of leverage detrended with a linear age trend. The right column presents our results when using a measure of leverage detrended with a linear and quadratic age trend. The solid black line presents the point estimate, the dark blue shaded area represents the 68\% confidence interval and the light blue shaded area represents the 90\% confidence interval.}
\end{figure}
\end{document}